# SUPERCONDUCTING MATERIALS – A TOPICAL OVERVIEW

**Roland Hott[1], Reinhold Kleiner[2], Thomas Wolf[1] & Gertrud Zwicknagl[3]**

[1] Forschungszentrum Karlsruhe, Institut für Festkörperphysik,
P. O. Box 3640, 76021 Karlsruhe, GERMANY

[2] Physikalisches Institut - Experimentalphysik II, Universität Tübingen,
72076 Tübingen, GERMANY

[3] Institut für Mathematische Physik, Technische Universität Braunschweig,
38106 Braunschweig, GERMANY

## 1. INTRODUCTION

The phenomenon of superconductivity has not lost its fascination ever since its discovery in 1911. The flow of electric current without friction amounts to the realization of the old human dream of a perpetuum mobile. The ratio of resistance between the normal-conducting and the superconducting ("SC") state has been tested to exceed $10^{14}$, i. e., it is at least as large as between a usual insulator and copper as the best normal-conducting material [1].

But superconductivity is more than just the disappearance of resistance: The *Meißner effect,* the expulsion of magnetic fields from a superconductor, discovered in 1933, shows that superconductivity is a true thermodynamical state of matter since, in contrast to the situation for a merely perfect conductor, the expulsion is independent of the experimental history [1]. As the progress of cooling technique gave access to lower and lower temperatures, superconductivity established as common low-temperature instability of most, possibly all metallic systems (see Fig. 1). As the apparent $T \rightarrow 0$ K state of metals, the zero-entropy postulate of thermodynamics for this limit points to its nature as a macroscopic quantum state.

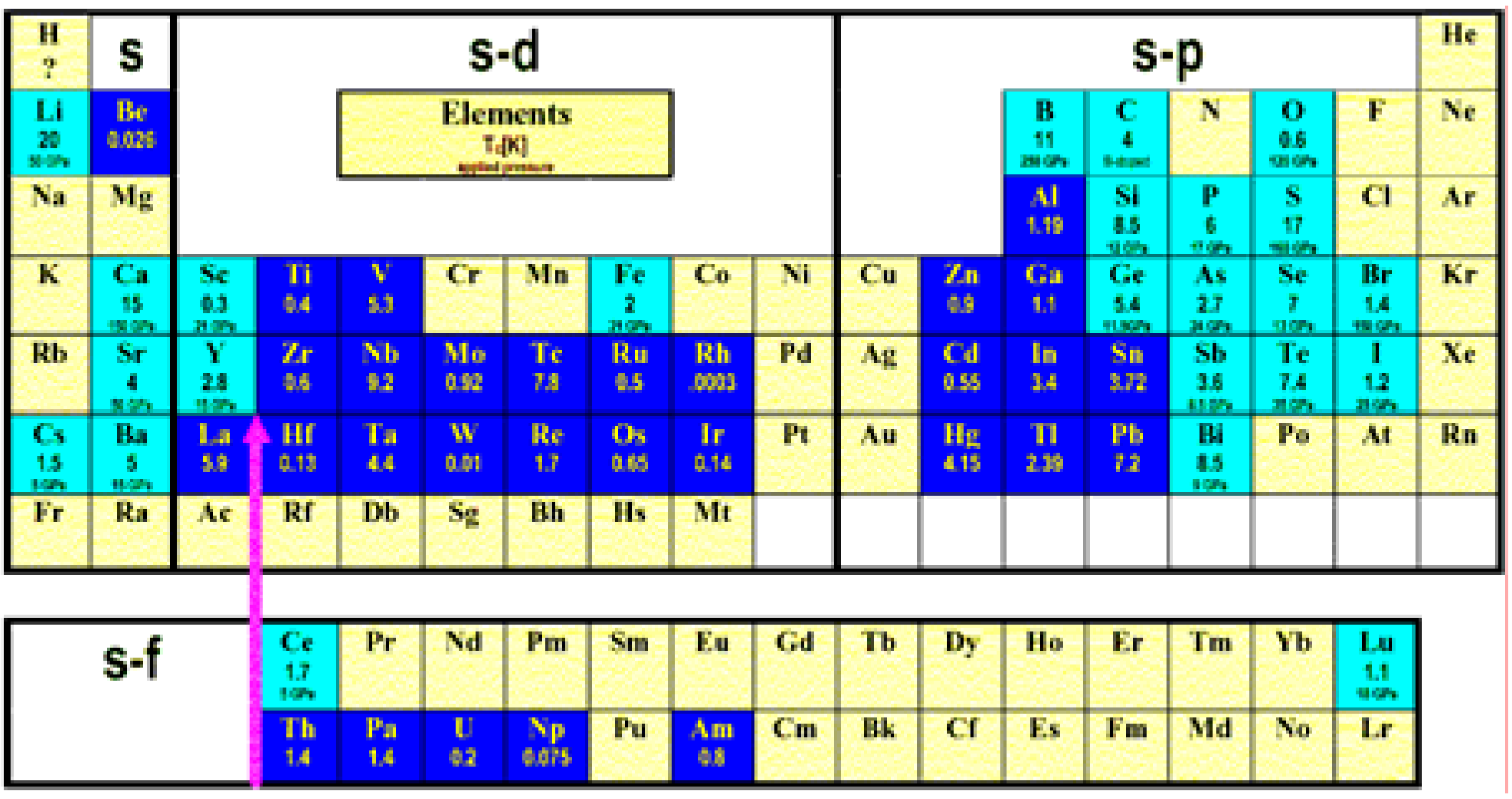

**Fig. 1** Periodic table with the distribution and $T_c$ [K] of chemical elements for which superconductivity has been observed with or without application of pressure [1,2,3,4].

The respective explanation of superconductivity by the *BCS theory* in 1957 [5] was a desperately awaited breakthrough of theoretical solid state physics. The starting point was the consideration that phonons introduce an attractive interaction between electrons close to the Fermi surface ("overscreening"). The key idea is that an (isotropic) attractive interaction leads here to bound electron pair states ("Cooper pairs"). These pair states are no longer obliged to obey the Fermi-Dirac statistics, which enforced the electrons to occupy high kinetic energy single particle states due to the Pauli principle. The energy gain of the SC state with respect to the normal state does not result from the small binding energy of the pairs but it is the condensation energy of the pairs merging into the macroscopic quantum state. It can be measured as an energy gap for electron excitations into single particle states.

In spite of the great impact of BCS theory[1], the discovery of *oxide "high-temperature superconductors" ("HTS")* in 1986 [6] made it very clear that new theoretical concepts will be required here. The problem is not the high $T_c$ of up to 138 K under normal pressure[2] [7], far above the pre-HTS record of 23 K [10]. However, in contrast to the "deep" Fermi sea of quasi-free electrons in the case of classical metals where the Cooper-pair condensed electrons amount only to a small part of the valence electron system ($k_B T_c << E_{Fermi}$), in these layered cuprate compounds there is only a "shallow" reservoir of charge carriers ($k_B T_c \sim E_{Fermi}$) which have to be introduced in the insulating antiferromagnetic (“AF”) stoichiometric parent compound by appropriate doping. The thus generated normal state corresponds to a “bad metal” in which Coulomb correlations strongly link the charge and spin degrees of freedom. This intrinsic proximity of metal-insulator, magnetic and SC transitions continues to present a great challenge to theory, which is sensibly more complicated than the classical superconductivity problem. The SC instability in cuprate HTS, as well as in the structurally and chemically related layered cobaltate and ruthenate compounds, is hence believed to stem predominantly from a magnetic and not from a phononic interaction as in the case of the classical metallic superconductors where magnetism plays only the role of an alternative, intrinsically antagonistic long range order instability.

---

[1] It had an impact not only on solid state physics but also on elementary particle physics where it was further developed to the idea of the Higgs mechanism of elementary particle mass generation.

[2] There is no theoretical argument why a textbook phonon BCS superconductor should not achieve such a high $T_c$. In the McMillan-Rowell formula [8], for example, as a commonly used theoretical-$T_c$ approximation, $T_c$ depends in a very sensitive way on the involved materials parameters. The HTS $T_c$ range is readily accessible with a still reasonable parameter choice [9].

A magnetic mechanism had been suspected already before the HTS discovery for *“Heavy-Fermion” superconductors*, discovered in 1979 [11]. In these intermetallic compounds the electronic degrees of freedom which are responsible for superconductivity are directly linked with magnetic moments of partially filled f-shells of Ce or U atoms[3]. The superconductivity below a typical $T_c \sim 1$ K seems to arise here from the delicate balance between the localized magnetic moments which try to imprint their magnetic signature on the shielding conduction electrons, and the conduction electrons which try to neutralize these magnetic moments by spin flipping, e. g., via the Kondo effect.

The search for *organic superconductors* had been boosted in the 1960s by the idea that conductive polymer chains with polarizable molecular groups may provide a highly effective Cooper pair coupling for electrons running along the polymer chains by means of an energy exchange via localized excitons [13]. Since the first discovery of an organic superconductor in 1980 [14] remarkable critical temperatures $T_c > 10$ K have been achieved [15]. However, the origin of superconductivity has turned out to be certainly far from the suggested excitonic mechanism. The electric conduction stems here from π-electrons in stacked aromatic rings, which form one- or two-dimensional delocalized electron systems. Similar to HTS, the restriction of the effective dimensionality[4] and strong Coulomb repulsion effects push the systems towards metal-insulator, magnetic and SC transitions.

*Fullerenes* ($C_{60}$, $C_{70}$, ...) attracted much attention since their discovery in 1985 as a third modification of elementary carbon. The superconductivity in $C_{60}$ introduced by doping and intercalation of alkali-metal atoms[5], with $T_c$ values up to 33 K at normal pressure [17], well above the pre-HTS record of 23 K [10] followed soon as another surprise. In spite of this high $T_c$, the explanation of superconductivity seems to be well within reach of conventional BCS theory based on intramolecular phonons [18].

---

[3] $PuCoGa_5$ is a recently found “Medium Heavy Fermion“ superconductor with $T_c = 18.5$ K [12].

[4] The verdict of the Mermin-Wagner theorem [16] that long-range order can not exist in two dimensions at finite temperature due to the influence of fluctuations has long been believed to restrict superconductivity to the physical dimension d = 3. However, HTS have shown that the limiting case $d = 2+\varepsilon$ ($\varepsilon \to 0$), i. e., a basically two-dimensional layer-oriented superconductivity scenario with a slight coupling of neighbouring layers, can be enormously beneficial for SC long-range order.

[5] The chemically appropriate denotation of metal-doped fullerenes is “fullerides“. The semiconductor-impurity based view of “doping“ amounts here to denoting common salt as “sodium-doped chlorine“.

*Borides* had been investigated systematically with respect to high-$T_c$ superconductivity already in the 1950s. The rationale was the BCS $T_c$-formula [8] where a high characteristic phonon frequency, as provided by the light boron atoms, was predicted to be particularly helpful with respect to a high $T_c$. In the 1990s, the borocarbide superconductors RE $Ni_2B_2C$ with $T_c$ up to 16.5 K [19] seemed to fulfill this promise at least halfway. However, phonons are here only one of the candidates of contributing superconductivity mechanisms[6].

The huge surprise came in 2001 with the discovery of superconductivity up to $T_c$ = 40 K in $MgB_2$, a compound which was well-known since the 1950s and which was in 2001 already commercially available in quantities up to metrical tons [20]. As for the fullerides, in spite of the high $T_c$ a phononic mechanism is highly plausible: Strong Coulomb correlation effects are not expected since the conduction electron system does neither involve inner atomic shells nor a reduction of the effective dimensionality. Nevertheless, new theoretical ingredients are required for a satisfactory explanation of the experiments [21].

This recent example demonstrates which scientific surprises can be encountered even in seemingly well-investigated research areas, and that superconductivity will certainly remain for a long time to come at the forefront of physics and materials research.

## 2. APPLICATIONS

Besides the scientific interest, the search for applications has always been a driving force for superconductor materials science [22]. Right from the discovery, it had been envisioned that SC coils with high persistent current might be used to generate strong magnetic fields. However, in the first generation of SC materials ("*type-I*") superconductivity was easily suppressed by magnetic fields: The magnetic self-field generated by the injected current prevented high-field as well as high-current applications. A first step towards this goal was the discovery of *type-II* superconductors where the magnetic penetration depth $\lambda$ exceeds the SC coherence length $\xi$. This enables a coexistence of superconductivity and magnetic fields, which are allowed to penetrate into the SC bulk in the quantized form of vortices. The concomitant substantial reduction of the loss of SC condensation energy that has to be paid for magnetic field penetration facilitates the survival of superconductivity even in strong magnetic fields, at least up to a certain critical field $H_{c2}$ where the SC state no longer survives the "vortex swiss cheesing". The last ingredient required for technically applicable "*hard*" superconductors was the discovery and engineering of pinning centers which fix penetrated magnetic flux and prevent its Lorentz force driven flow through the superconductor that otherwise generates power dissipation.

Today, NbTi and $Nb_3Sn$ conductors are the basis of a billion Euro SC wire industry which delivers magnets that cannot be realized by means of conventional metal wire conductors, e. g., for Magnetic Resonance Imaging (MRI) systems and High-Energy Physics (HEP) particle accelerators[7] [1,22,23]. The enormously high critical fields $H_{c2}$ ~ 100 T of HTS [24] indicate their potential for extremely high-field applications. However, HTS vortex physics has turned out to be much more complex than what had been known from classical superconductors [25]. This implies strong restrictions for high-field, high-temperature HTS magnet hopes[8]. Nevertheless, in spite of earlier concerns about the ceramic nature of HTS, flexible[9] HTS-based conductors are steadily progressing towards applications where a substantial size decrease justifies the cryogenic efforts. HTS current leads are just being introduced worldwide in HEP accelerators to transport kA-sized feed currents at a substantially reduced heat leakage from a liquid-nitrogen ("$LN_2$") temperature region to LHe cooled SC NbTi coil systems [28][10].

---

[6] For Rare Earth borocarbides there is additional magnetism due to localized $RE^{3+}$ 4f-electrons which is weakly interacting with superconductivity associated with the 3d-electrons of the $Ni_2B_2$ layers.

Nb-based SC rf-cavities represent another recent technological progress of HEP accelerators: An extremely high quality factor provides here a much better transfer of acceleration energy to the particle bunches than in conventional cavities[11]. Miniaturized microwave filters, e. g., for mobile phone base stations, are at present the most advanced HTS electronics application [31]: The low loss of thin film HTS resonator stripes with a typical size 50 μm × 1 cm (see Fig. 17 of this article) allows a complex coupling of a large number of such resonators on a chip which enables filters with sharp frequency cut-offs.

Josephson junctions [1], well-defined weak links of SC regions, can be coupled to "Superconducting QUantum Interferometric Devices" ("SQUIDs") [32,33], magnetic flux detectors with quantum accuracy that are the most sensitive magnetic field detectors presently available. SQUIDs based on Nb/$AlO_x$/Nb Josephson junctions achieve today at LHe temperature a magnetic noise floor ~ 1 fT/√Hz which enables diagnostically relevant magnetic detection of human brain signals ("magnetoencephalography", "MEG" [34]). HTS SQUIDs at liquid nitrogen operation have approached this magnetic sensitivity within one order of magnitude [35] and are already in commercial use for the nondestructive evaluation (NDE) of defects in complex computer chips [36] and aircrafts [37].

---

[7] Only persistent-current mode operation fulfills the MRI temporal field stability requirements of ~ 1 ppm / hour. This can not be achieved by active stabilisation. Moreover, conventional multi-Tesla magnet systems would require a much more complicated and much more expensive cooling systems than the liquid He ("LHe") cooling of SC coils. The Large Hadron Collider (LHC), presently under construction at CERN, will employ SC 9 T dipole coils to keep the high-energy protons on their orbit in a 9 km diameter circular tunnel. The high cost of the in total more than 8000 SC beam guiding magnets on the order of ~ 500 million Euro is still a bargain compared to the alternative of building a ~ 100 km diameter circular tunnel that would be required by conventional magnet technology.

[8] The generation of an extra 5 T field by an HTS coil at 4.2 K on top of a 20 T background field, adding up to a total field of the SC coil system of > 25 T has just been demonstrated [26]. At low temperature, the actual field limit of HTS will always be determined by the mechanical reinforcement that has to be provided to prevent the HTS wire or bulk from being disrupted by the Lorentz forces. At liquid nitrogen temperature, YBCO tapes are promissing candidates for field levels up to ~ 10 T.

[9] Bending such HTS tapes around a finger does not deteriorate their critical current capacity [27].

[10] Superconductors are not restricted by the Wiedemann-Franz coupling of electric and thermal transport since supercurrents do not transport entropy.

[11] At CERN, this energy increase has allowed the recently discussed "glimpse at the Higgs Boson" [29]. For the next generation of Nb cavities, the rf-magnetic fields will be increased up to 50% of $H_{c2}$ of Nb [30].

In the 1970s and 1980s, IBM as well as a Japanese consortium including Fujitsu, Hitachi, and NEC tested in large projects the fast switching of Josephson junctions from the SC to the normal state with respect to a post-semiconductor computer generation [38]. Unfortunately, the switching from the normal to the SC state turned out to limit the practical performance to several GHz instead of the theoretical ~1 THz[12]. Meanwhile, new device concepts based on the transport of single magnetic flux quanta reestablished the feasibility of THz operation [1]. The hottest topic of present Josephson circuit investigations[13] is the realization of quantum computing [39] with "Qubits" encoded by the SC wave function around µm-sized loops containing single [40] or even half [41] flux quanta[14]. At present, among all demonstrated Qubit realizations a SC electronics implementation appears to have the largest potential of upscalability to the size of several kQubit, which is required for first real applications: The lithographic requirements of ~ 1 µm minimum feature size are already common practice in present semiconductor circuits.

For all these applications of superconducitvity, the necessity of cryogenics is at least a psychological burden. Nevertheless, with the present progress of small cryocoolers [42] SC devices may evolve within foreseeable future to push-button black-box machines that may be one day as common practice as nowadays vacuum tube devices in ordinary living rooms[15].

## 3. CUPRATE HIGH-TEMPERATURE SUPERCONDUCTORS

Cuprate High-Temperature Superconductors ("HTS") play an outstanding role in the scientific development and for the present understanding of superconductivity. Except for semiconductors, no other class of materials has been investigated so thoroughly by thousands and thousands of researchers worldwide:

- A huge number of samples has been produced, in quantities of the order of metrical tons.
- Details of materials science have been diligently elaborated.
- High reproducibility has thus been achieved taking the materials complexity into account.
- The whole tool-set of experimental solid-state physics has been applied. For some techniques such as photoelectron spectroscopy [43,44], inelastic neutron scattering [45] or scanning tunneling microscopy [46,47] HTS have become a "drosophila"-like favorite object of investigation which still challenges further methodological development.

---

[12] The practical problem is to avoid a "punch through", i. e., the jump from a $V = +V_0$ to the $V = -V_0$ instead of the targeted $V = 0$ voltage state.

[13] which may even be combined with the former

[14] There are even several possibilities for such an encoding of a Qubit, e. g., it can represented in terms of the polarity of the flux quanta.

[15] such as old-fashioned TV sets

- HTS still represent a great challenge to theoretical solid-state physics since not only the superconducting but even more the "normal" conducting state of HTS is awaiting a satisfactory explanation.

It is due to this extraordinary importance that a large part of this book on superconductors and of this topical overview on SC materials is dedicated to cuprate HTS.

## 3A. INTRODUCTORY REMARKS

The discovery of superconductivity above liquid-nitrogen temperature in cuprate materials ("High-Temperature Superconductors", "HTS") [6] raised an unprecedented scientific euphoria[16] and challenged research in a class of complicated compounds which otherwise would have been encountered on the classical research route of systematic investigation with gradual increase of materials complexity only in a far future. The plethora of preparational degrees of freedom, the inherent tendency towards inhomogeneities and defects, in combination with the very short SC coherence length of the order of the dimensions of the crystallographic unit cell did not allow easy progress in the preparation of these materials. Nevertheless, after enormous preparation efforts HTS arrived meanwhile at a comparatively mature materials quality [49] that allows now a clearer experimental insight in the intrinsic physics which is still awaiting a satisfactory theoretical explanation [50,51,52].

The present situation of HTS materials science resembles in many aspects the history of semiconductors half a century ago [53]. The new dimension in the development of HTS materials, in particular in comparison with the case of silicon, is that HTS are *multi-element* compounds based on complicated sequences of *oxide layers* [54,55,56]. In addition to the impurity problem due to undesired additional elements, which gave early semiconductor research a hard time in establishing reproducible materials properties, intrinsic local stoichiometry defects arise in HTS from the insertion of cations in the wrong layer and defects of the oxygen sublattice. As additional requirement for the optimization of the SC properties, the oxygen content has to be adjusted in a compound-specific off-stochiometric ratio [57], but nevertheless with a spatially homogeneous microscopic distribution of the resulting oxygen vacancies or interstitials [58]. Today, reproducible preparation techniques for a number of HTS material species are available which provide a first materials basis for applications. As a stroke of good fortune, the optimization of these materials with respect to their SC properties seems to be in accord with the efforts to improve their stability in technical environments in spite of the only metastable chemical nature of these substances under such conditions [59,60,61].

---

[16] It has been estimated that in the hottest days of this "Woodstock of physics" in 1987 [48] about one third of *all* physicists all over the world tried to contribute to the HTS topic.

## 3B. STRUCTURAL ASPECTS

The structural element of HTS compounds related to the location of mobile charge carriers are stacks of a certain number n = 1, 2, 3, ... of $CuO_2$ layers which are "glued" on top of each other by means of intermediate Ca layers (see Fig. 2&3) [54,55,62,63]. Counterpart of these "*active blocks*" of $(CuO_2/Ca/)_{n-1}CuO_2$ stacks are "*charge reservoir blocks*" $EO/(AO_x)_m/EO$ with m = 1, 2 monolayers of a quite arbitrary oxide $AO_x$ (A = Bi [55], Pb [62], Tl [55], Hg [55], Au [64], Cu [55], Ca [65], B [62], Al [62], Ga [62]; see Table 1)[17] "wrapped" on each side by a monolayer of alkaline earth oxide EO with E = Ba, Sr (see Fig. 2&3). The HTS structure results from alternating stacking of these two block units. The choice of BaO or SrO as "wrapping" layer is not arbitrary but depends on the involved $AO_x$ since it has to provide a good spatial adjustment of the $CuO_2$ to the $AO_x$ layers.

The general chemical formula[18] $\mathbf{A_m}E_\mathbf{2}Ca_\mathbf{n-1}Cu_\mathbf{n}O_{2n+m+2+y}$ (see Fig. 3) is conveniently abbreviated as **A-m2(n-1)n** [63] (e. g. $Bi_2Sr_2Ca_2Cu_3O_{10}$: Bi-2223) neglecting the indication of the alkaline earth element[19] (see Tab.1). The family of all n = 1, 2, 3, ... representatives with common $AO_x$ are often referred to as "A-HTS", e. g. Bi-HTS. The most prominent compound **Y**$Ba_\mathbf{2}Cu_\mathbf{3}\mathbf{O}_7$ (see Fig. 2), the first HTS discovered with a critical temperature $T_c$ for the onset of superconductivity above the boiling point of liquid nitrogen [67], is traditionally abbreviated as "**YBCO**" or "**Y-123**" ($\mathbf{Y_1}Ba_\mathbf{2}Cu_\mathbf{3}O_{7-\delta}$). It also fits into the general HTS classification scheme as a modification of Cu-1212 where Ca is completely substituted by Y. This substitution introduces extra negative charge in the $CuO_2$ layers[20] due to the higher valence of Y (+3) compared to Ca (+2). The HTS compounds $REBa_2Cu_3O_{7-\delta}$ ("RBCO", "RE-123") where RE can be La [55] or any rare earth element [68] except for Ce or Tb [69] can be regarded as a generalization of this substitution scheme. The lanthanide contraction of the RE ions provides an experimental handle on the distance between the two $CuO_2$ layers of the active block of the doped Cu-1212 compound [70,71,72]. $\mathbf{Y_1}Ba_2\mathbf{Cu_4}O_8$ ("**Y-124**") is the m = 2 counterpart Cu-2212 of YBCO.

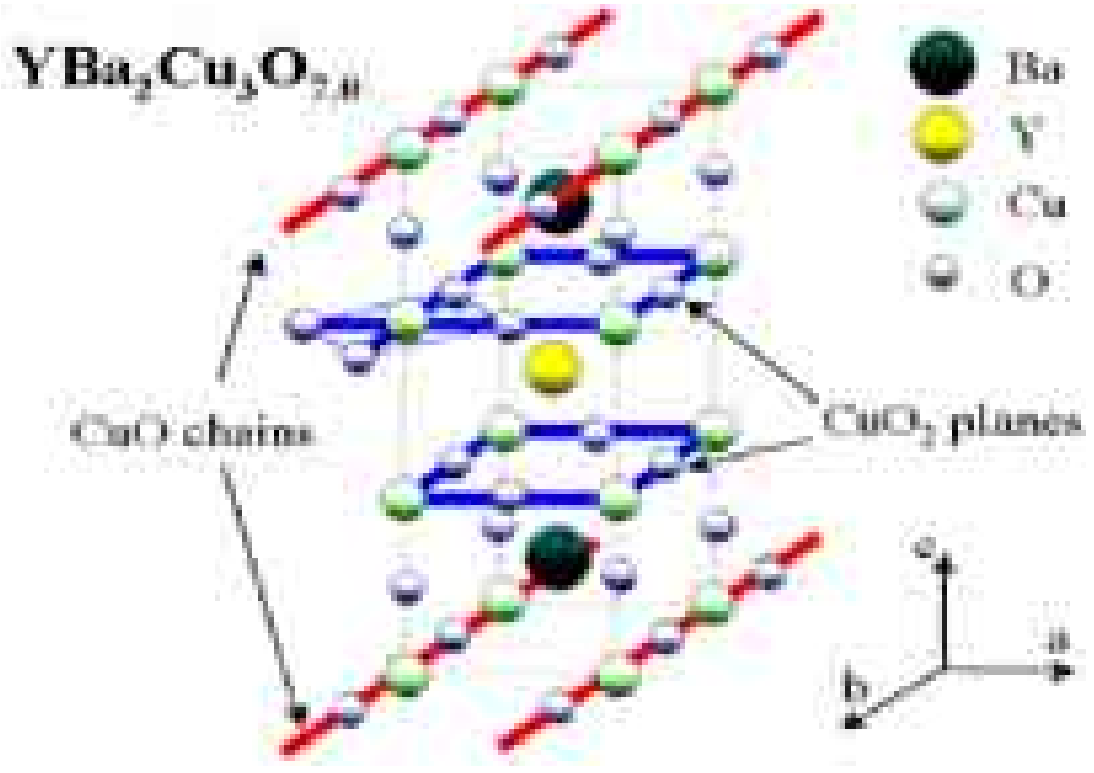

**Fig. 2** Crystal structure of $YBa_2Cu_3O_7$ ("YBCO"). The presence of the CuO chains introduces an orthorhombic distortion of the unit cell (a = 0.382 nm, b = 0.389 nm, c = 1.167 nm [66]).

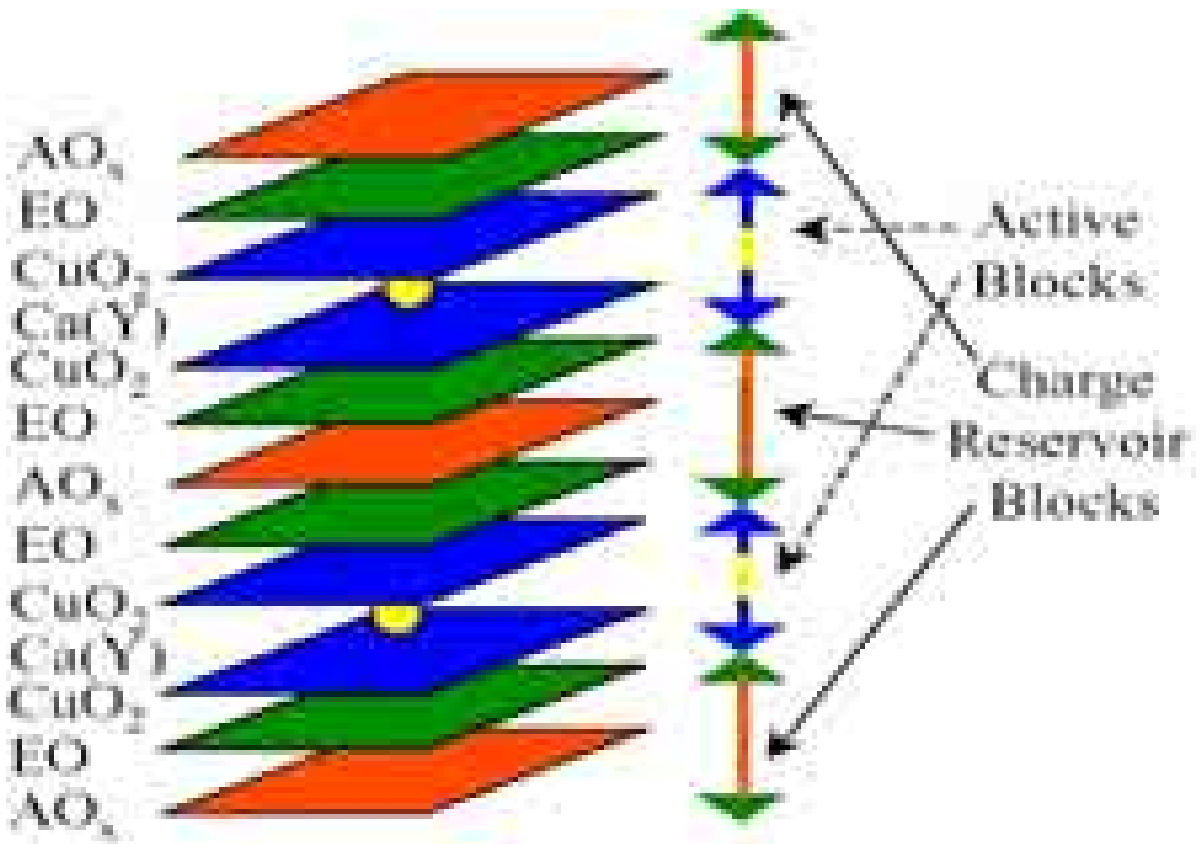

**Fig. 3** General structure of a cuprate HTS A-m2(n-1)n ($A_mE_2Ca_{n-1}Cu_nO_{2n+m+2+y}$) for m = 1. For m = 0 or m = 2 the missing (additional) $AO_x$ layer per unit cell leads to a (a/2, b/2, 0) "side step" of the unit cells adjoining in c-axis direction.

---

[17] HTS compounds based on "oxide mixtures" $A^{(1)}_s A^{(2)}_{1-s} O_x$ are omitted in this consideration.

[18] y = m (x – 1) in the oxygen stoichiometry factor.

[19] A more precise terminology A-m$^{(E)}$ 2(n-1)n has been suggested [63].

The "**214**" HTS compounds $E_\mathbf{2}Cu_\mathbf{1}O_\mathbf{4}$ (s. Tab1), e. g. $La_{2-x}Sr_xCuO_4$ ("LSCO") or $Nd_{2-x}Ce_xCuO_4$ ("NCCO") are a bit exotic in this ordering scheme but may also be represented here as “0210” with m = 0, n = 1 and $E_2 = La_{2-x}Sr_x$ and $E_2 = Nd_{2-x}Ce_x$, respectively.

Further interesting chemical modifications of the basic HTS compositions are the introduction of fluorine [73,74] or chlorine [75] as more electronegative substituents of oxygen. For Hg-1223, $T_c$ = 135 K can thus be raised to 138 K [7], the highest $T_c$ reported by now under normal pressure conditions (164 K at 30 GPa [76]).

## 3C. METALLURGICAL ASPECTS

Within 18 years since their discovery, cuprate HTS samples have greatly improved towards high materials quality and can in fact nowadays be prepared in a remarkably reproducible way. The enormous worldwide efforts for this achievement may be estimated from the plain number of more than 100 000 articles which meanwhile have been published on cuprate high-$T_c$ superconductivity, strongly outnumbering the only about 15 000 publications [77, 78,79] that appeared within a whole century on the remaining superconducting materials.

Nevertheless, the HTS materials quality level is still far from the standards of classical superconductors. The reason is the larger number of at least four chemical elements from which the various cuprate HTS phases have to be formed: The majority of the classical SC compounds, e. g. B1 or A15 compounds, is made of only two elements. Each element contributes an additional degree of freedom to the preparation route towards new compounds. For the metallurgist, this translates roughly into one order of magnitude more elaborate exploratory efforts. A new cuprate HTS compound requires therefore typically as much metallurgical optimization work as 100 binary compounds.

As a further complication, the large number of adjacent phases in the phase diagram constituted by the contributing elements hamper the preparation of phase pure quaternary or quinary cuprates. The HTS layer structure based on $(CuO_2/Ca/)_{n-1}CuO_2$ stacks with a certain $CuO_2$ layer number $n$ (see Fig. 3) introduces another dimension of preparation challenges: As the formation enthalpy of a compound, e. g. with a single $CuO_2$ layer ($n$ = 1), differs only little from that of the ($n$ = 2)-compound with two adjacent $CuO_2$ layers, these materials tend to form polytypes [80]. A sophisticated process control during sample preparation is essential in order to reduce the amount of such stacking faults.

[20] Since the $CuO_2$ layers in YBCO are hole-doped this amounts to a reduction of the hole doping compared to the (hypothetical) canonical Cu-1212 compound.

The big metallurgical challenge introduced by the cuprate HTS materials in a previously unknown extent is the large number of chemical degrees of freedom of such materials. Binary compounds may have a 1-dimensional homogeneity range, e. g., $NbN_{1-x}$. In quaternary compounds, e. g., for RE-123 this range is in general 3-dimensional! In all cuprate HTS materials the oxygen degree of freedom $O_x$ can be adjusted quite arbitrarily far from the stoichiometric ratio. The issue how to do this has been the subject of intensive investigations. As will become clear in chapter 3G, the oxygenation is an extraordinarily important experimental handle to introduce even radical changes of the physical properties of the material, e. g., to turn it from an insulator into a high-temperature superconductor! [81] The critical current capability steering the ampacity of technical conductors is another vital issue with respect to applications that depends critically not only on the content but also on the atomic-scale distribution of the oxygen atoms which are in the form of oxygen vacancy clusters probably the most important pinning centers [82] (see chapter 3F).

With respect to the cation stoichiometry, for the RE-123 phase as a particularly well-investigated HTS example the RE/Ba ratio represents the second chemical degree of freedom which may also deviate from the stoichiometric value 1/2. Y-123 (YBCO) represents here a remarkable exception. However, RE-123 compounds where the Y ions are replaced by larger RE ions like Gd or Nd exhibit a pronounced homogeneity range. For Nd-123, the Nd/Ba ratio encountered in such solid solutions may vary from 0.49 to 2, with the oxygen content being still a fully independent chemical variable. The Cu/(EA=Ba or Sr) ratio as chemical degree of freedom of all "hole-doped" HTS compounds (see Table 1) has up to now not been studied in great detail. For the 123 phase as a particularly complicated HTS example with respect to the Cu stoichiometry featuring two $CuO_2$ layers and an additional CuO chain structure in the unit cell, the present assumption is that the Cu/Ba ratio sticks to the stoichiometric value 3/2. For samples prepared close to their peritectic temperatures or at very low temperatures close to the boundary of the stability field of the 123 phase this may no longer be the case.

Chemical purity of the starting material is still a topical issue for a reproducible preparation of cuprate HTS. The use of chemicals with a purity of 99.99% and better is mandatory but still not sufficient. The frequently used Ba source material $BaCO_3$ is usually not completely reacted and may thus lead to the incorporation of carbonate ions into the HTS cuprate phase. During the preparation procedure, the formation of even a small amount of liquid has to be carefully avoided since this may corrode the substrate or crucible and may thus introduce impurities into the sample. This corrosion process is particular harmful for crystal growth experiments where the complete melt encounters the crucible wall. BaO/CuO melts are highly corrosive and attack all conventional types of crucible materials forming new solid phases. Some of these reaction products have turned out to be well suited as materials for corrosion resistant crucibles for the preparation of cuprate HTS. $BaZrO_3$ is here the best-known example [49], $BaHfO_3$ and $BaSnO_3$ are promissing candidates.

Bearing in mind all these drawbacks and problems, the surprisingly high quality achieved nowadays for HTS cuprate materials, in particular for single crystals with respect to low impurity content, composition close to the desired stoichiometry, low concentration of defects such as inclusions, stacking faults or dislocations is rather amazing. However, a lot still remains to be done before the HTS sample quality becomes really comparable to that of binary classical superconductors.

### 3D. STRUCTURE AND $T_C$

Experimentally, for all HTS families A-m2(n-1)n the optimized $T_c$ is found to increase from n = 1 to n = 3, 4 and to decrease again for higher n (s. Tab.1). It is still unclear whether this $T_c$ maximum is an intrinsic HTS property since the synthesis of higher-n members of the HTS families turns out to be more and more complicated [107,108,109,110]. In particular, there is at present no preparation technique that alllows to adjust here a sufficiently high oxygen content[21] or to provide otherwise sufficient electronic doping that would allow to clarify the possible range of $T_c$ optimization for higher n HTS A-m2(n-1)n[22] [107,110]. Hence the question is still open if such an optimized $T_c$ may eventually continue to increase towards higher n, possibly up to $T_c \sim 200$ K as can be estimated for optimized infinite-layer HTS based on model-independent theoretical considerations extrapolating from $T_c$ (n = 1) ~ 100 K of Hg-1201 [121].

Another well-investigated experimental $T_c$ trend is the slight increase of the optimized $T_c$ of the RE-123 HTS with increasing distance between the two $CuO_2$ layers in the active block[23]. It has been explained in terms of a higher effective charge transfer to the $CuO_2$ layers [68,70]. For RE-123 HTS with larger RE ions (La, Pr, Nd,), sufficient oxygenation with respect to $T_c$ optimization becomes increasingly difficult. In this respect, La-123 is apparently beyond the present practical limit [49]: The intrinsic $T_c \sim 100$ K, estimated from optimized $T_c$ values of the other RE-123 as a function of oxygenation [49], has not been achieved yet.

As a further complication, these larger RE ions are comparable in size with the Ba ions. This favors cation disorder with respect to the RE and Ba lattice sites [123] which leads to substantial $T_c$ degradation [124]. This disorder effect, oxygen deficiency and / or impurities had been suggested as reasons for the non-appearance of superconductivity in Pr-123 [125] and Tb-123 as the only non-SC members of the 123-HTS family. However, in 1998 first superconducting Pr-123 samples with $T_c \sim 80$ K [126] (105 K at 9.3 GPa [127]) could be synthesized which lost their SC properties within few days after their preparation. The nature of this SC Pr-123 phase is still unclear. It has been suggested that it consisted of Ba-rich Pr-123 which is unstable at room temperature with respect to spinodal decomposition. The c-axis lattice parameter of these SC Pr-123 samples fits well in the series of the other SC RE-123 compounds, in remarkable contrast to the c-axis lattice parameter of the usual non-SC Pr-123 samples [126,127] where the absence of superconductivity is attributed to a hybridisation of the Pr 4f electrons with O $2p_\pi$ electrons depriving mobile charge carriers [128,129].

---

[21] The microscopic distribution of oxygen [111] is neither easily observed [112,113,114] nor easily controlled [115]: Experimentally, oxygen content and distribution are adjusted by means of (oxygen partial) pressure, temperature and sufficient holding time to achieve thermal equilibrium [116].

[22] The $T_c$ depression is believed to arise from a redistribution of doping charge in the $CuO_2$ layer stacks from the interior to the outer $CuO_2$ layers [117] which has been confirmed by NMR measurements [118,119]. Based on these experimental results, a reasonable theoretical description of $T_c$(n) has recently been modeled in a phenomenological Ginzburg-Landau free energy approach [120].

[23] On hydrostatic pressure application, $dT_c/dP$ shows the same trend which further increases the $T_c$ differences of the RE-123 [68]. On uniaxial pressure application, $T_c$ increases if the pressure effect is to reverse the tetragonal-to orthorhombic lattice distortion [122].

| HTS Family | Stochiometry | Notation | Compounds | Highest $T_c$ |
|---|---|---|---|---|
| Bi-HTS | $Bi_mSr_2Ca_{n-1}Cu_nO_{2n+m+2}$ | Bi-m2(n-1)n, BSCCO | Bi-1212 | 102 K [83] |
| | m = 1, 2 | | Bi-2201 | 34 K [84] |
| | n = 1, 2, 3 . . . | | Bi-2212 | 96 K [85] |
| | | | Bi-2223 | 110 K [55] |
| | | | Bi-2234 | 110 K [86] |
| Pb-HTS | $Pb_mSr_2Ca_{n-1}Cu_nO_{2n+m+2}$ | Pb-m2(n-1)n | Pb-1212 | 70 K [87] |
| | | | Pb-1223 | 122 K [88] |
| Tl-HTS | $Tl_mBa_2Ca_{n-1}Cu_nO_{2n+m+2}$ | Tl-m2(n-1)n, TBCCO | Tl-1201 | 50 K [55] |
| | m = 1, 2 | | Tl-1212 | 82 K [55] |
| | n = 1, 2, 3 . . . | | Tl-1223 | 133 K [89] |
| | | | Tl-1234 | 127 K [90] |
| | | | Tl-2201 | 90 K [55] |
| | | | Tl-2212 | 110 K [55] |
| | | | Tl-2223 | 128 K [91] |
| | | | Tl-2234 | 119 K [92] |
| Hg-HTS | $Hg_mBa_2Ca_{n-1}Cu_nO_{2n+m+2}$ | Hg-m2(n-1)n, HBCCO | Hg-1201 | 97 K [55] |
| | m = 1, 2 | | Hg-1212 | 128 K [55] |
| | n = 1, 2, 3 . . . | | Hg-1223 | 135 K [93] |
| | | | Hg-1234 | 127 K [93] |
| | | | Hg-1245 | 110 K [93] |
| | | | Hg-1256 | 107 K [93] |
| | | | Hg-2212 | 44 K [94] |
| | | | Hg-2223 | 45 K [95] |
| | | | Hg-2234 | 114 K [95] |
| Au-HTS | $Au_mBa_2Ca_{n-1}Cu_nO_{2n+m+2}$ | Au-m2(n-1)n | Au-1212 | 82 K [64] |
| 123-HTS | $REBa_2Cu_3O_{7-\delta}$ | RE-123, RBCO | Y-123, YBCO | 92 K [70] |
| | RE = Y, La, Pr, Nd, Sm, | | Nd-123, NBCO | 96 K [70] |
| | Eu, Gd, Tb, Dy, Ho, | | Gd-123 | 94 K [96] |
| | Er, Tm, Yb, Lu | | Er-123 | 92 K [49] |
| | | | Yb-123 | 89 K [68] |
| Cu-HTS | $Cu_mBa_2Ca_{n-1}Cu_nO_{2n+m+2}$ | Cu-m2(n-1)n | Cu-1223 | 60 K [55] |
| | | | Cu-1234 | 117 K [97] |
| | m = 1, 2 | | Cu-2223 | 67 K [55] |
| | n = 1, 2, 3 . . . | | Cu-2234 | 113 K [55] |
| | | | Cu-2245 | < 110 K [55] |
| Ru-HTS | $RuSr_2GdCu_2O_8$ | Ru-1212 | Ru-1212 | 72 K [98] |
| B-HTS | $B_mSr_2Ca_{n-1}Cu_nO_{2n+m+2}$ | B-m2(n-1)n | B-1223 | 75 K [99] |
| | | | B-1234 | 110 K [99] |
| | | | B-1245 | 85 K [99] |
| 214-HTS | $E_2CuO_4$ | LSCO | $La_{2-x}Sr_xCuO_4$ | 51 K [100] |
| | | "0201" | $Sr_2CuO_4$ | 25 (75)K [101] |
| | | *Electron-Doped HTS* | $La_{2-x}Ce_xCuO_4$ | 28 K [102] |
| | | PCCO | $Pr_{2-x}Ce_xCuO_4$ | 24 K [103] |
| | | NCCO | $Nd_{2-x}Ce_xCuO_4$ | 24 K [103] |
| | | | $Sm_{2-x}Ce_xCuO_4$ | 22 K [104] |
| | | | $Eu_{2-x}Ce_xCuO_4$ | 23 K [104] |
| | $Ba_2Ca_{n-1}Cu_nO_{2n+2}$ | "02(n-1)n" | "0212" | 90K [105] |
| | | | "0223" | 120K [105] |
| | | | "0234" | 105K [105] |
| | | | "0245" | 90K [105] |
| *Infinite-Layer* HTS | $ECuO_2$ | *Electron-Doped I. L.* | $Sr_{1-x}La_xCuO_2$ | 43 K [106] |

**Table 1** Classification and reported $T_c$ values of HTS compounds.

In Bi-HTS, cation disorder at the Sr crystallographic site is inherent and strongly affects the value of $T_c$. In Bi-2201, partial substitution of Sr by RE = La, Pr, Nd, Sm, Eu, Gd, and Bi was shown to result in a monotonic decrease of $T_c$ with increasing ionic radius mismatch [85]. For Bi-2212, partial substitution of Ca by Y results in a $T_c$ optimum of 96 K at 8% Y doping, apparently due to a trade-off between the respective disorder effect and charge doping [85].

In 214-HTS without BaO or SrO “wrapping” layers around the $CuO_2$ layers (see Tab. 1), $T_c$ seems to be particularly sensitive with respect to oxygen disorder [130]. The electron doped 214-HTS such as $Nd_{2-x}Ce_xCuO_4$ ("NCCO") have here the additional complication of a different oxygen sublattice where oxygen ions on interstitial lattice positions in the $Nd_{2-x}Ce_xO_2$ layer (which are yet for hole doped 214-HTS the regular oxygen positions in the non-$CuO_2$ layer!) tend to suppress $T_c$ [131].

With respect to the $T_c$ dependence of the A-m2(n-1)n HTS families on the cation A of the charge reservoir blocks, there is an increase moving in the periodic table from Bi to Hg (see Tab. 1). However, continuing to Au, the reported $T_c$ is already substantially lower[24]. This $T_c$ trend seems to be related to the chemical nature of the A-O bonds in the $AO_x$ layers [94].

The correlation of $T_c$ with the buckling angles of the $CuO_2$ layer is less pronounced [132]. Such and other deviations from a simple tetragonal crystal structure are found in most of the HTS compounds as a chemical consequence of the enforced $AO_x$ layer arrangement in the cuprate HTS structure. These structural modifications certainly influence the SC properties.

However, these peculiarities can not be essential for high-temperature superconductivity, since there *are* cuprate HTS with a simple tetragonal crystal structure *and* high $T_c$: Tl-2201 [133,134] and Hg-1201 [135] are such HTS model compounds based on a single $CuO_2$ layer in the unit cell[25] with $T_c$ values of 90 K and 97 K, respectively. Tl-2201 has already been used several times for experimental falsification of HTS theory predictions [136,137]. These simple model compounds represent also a much more natural starting point for the theoretical ab-initio understanding of HTS than YBCO and LSCO which are for historic reasons still favorites of present computational HTS materials science.

The rationale that the phenomenon of superconductivity in HTS can be conceptually reduced to the physics of the $CuO_2$ layers [51] has evolved to a more and more 2-dimensional view in terms of $CuO_2$ *planes*. The superconductive coupling between these planes within a given $(CuO_2/Ca/)_{n-1}CuO_2$ stack ("interplane coupling") is much weaker than the intraplane coupling, but still much stronger than the superconductive coupling between the $(CuO_2/Ca/)_{n-1}CuO_2$ stacks which can be described as Josephson coupling (see Fig. 4).

---

[24] Earlier eports of Ag-HTS with $AgO_x$ integrated in the HTS lattice have not been confirmed [62].

[25] Strictly speaking, the primitive unit cell of Tl-2201 has only rhombohedral crystal symmetry due to the “diagonal side step“ of the $EO/(AO_x)_m/EO$ blocks common to all A-m2(n-1)n compounds with even m, i. e., m = 0, 2 (see Fig. 2). The tetragonal symmetry can be conceptually recovered enlarging the unit cell to include two primitive cells on top of each other.

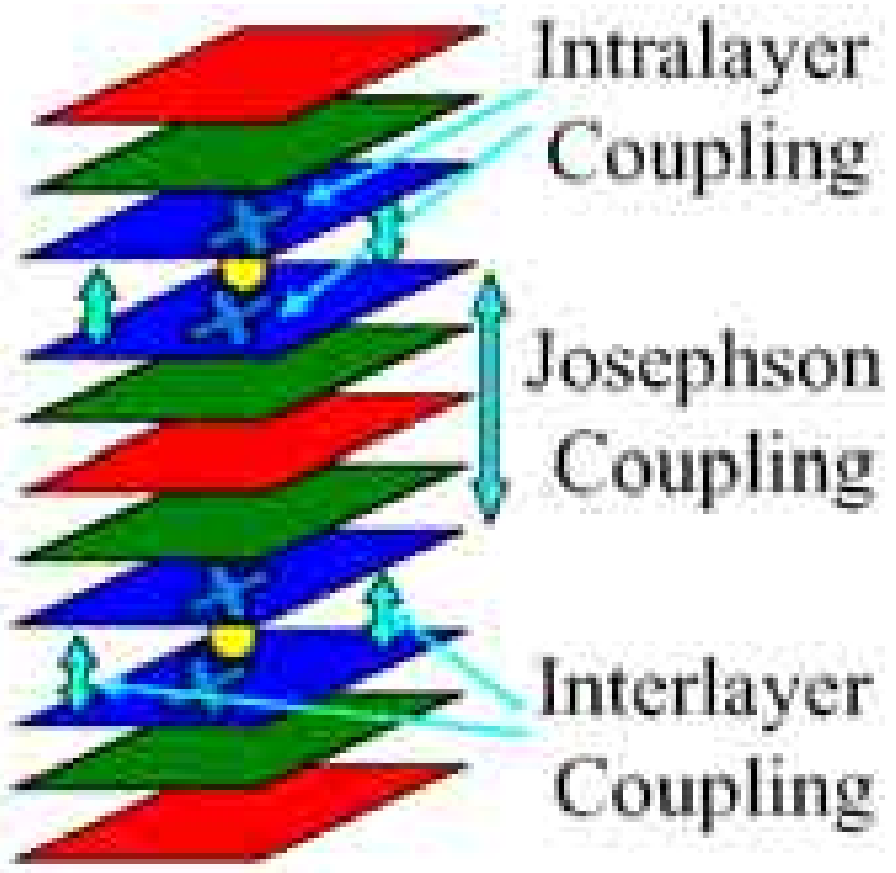

**Fig. 4** Hierarchy of the SC coupling between the different structural elements of cuprate HTS.

The charge reservoir blocks $EO/(AO_x)_m/EO$ play in this idealized theoretical picture only a passive role in providing the doping charge as well as the "storage space" for extra oxygen ions and cations introduced for additional doping. However, the huge pressure dependence of $T_c$ [76] in combination with the large quantitative variation of this effect for the various A-m2(n-1)n HTS families points to a more active role where the cations change not only their valency but also their transmission behavior for the interstack tunneling of Cooper pairs [138].

This becomes most evident for the Cu-HTS family, in particular for YBCO or the RE-123 HTS where the $AO_x$ layer is formed by 1-d CuO chain structures (see Fig. 2). There is experimental evidence that these CuO chains become SC, probably via proximity effect. The intercalation of superconductive CuO chain layers in-between the $CuO_2/Ca/CuO_2$ bilayer stacks is most likely the origin of the strong Josephson coupling between these bilayer stacks. This explains the remarkable reduction of the superconductive anisotropy in c-axis direction as compared to the other HTS families. Moreover, in contrast to the usually isotropic SC behavior within the a-b-plane, the CuO chains seem to introduce a substantially higher SC gap in b- compared to the a-direction [139]. This particular SC anisotropy renders YBCO and the RE-123 HTS exceptional among the cuprate HTS.

## 3E. SYMMETRY OF THE SUPERCONDUCTING ORDER PARAMETER

The long search for superconductors with a nonisotropic energy gap $\Delta$ arrived finally in the 1980s with a small list of plausible candidates. The observation of power law temperature dependencies of physical properties such as specific heat and thermal transport coefficients instead of the $\exp(-2\Delta_0 / k_B T)$ dependence of “s-wave” superconductors gave here the only clue to the structure of the energy gap nodes and their topology, and hence to the symmetry of the "unconventional" order parameter. Meanwhile, cuprate HTS have been established as a textbook example of a d-wave symmetric SC order parameter which can be observed directly by means of tricky “Superconducting QUantum Interference (Device)“ ("SQUID") circuits[26]. Essential prerequisite for the applicability of these experimental techniques is a high quality single crystal and epitaxial thin film preparation.

First convincing experimental evidence of d-wave superconductivity of cuprate HTS was provided by a SQUID configuration based on YBCO-Au-Pb Josephson junctions which were fabricated on the a and b edges and straddling the a-b corners of a YBCO single crystal (”corner-SQUID”) [140,141]. The most striking proof came from experiments performed on epitaxial c-axis oriented HTS thin films on “tricrystal substrates”, substrates composed of three single-crystalline parts with different in-plane orientations which enforce the HTS films to a corresponding alignment of the a-b-crystal axes [142,143]. By suitable choice of the relative angles between these in-plane orientations, the boundary conditions for the SC order parameter connecting the three $d_{x^2-y^2}$-wave subsystems impose a $\pi$-shift on any path circulating around the meeting point of these three film areas[27] (see Fig. 5) [144,145]. Since the SC wave function has to return to the same (complex) value, a compensating odd number of half-integer multiples of the magnetic flux quantum $\Phi_0$, i. e. $\pm\ \Phi_0/2$ in the ground state, must be encountered at this meeting point in the case of a $d_{x^2-y^2}$-wave SC order parameter, whereas an s-wave order parameter symmetry requires here integer multiples of the full magnetic flux quantum, i. e. $\pm\ n\ \Phi_0$ [146,147]. Magnetic microscopy experiments[28] with a flux resolution of a minute fraction of $\Phi_0$ clearly resolved at this tricrystal meeting point for all investigated cuprate HTS systems $\Phi_0/2$ vortices[29] (see Fig. 6) [142,143]. A particular SQUID circuit based on a tricrystal configuration (”$\pi$-SQUID”) allows determining admixtures of other symmetries to the $d_{x^2-y^2}$ SC order parameter [148,149,150,151]. Up to now, no s-admixture has been observed within experimental error.

In contrast to the clear-cut experimental situation for hole-doped HTS representing the majority of the cuprate HTS compounds (see Table 1), the results for electron-doped HTS are still contradictory. Magnetic microscopy of NCCO and PCCO thin films on tricrystal substrates revealed $\Phi_0/2$ vortices at the tricrystalline "meeting point" demonstrating $d_{x2-y2}$-symmetry [142,152]. π-SQUID experiments on $La_{2-x}Ce_xCuO_4$ [153] and photoelectron spectroscopy on NCCO [154] arrived at the same result. However, the temperature dependence of the London penetration depth $\lambda_{ab}(T)$ [103,155,156] as well as tunneling spectra derived from microbridges connecting epitaxial c-axis film areas with different in-plane orientations [157,158] seem to speak more in favor of an s-wave-symmetric SC order parameter: Tunneling spectra on hole-doped HTS show a distinct zero-(voltage-)bias anomaly (ZBA) with an amplitude that is directly proportional to the in-plane misorientation angle [159]. This can be readily explained as a ("Andreev") bound state located in the transition region between the two film areas [157,159,160,161,162]. The d-wave related π-phase shift (see Fig. 5) is an essential quantization condition for the existence of such a bound state. For the electron-doped HTS NCCO and PCCO clear tunneling spectra have been obtained without any indication of such a ZBA [157,158]. At present, no reconciling explanation of these contradictory experimental facts has been found yet[30]. Experimentally, the preparation of electron-doped HTS is much more demanding than of hole-doped HTS: Superconductivity occurs in a narrower doping range ($Nd_{2-x}Ce_xCuO_4$: $0.14 < x < 0.17$, $Pr_{2-x}Ce_xCuO_4$: $0.13 < x < 0.2$ [142]; $La_{2-x}Sr_xCuO_4$ : $0.05 < x < 0.27$ ; $YBa_2Cu_3O_{6+x}$: $0.35 < x < 1.0$ [167]). Photoelectron spectroscopy revealed a switching from electron to hole-like states on doping [168] which may match the switching from s- to d-wave superconductivity concluded from London penetration depth measurements[31] [169].

---

[26] Recent success with respect to YBCO-Nb Josephson junctions [41] enables now even the use of d-wave symmetry as a new functional principle for the construction of novel electronic devices.

[27] Actually, due to the technical problem of providing a high quality substrate environment around the tricrystal point, tetracrystal substrates with identical crystal orientations of two neighboring single-crystal pieces of the four substrate parts have been used in most of these experiments.

[28] LTS SQUIDs are used in these experiments as external, at present most powerful research tools for the magnetic field sensing of microscopic samples with a spatial resolution ~ 1 µm.

[29] in external fields surrounded by regular Abrikosov $\Phi_0$ vortices in the remaining film (see Fig. 6)

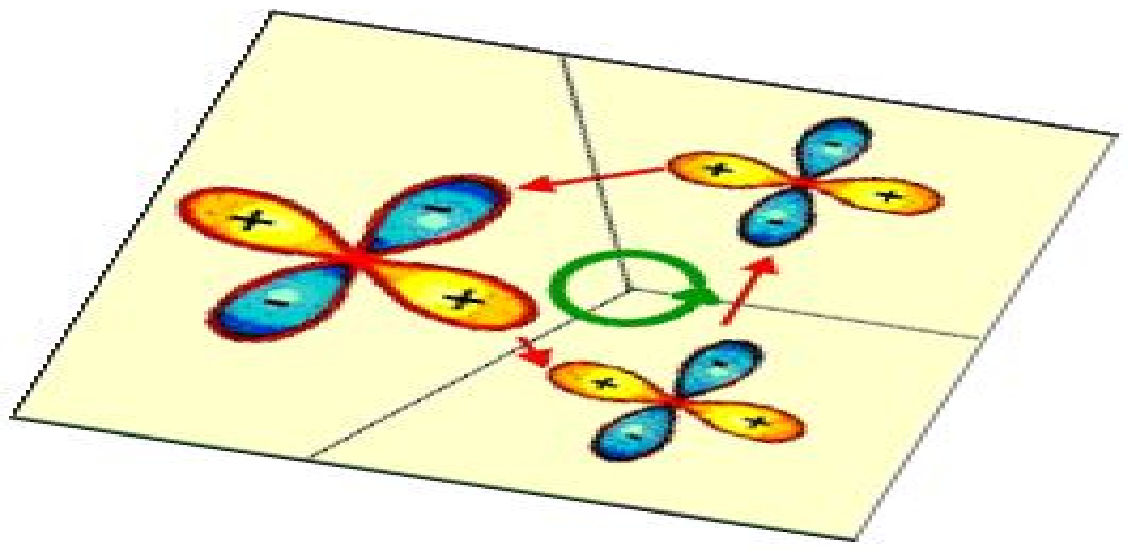

**Fig. 5** Basic idea of tricrystal experiments for the verification of the $d_{x2-y2}$ symmetry of the SC order parameter of cuprate HTS.

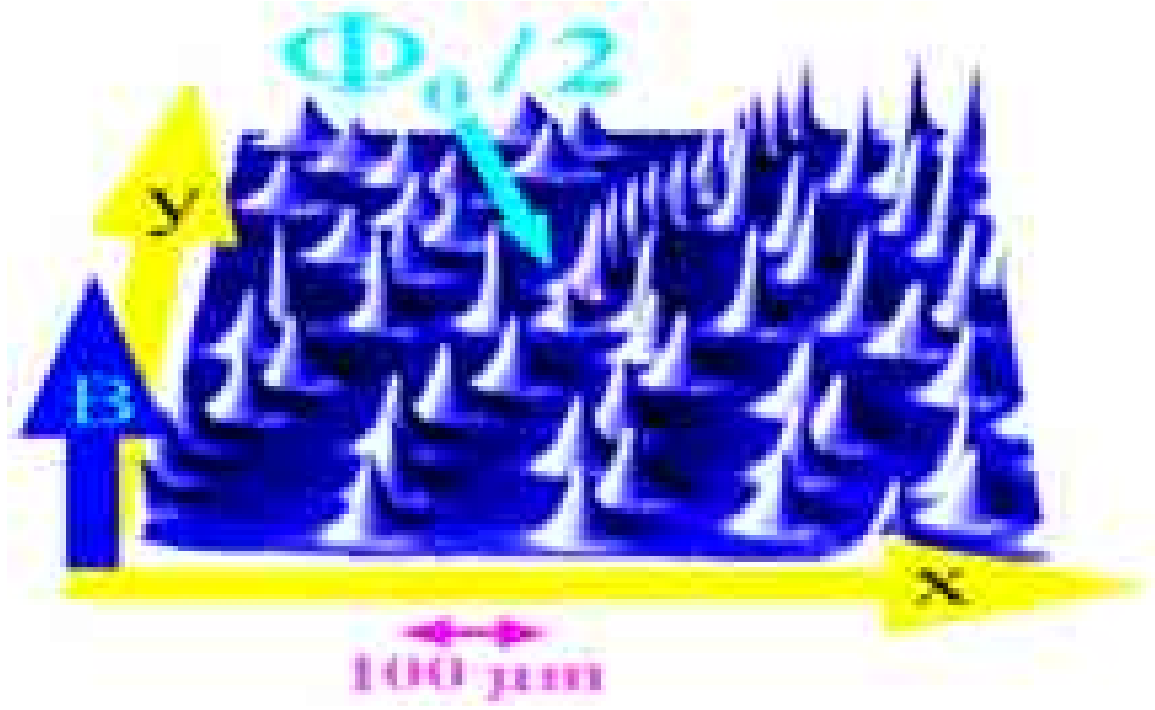

**Fig. 6** Spatial distribution of magnetic flux in a Bi-2212 thin film deposited on such a tricrystal substrate in an applied field of 3.7 mG (adapted from [143]).

---

[30] In NCCO low-energy excitations are observed for $T < 1$ K which presumably have their origin in the interaction of the magnetic Nd ions [163,164]. However, this additional physics is not an intrinsic pecularity of electron-doped HTS: $GdBa_2Cu_3O_{7-\delta}$ shows similar effects attributed to the magnetic Gd ions [165,166] but the same "high energy" superconducting behavior as other RE-123 HTS.

[31] A possibly related quantum phase transition of $Pr_{2-x}Ce_xCuO_4$ at $x_c = 0.165$ has been reported [170]. The presence of both charge carriers suggested by magnetothermopower measurements [171] could be due to local doping inhomogeneities and the strong doping sensitivity of the SC properties.

## 3F. SUPERCONDUCTIVE COUPLING

HTS are extreme type-II superconductors [172] with $\lambda > 100$ nm and $\xi \sim 1$ nm. Superconductivity in HTS is believed to have its origin in the physics of the $CuO_2$ layers where the mobile charges are located [50,51]. The superconductive coupling between these $CuO_2$ layers within a given $(CuO_2/Ca/)_{n-1}CuO_2$ stack ("*interlayer coupling*") is much weaker than the *intralayer coupling* within the $CuO_2$ layers, but still much stronger than the coupling between the $(CuO_2/Ca/)_{n-1}CuO_2$ stacks which can be described as Josephson coupling (see Fig. 4). This quasi-2-dimensional nature of superconductivity in HTS leads to a pronounced anisotropy of the SC properties with much higher supercurrents along the $CuO_2$ planes than in the perpendicular direction, a property which is not appreciated for technical applications but which can be dealt with by additional engineering efforts [173,174].

However, being type-II superconductor is not enough for a flow of strong currents without dissipation: The magnetic vortices that are introduced into the SC material by a magnetic field, in particular by the self-field generated by injected current, need to be fixed by *pinning centers,* or else dissipation by *flux flow* of the vortices is induced. In technical "hard" superconductors, material imperfections of the dimension of the coherence length do this job by blocking superconductivity from these regions. This provides a natural "parking area" for vortex cores where no SC condensation energy needs to be "paid".

Material imperfections of the dimension of the coherence length are easily encountered in HTS due to their small coherence lengths, e. g., for optimally doped YBCO $\xi_{ab} = 1.6$ nm, $\xi_c = 0.3$ nm for $T \to 0$ K [175] which are already comparable to the lattice parameters (YBCO: $a = 0.382$ nm, $b = 0.389$ nm, $c = 1.167$ nm [66]). However, the low $\xi_c$, i. e. the weak superconductive coupling between the $(CuO_2/Ca/)_{n-1}CuO_2$ stacks causes new problems. The thickness of the charge reservoir blocks $EO/(AO_x)_m/EO$ in-between these stacks is larger than $\xi_c$ with the result that due to the low Cooper pair density vortices are here no longer well-defined (see Fig. 7). This leads to a quasi-disintegration of the vortices into stacks of "*pancake vortices*" which are much more flexible entities than the continuous quasi-rigid vortex lines in conventional superconductors and therefore require individual pinning centers.

The extent of this quasi-disintegration is different for the various HTS compounds since $\xi_c$ is on the order of the thickness of a single oxide layers: Hence the number of layers in the charge reservoir blocks $EO/(AO_x)_m/EO$ makes a significant difference with respect to the pinning properties and thus to their supercurrents in magnetic fields. This is one of the reasons why YBCO ("Cu-**1**212") has a higher supercurrent capability in magnetic fields than the Bi-HTS Bi-**2**212 and Bi-**2**223 which for manufacturing reasons are still the most prominent HTS conductor materials[32].

Beside these intrinsic obstacles for the transport of supercurrent in single-crystalline HTS materials there are additional hurdles since HTS materials are not a homogeneous continuum but rather a network of linked grains (see Fig. 8). The mechanism of crystal growth is such that all material that cannot be fitted into the lattice structure of the growing grains is pushed forward into the growth front with the consequence that in the end all remnants of secondary phases and impurities are concentrated at the boundaries in-between the grains. Such barriers impede the current transport and have to be avoided by careful control of the growth process, in particular of the composition of the offered material.

---

[32] In addition, in the Cu-HTS family the $AO_x$ layer is formed by CuO chain structures (see Fig. 2). There are indications that the CuO chains become superconducting via proximity effect. This leads to stronger Josephson coupling in c-axis direction and thus to the smallest superconductive anisotropy among all HTS families [177].

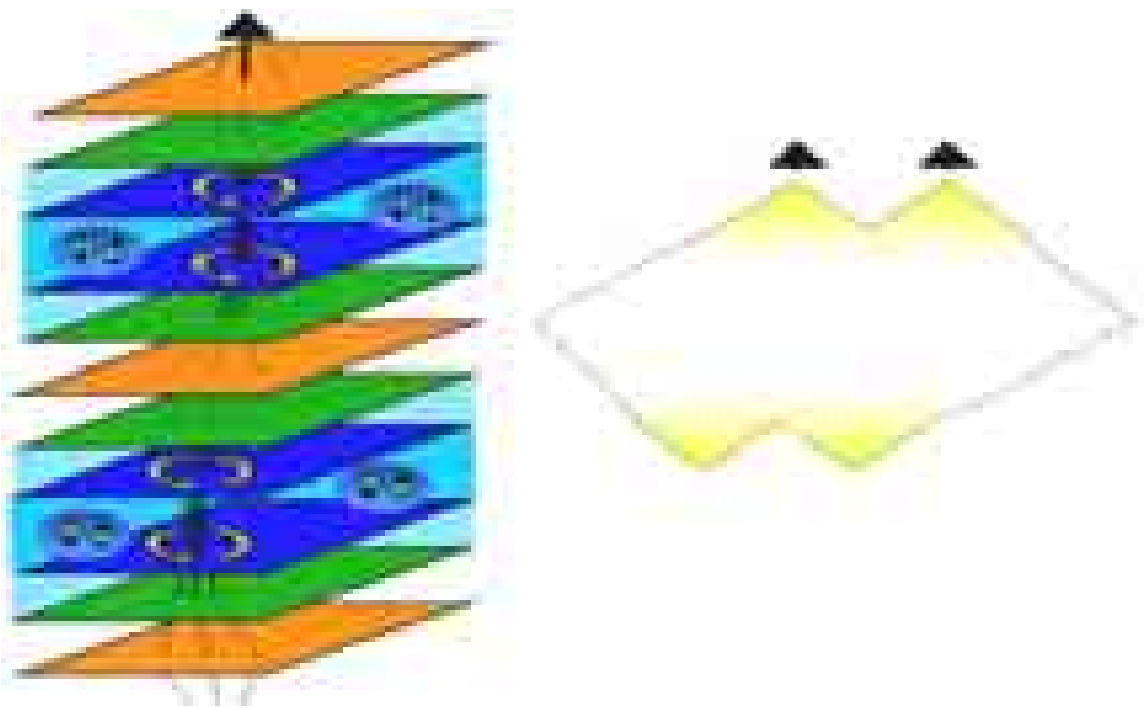

**Fig. 7** Quasi-disintegration of magnetic vortex lines into "pancake" vortices due to weak SC interlayer coupling and schematic overlap of neighboring vortices [176].

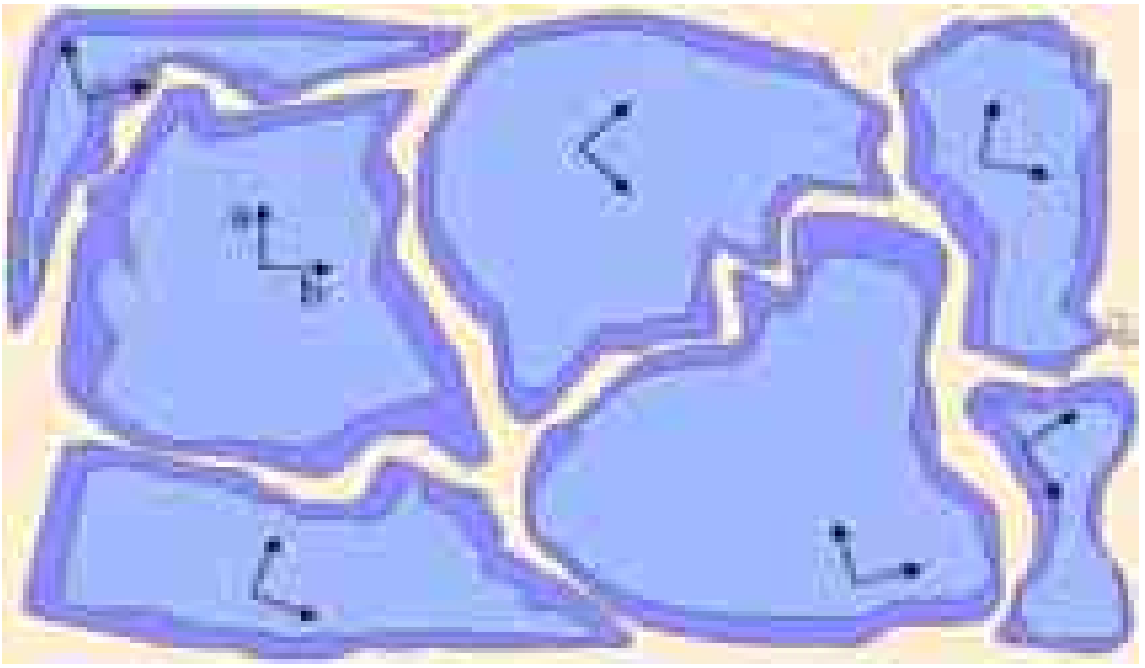

**Fig. 8** Schematics of the HTS microstructure: Differently oriented single crystal grains are separated by regions filled with secondary phases. In addition, oxygen depletion may occur at grain boundaries.

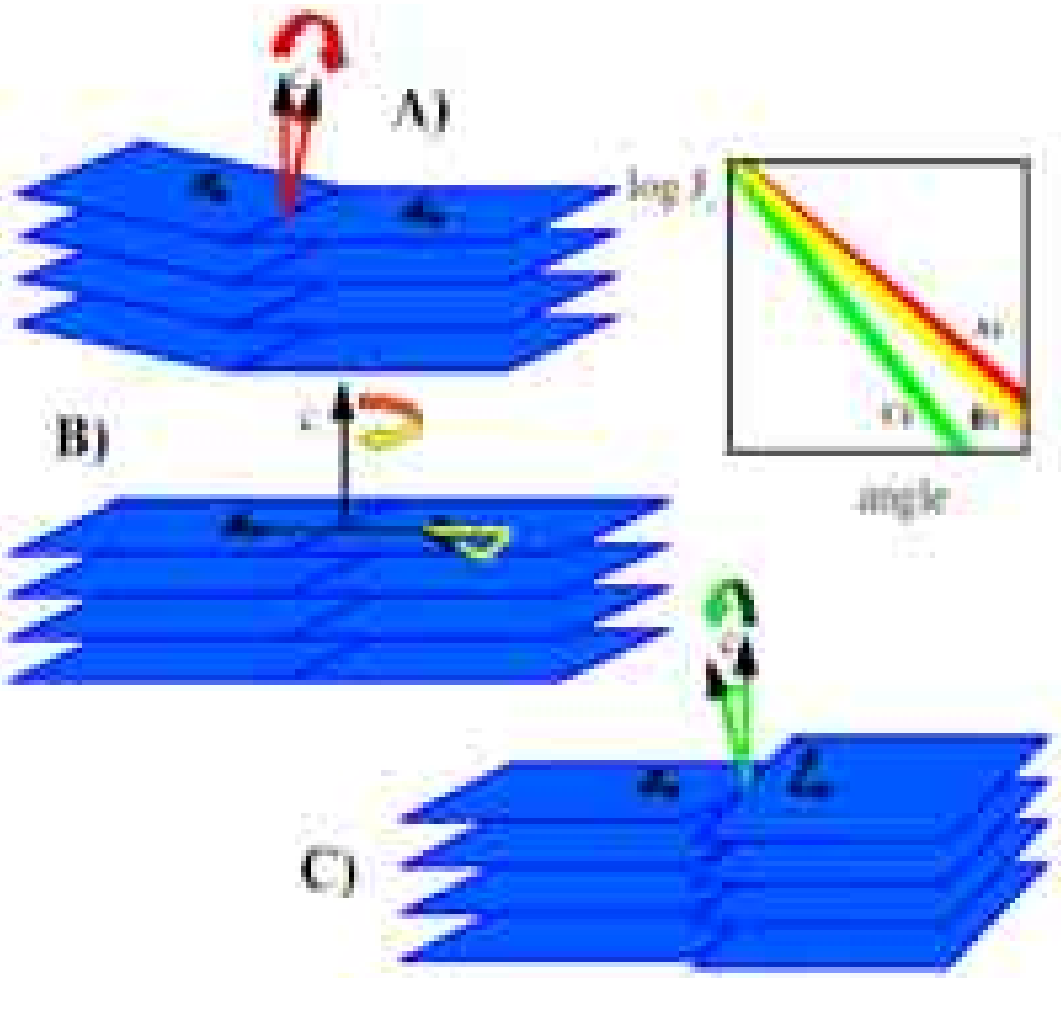

**Fig. 9** Basic grain boundary geometries and experimentally observed $J_c$ reduction $J_c \sim e^{-\alpha/\alpha_0}$ as function of the misalignment angle $\alpha$: $\alpha_0 \approx 5°$ for A) and B), $\alpha_0 \approx 3°$ for C) independent of temperature [178,179].

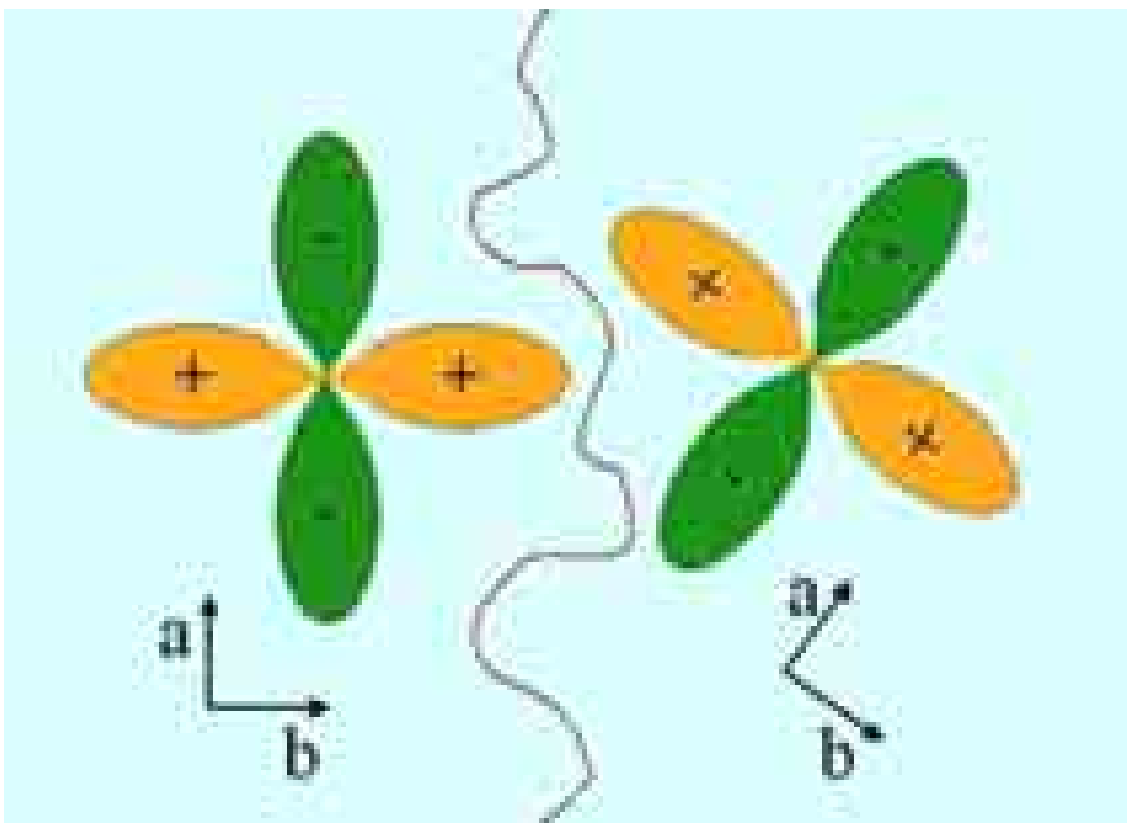

**Fig. 10** Schematics of a boundary between HTS grains. Misorientation of the SC d-wave order parameter leads to partial cancellation of the supercurrents modified by the faceting of the grain boundaries.

Another obstacle for supercurrents in HTS is misalignment of the grains: Exponential degradation of the supercurrent transport is observed as a function of the misalignment angle (see Fig. 9). One of the reasons for this behavior is the d-symmetry of the SC order parameter (see Fig. 10) [142]. However, the $J_c$ reduction as a function of the misalignment angle $\alpha$ turns out to be much larger than what is expected from d-wave symmetry alone [178,179,180,181]. This extra $J_c$ degradation as well as the change of the current-voltage characteristics of the transport behavior[33] [182] are believed to arise from structural defects such as dislocations [183] and deviations from stoichiometry. In particular, the loss of oxygen at the grain surfaces [184] leads to a decrease of doping with respect to the grain bulk value and thus to a local degradation of the SC properties according to the temperature-doping phase diagram of HTS (see Fig. 13) [185]. For YBCO this can be partially eliminated by means of Ca-doping [186], which widens the range of acceptable grain misalignment in technical HTS material.

### 3G. BASIC PHYSICS

The $CuO_2$ layers are believed to be the location of the mobile charge carriers in the cuprate HTS compounds [50]. This is in agreement with band structure calculations (see Fig. 11 a) [187,188]. However, in contrast to these results based on the independent electron approximation the stoichiometric compounds are actually antiferromagnetic ("AF") insulators due to strong correlation effects: In HTS cuprates, the copper 3d-shells are filled with nine electrons, resulting in $Cu^{2+}$ ions in $d^9$ configuration which are subject to a strong Jahn-Teller effect[34]. The deformation of the oxygen octahedra which surround each Cu ion make the $d_{x2-y2}$ orbital the only unoccupied Cu 3d orbital, i. e., it accomodates the single hole [190]. This copper $d_{x2-y2}$ state hybridizes with the p$\sigma$-orbitals of the surrounding four oxygen atoms (see Fig. 12). Due to the strong Coulomb repulsion for adding another hole at a Cu site these holes cannot move to neighboring Cu sites. This Coulomb correlation effect prevents a metallic behavior and turns the stoichiometric HTS compounds into insulators[35] (see Fig. 11 b). Virtual charge fluctuations generate a "superexchange" interaction, which favors antiparallel alignment of neighboring spins [192]. The result is long-range AF order up to rather high Néel temperatures $T_N$ = 250 - 400 K [50].

---

[33] In addition to the quantitative effect on increasing misalignment angles of a reduction of the supercurrents, the current-voltage characteristics shows in addition a qualitative change from a flux-flow behavior for smaller misalignment angles [25,178,179] to overdamped Josephson junction behavior for larger misalignment angles [32].

[34] The starting point of K. A. Müller's investigations on copper oxides were considerations that this might be used to achieve a large electron-phonon coupling with beneficial effects for an onset of superconductivity.

[35] Actually, the metal-insulator transition of copper oxides is more complicated than this classical Mott scenario. The gap for electronic excitations is here not determined by the mutual Coulomb repulsion of copper 3 $d_{x2-y2}$ holes but by the energy required to transfer a charge from an oxygen ion to the copper ion since the highest occupied electronic states are derived from oxygen 2p orbitals (see Fig. 11 b) with only a small admixture of Cu character (see Fig. 11 c) [189]. Hence, undoped copper oxides are not "Mott(-Hubbard) insulators" but "Charge-Transfer insulators".

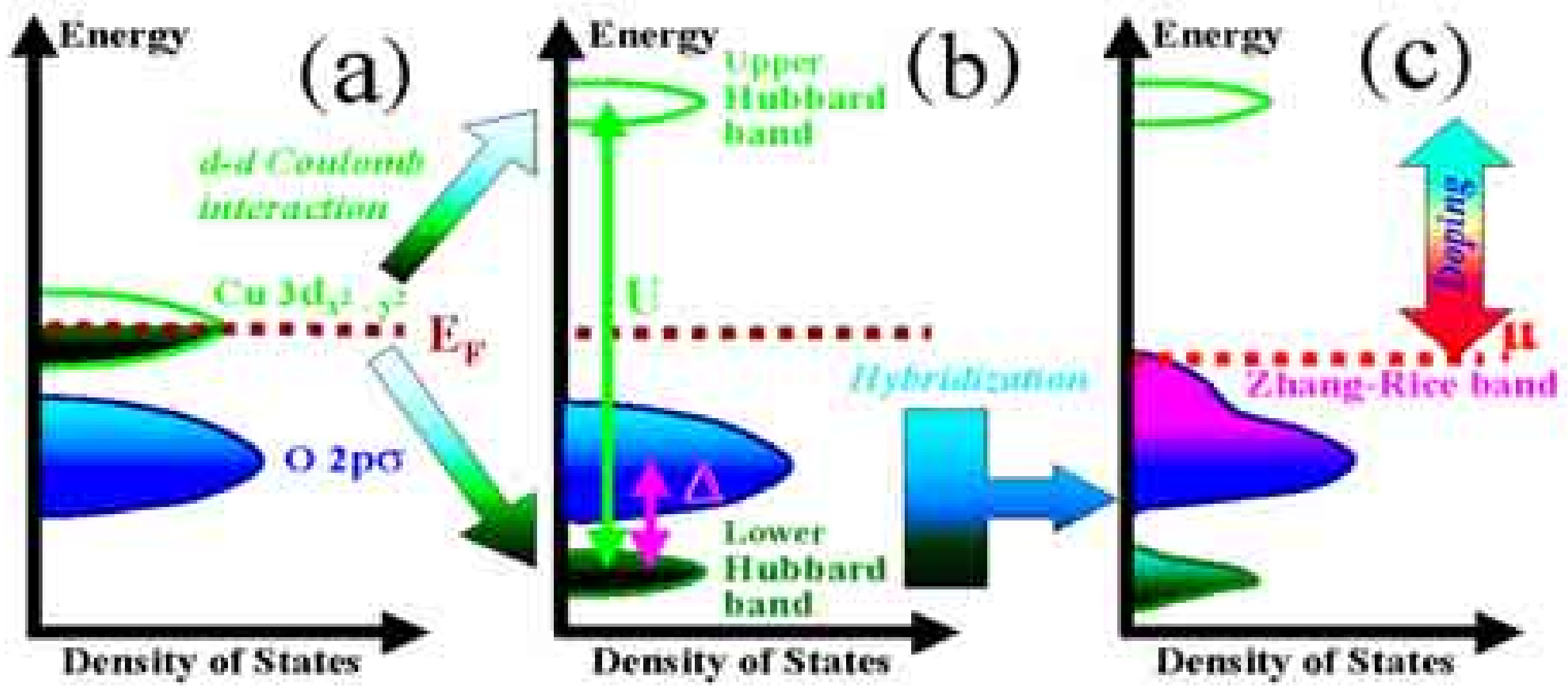

**Fig. 11** Schematics of the electronic energy states involved in the charge transport in HTS
(a) in an independent electron model,
(b) including on-site Coulomb repulsion of the Cu $3d_{x2-y2}$ states,
(c) and hybridization of oxygen 2pσ states and the lower Hubbard Cu 3d band [191].

What is special about copper oxides in the big zoo of conducting oxides is that they are composed of $CuO_2$ layers which form 2-dimensional spin-1/2 Heisenberg AF subsystems that are subject to particularly strong quantum spin fluctuations[36].

---

[36] This is the starting point for many theoretical description of the High-Temperature Superconduvtivity in cuprate compounds such as the Resonating Valence Bond ("RVB") [51,52] or the Quantum Critical Point ("QCP") scenario [193].

In the RVB picture, the antiferromagnetic lattice is perceived as a network of spin singlets made up of pairs of antiparallel Cu spins of the AF lattice. The ground state of the antiferromagnetic lattice is in this view a "resonating" superposition of such configurations of "valence bonds" in close analogy to the chemical picture of resonating double bonds in benzene. Putting an additional oxygen hole on a certain lattice site compensates the corresponding Cu spin by means of a singlet formation and leaves somewhere an unpaired spin of the related broken valence bond. The same holds true for the removal of a Cu hole (along with its spin). The many-body ground state of such a system with broken valence bonds is suggested to be a liquid-like state made up of the remaining resonating valence bonds and the decoupled charges ("holons") and unpaired spins ("spinons") as independent degrees of freedom.

Mathematically, the RVB wave function can be generated by projecting out double (lattice site) occupations from a d-wave BCS wavefunction used as a variational testfunction. However, due to the mathematical complexity of this approach, practical computations could be performed only on the basis of rather crude approximations. A recent refinement [194, 195] was claimed to show that RVB is able to explain many experimental features, but the criticism that the RVB concept is rather vague did not fall silent [196]. A recent RVB derivative is the "gossamer" superconductor wavefunction where the double occupations are only partially removed [197]. The new idea is that in the limit of small doping HTS do not become real insulators but remain superconductors, however, with an extremely large gap and an extremely rarefied superfluid density [198].

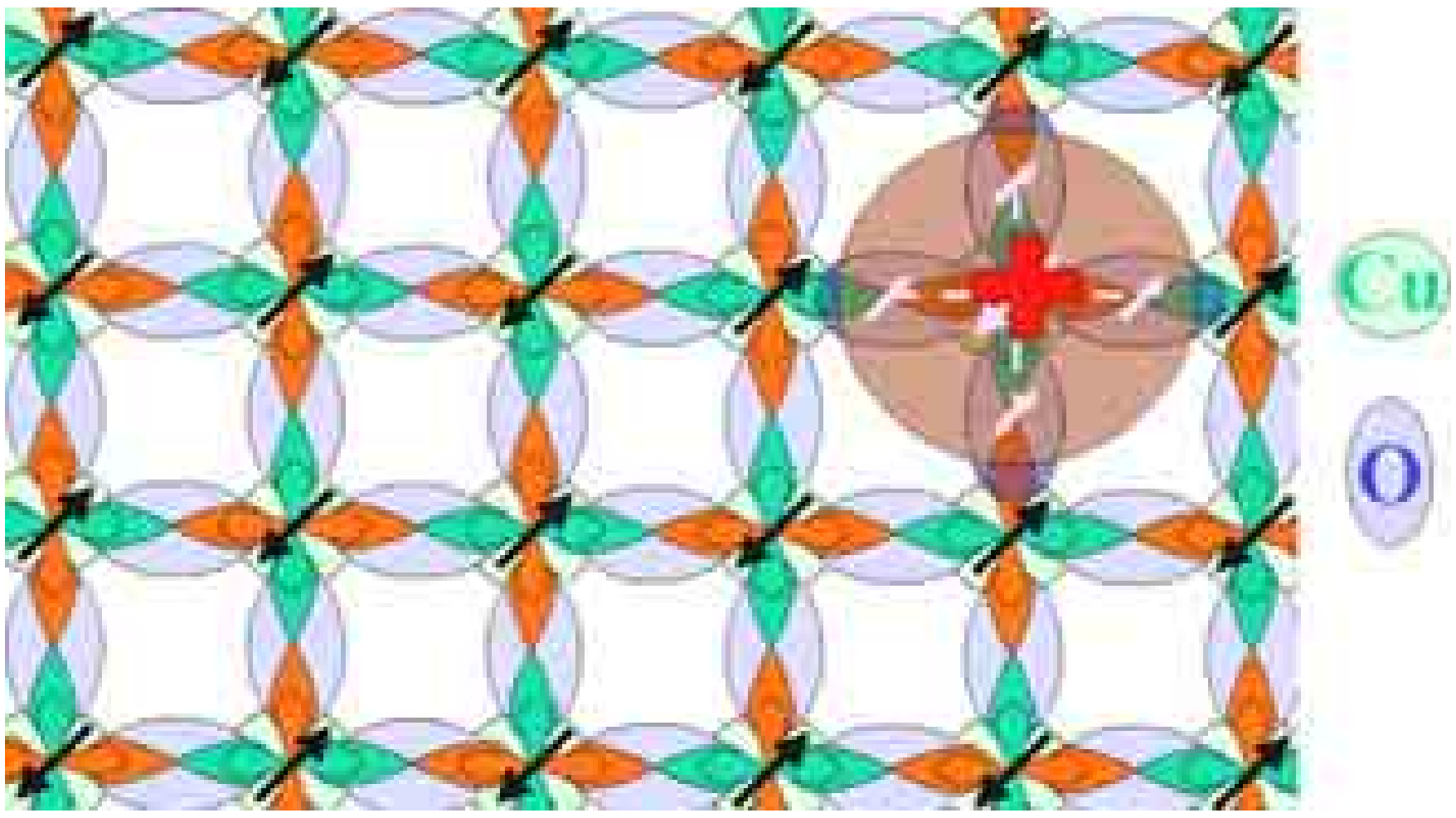

**Fig. 12** Schematics of the spin-charge distribution atomic orbitals in the $CuO_2$ planes[37] with an additional hole charge. Doped holes are believed to enter a bonding combination of four oxygen pσ-orbitals surrounding a copper ion. Antiparallel orientation of the involved spins leads to the formation of a "Zhang-Rice" singlet state with charge Q = + e and spin S = 0 [199,200]. Doped electrons, in contrast, are believed to go to the copper sites and to eliminate there directly the hole in the $3d_{x2-y2}$ state along with its spin.

The following scenario applies for *hole-* [50] (see Fig. 12) as well as for *electron-doping* [201]: Adding charge carriers by variation of the oxygen content or by suitable substitution of cations relaxes the restrictions of spin alignment due to the interaction of these additional spin-1/2-particles with the spin lattice. $T_N$ decreases and the insulator turns into a "bad metal": The room-temperature mobility of the charge carriers increases gradually, in direct inverse proportionality to the AF correlation length [202]. At low temperature, however, the electric transport shows a dramatic change within a small doping range from an insulating behavior with an divergent upturn of the resistivity on decreasing temperature, to a superconducting behavior with a respective sudden downturn of resistivity next to $T_c$ [202]. For $La_{2-x}Sr_xCuO_4$ this happens at a critical hole concentration x = 0.05 in the $CuO_2$ planes[38] (see Fig. 13). On stronger doping, superconductivity can be observed up to an increasingly higher critical temperature until for "optimal doping" ( x ≈ 0.16 for $La_{2-x}Sr_xCuO_4$ ) the maximum $T_c$ is

---

[37] The indicated in-plane spin orientations are encountered in LSCO [204]. Most interestingly, in lightly doped antiferromagnetic LSCO this in-plane spin orientation is manifestly linked with the orthorhombic lattice distortion: LSCO can be magnetically detwinned applying a 10 T in-plane magnetic field [203]. This gives evidence of a close connection of spin and lattice degrees of freedom in cuprate HTS.

In YBCO the spins are in-plane aligned parallel to the CuO bonds [205].

[38] For $La_{2-x}Sr_xCuO_4$, a spin-glass like state has been reported at this critical doping x ~ 0.05. However, due to the tendency of HTS towards spatial doping inhomogeneites it is unclear whether this kind of decay of the long-range AF order is intrinsic.

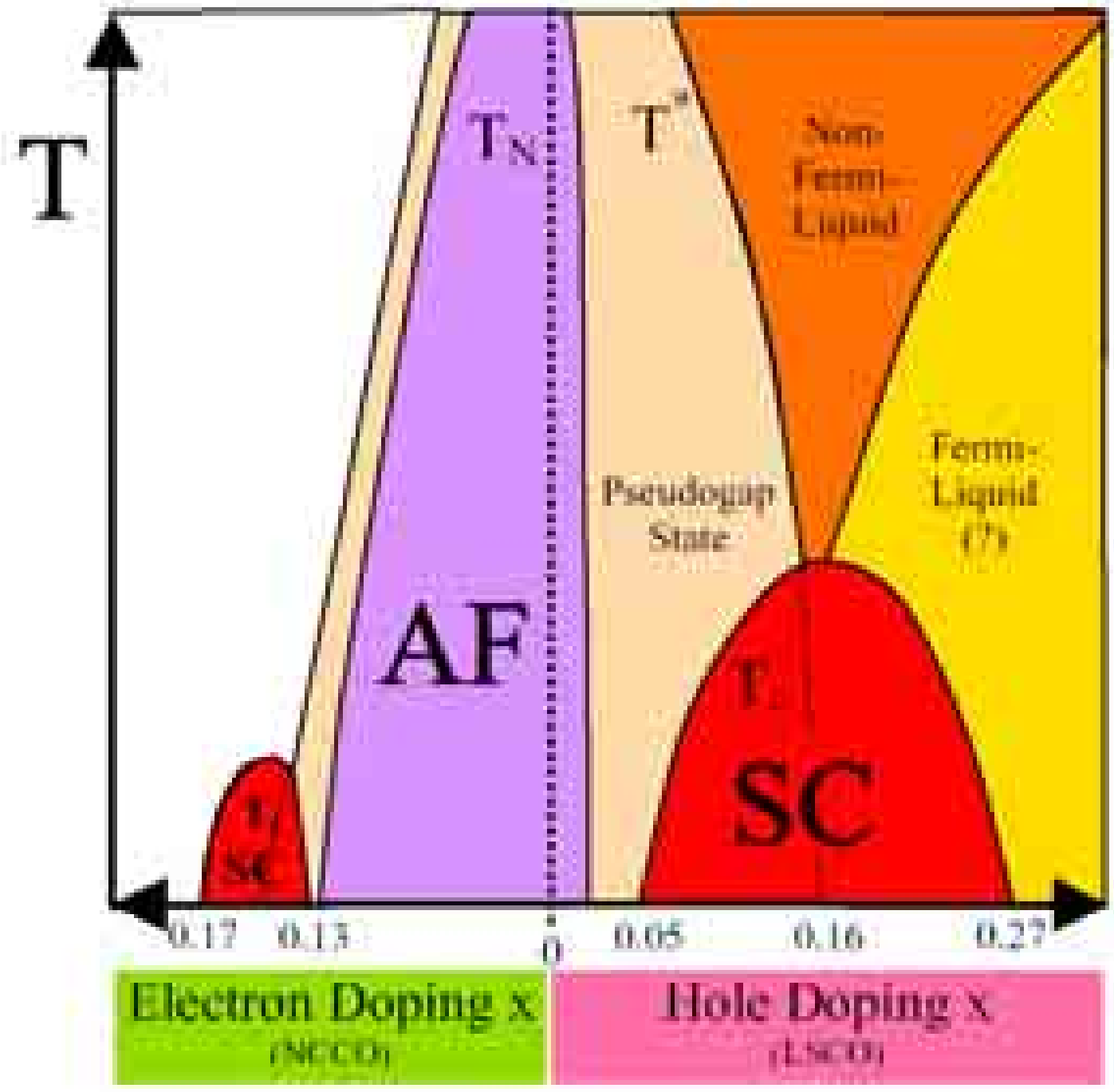

**Fig. 13** Schematic HTS temperature-doping phase diagram dominated by the interplay of antiferromagnetism ("AF") and superconductivity ("SC") [50,206,207,208].

achieved. On further doping, the critical temperature decreases again until finally (x ≥ 0.27 for $La_{2-x}Sr_xCuO_4$) only normal conducting behavior is observed[39].

The rise and fall of $T_c$ as a function of doping leads to the classification of the corresponding regions of the chemical phase diagram (see Fig. 13) as "underdoped" and "overdoped" regimes, respectively. At a first glance, the physical properties of these two regimes seem to be not so different for the SC state[40], but they are dramatically different for the normal state: In the underdoped regime for temperatures far above $T_c$, even up to room temperature, a peculiar redistribution of electronic states in the energy spectrum is observed in the vicinity of the Fermi energy resembling the features of a SC energy gap [214,215]. The critical temperature $T^*$ associated with this "*pseudogap*" energy decreases monotonically on doping, in stark contrast to the monotonic increase of $T_c$ in this doping regime (see Fig. 13). In the overdoped regime, in electron spectroscopy only a regular SC gap is identified. This is widely interpreted as coincidence of the SC gap and the pseudogap, which is hence regarded as a precursor of the SC gap [216]. However, from other experiments it has been concluded that $T^*(x)$ may actually continue its rapid decrease as a function of doping x "inside of the SC dome", i. e., below $T_c$, and that the pseudogap actually vanishes at a critical value $x_c$ close or even identical to the optimal doping [206].

---

[39] This doping scenario has bee suggested to be even quantitatively universal with respect to the concentration x of holes or electrons in the $CuO_2$ planes [206] albeit with a large asymmetry between hole- and electron-doping (actually, there is no principal reason for electron-hole doping symmetry since different electronic states are affected; see Fig. 11), and with individual $T_c$ and $T_N$ parameters for each HTS compound (see Fig. 13). However, at least for the R-123 HTS family there is a substantial deviation of $T_c(x)$ from this "Tallon" parabola, with a $T_c$ maximum around x ~ 0.25 and a "dent" or "plateau" at x ~ 0.15 [209,210,211].

[40] The SC condensation energy is lower [212] and the SC anisotropy is stronger in the underdoped regime [213], featuring an even more 2-dimensional character with stronger SC fluctuations.

The so-called normal state of HTS is not at all "normal" according to the laws of conventional metal physics. In fact, it is now considered to be even more astonishing than the SC state. Early on, the dependencies of its physical properties were found to be highly unusual [217,218]. For *optimally doped* HTS, a linear temperature dependence of the electrical resistance $\rho(T) = \rho_0 + \rho_1 T$ is observed over the entire accessible temperature range up to 1000 K with essentially the same slope $\rho_1$ for all HTS compounds[41] [217]. In spite of a well-developed Fermi surface [219,220,221,222], a Landau Fermi-liquid ("FL") description of these dependencies necessitates the assumption of a strangely modified excitation spectrum ("Marginal Fermi Liquid") [223]. As another indication of the breakdown of the Landau FL theory, in optimally (electron-)doped $Pr_{2-x}Ce_xCuO_4$ with the SC state quenched by a magnetic field the Wiedemann-Franz relation of charge and heat transport seeems to be not obeyed even in the limit $T \rightarrow 0$ K[42] [224].

In photoelectron spectroscopy, in the energy spectrum of the normal state no sharp quasiparticle peaks are observed, but only very broad features [50,230,231]. This can no longer be attributed to insufficient sample quality[43] or intrinsic problems of photoelectron spectroscopy [233,234]: In the SC state, already even slightly above $T_c$, on top of these broad features distinct quasiparticle peaks appear at energies right above the pseudogap. These peaks become continuously sharper towards lower temperatures[44] [237]. This looks quite like the opposite of the familiar BCS scenario of a condensation of elementary normal state quasi-particles into a complex SC many-particle state [5,238] and gave rise to the speculation that charge may not be a good quantum number for the quasiparticles associated with the pseudogap state[45].

---

41 $\rho(T) \approx \rho_0 + \rho_1 T^\alpha$ with $\alpha < 1$ for underdoped HTS. $\alpha \rightarrow 2$ for overdoping indicates the "normalisation" towards a regular Fermi liquid description of the normal state resistivity.

42 PCCO was selected for this experiment since a comparatively low, technically readily accessible $H_{c2} \leq 14$ T is believed to allow a magnetic suppression of superconductivity [224]. However, even if such a magnetic field is sufficient to destroy the phase coherence of the Cooper pairs in HTS, the SC pairing appears to be rather insensitive to these fields [225]. This argument, derived for hole-doped HTS, may also apply to electron-doped HTS such as PCCO where magnetic fields of the order of 10 T have a much more dramatic effect on the physical properties than for hole-doped HTS [208]. Nevertheless, measurement of the voltage induced by vortex flow in a driving temperature gradient (Nernst effect [226]) in magnetic fields up to 45 tesla seems to allow a more precise determination of $H_{c2}$ [24]. Such measurements indicate in fact, e. g., $H_{c2}$ $(T \rightarrow 0) = 10$ T for optimally electron-doped $Nd_{2-x}Ce_xCuO_4$ ("NCCO"; $x_{opt} = 0.15$) in contrast to $H_{c2}$ $(T \rightarrow 0) \sim 60$ T for optimally hole-doped Bi-2212. Recent experiments on overdoped Tl-2201 [227], LSCO [228] and optimally doped Bi-2201 [229] indicate the validity of the Wiedemann-Franz relation for these hole-doped HTS compounds.

43 Bi-2212 is the HTS best-suited for photoelectron spectroscopy experiments [232] since the lattice can easily be cleaved in-between two adjacent BiO layers which are only weakly ("van-der-Waals") bonded. This leads to a micaceous material consistency (pure single-crystalline films can easily be wiped away by cotton wool) which offers good prospects for undegraded sample surfaces after cleaving in vacuum. Detwinned YBCO single crystals have meanwhile reached sufficient surface perfection as well [139]. However, the analysis of the quasiparticle peaks is complicated here by a low-energy surface state which is probably due to the cleaving of the lattice in-between a CuO chain layer and a BaO layer. Similar concerns have to be borne in mind for other HTS materials without a suitable cleavage plane.

Another amazing experimental finding was the discovery of vortex-like excitations in the pseudogap phase of underdoped HTS well above $T_c$ [240,241,242,243], which indicate the presence of strong SC fluctuations [216,225,244] even in this normal state. This points to the idea of "preformed (Cooper) pairs" [245] which exist already above $T_c$ as noncoherent fluctuations and achieve coherent behavior only at lower temperature after condensation into the SC state. For spatially localized pairs as suggested for HTS by their small coherence length (e. g. for (optimally doped) YBCO: $\xi_{ab}$ = 1.6 nm, $\xi_c$ = 0.3 nm for T $\rightarrow$ 0 K [175]; lattice parameters: a = 0.382 nm, b = 0.389 nm, c = 1.167 nm [66]), such a preforming of pairs is much easier to imagine than for the cloud of spatially strongly overlapping pair states associated with the BCS description of conventional superconductors.

In this spirit, the difference between the underdoped and the overdoped regime has been attributed to a doping-induced crossover from the Bose-Einstein (“BE”) condensation of preformed pairs in underdoped HTS, to a BCS pair-state formation in overdoped HTS [246,247]. The assumption that the pseudogap is a redistribution of the electronic states due to incoherent pair formation is tantamount to identifying the pseudogap as precursor of the BCS gap $\Delta$ which directly measures the binding of electrons into Cooper pairs. Even in the description of conventional superconductors the superfluid density $\rho_s$ introduces a second temperature scale besides $\Delta$ [246]: $\rho_s$ plays the role of a phase stiffness which measures the ability of the SC state to carry a supercurrent. In conventional superconductors, $\Delta$ is much smaller than the energy associated with $\rho_s$, and the destruction of superconductivity begins with the breakup of electron pairs. However, in cuprates the two energy scales seem to be more closely balanced. Actually, in underdoped materials the ordering is apparently reversed, with the phase stiffness now being the weaker link. When the temperature exceeds the critical temperature associated with $\rho_s$, thermal agitation destroys the ability of the superconductor to carry a supercurrent while the pairs continue to exist [248,249]. According to this phenomenological interpretation of the pseudogap, $T_c$ would be defined here by $\rho_s$ and not by $\Delta$ as in the case of conventional superconductors.

---

[44] The peak amplitude is apparently proportional to the superfluid density: In experiments on Bi-2212 samples with a broad variety of doping levels [235], the amplitude of these quasiparticle peaks measured at low temperature has been determined to show more or less a $T_c$-parabola-like behavior as a function of doping, very similar to the behavior of the low temperature superfluid density measured in μSR, or of the jump of the specific heat coefficient at $T_c$ [236].

[45] The charge-detection principle of photoelectron spectroscopy would then imply the observation of superpositions of quasiparticle states with the consequence of a respective energy broadening. This argument has been discussed in the context of spin-charge separation as predicted by RVB theory [51,52,167,239].

Another suggestion is that the pseudogap is due to a "hidden order" [250]. The favorite candidate for HTS cuprates is a $d_{x2-y2}$ density wave state (see Fig. 14) which was already discussed in the late 1980s as ground state in HTS-oriented model calculations[46] [251,252], and which is also in agreement with the d-wave characteristics of the pseudogap as observed in photoelectron spectroscopy [253,254]. This $d_{x2-y2}$ density wave state is a particular example for a recently suggested extension of charge density waves (CDWs) to more general density waves based on the condensation of particle-hole pairs of a certain angular momentum into a BCS-like many-particle state [255]. This results in an energy gap for single-particle excitations, in close analogy with the usual BCS formalism. Conventional CDWs are conceived in this framework as s-density wave states where the charges are subject to a coherent, but in this case merely translatory motion. For non-zero angular momentum density wave states, additional synchronization of orbital motion comes into play. On a lattice, non-zero angular momentum density wave states correspond to patterns of circular bond currents (see Fig. 14), which describe perfectly synchronized hopping processes. This mechanism enables a charge transport without strong hindrance by mutual Coulomb repulsion.

Another attractive feature of the $d_{x2-y2}$ density wave state approach to the pseudogap is that the universal HTS-$T_c$ parabola as a function of doping can be explained quantitatively in a Ginzburg-Landau approach in terms of a competition between $d_{x2-y2}$ superconductivity based on particle-particle (hole-hole) pairing, and the $d_{x2-y2}$ density wave state based on particle-hole pairing [250]. A major difference between the respective energy gaps is that the SC gap remains fixed at the Fermi energy while the $d_{x2-y2}$ density wave depends on the doping. In mean-field approximation it is found to be fully developed only at half filling of the energy band. It becomes reduced for both doping directions and vanishes towards either complete filling or full depleting. Physically this can be understood from the discussed property of the $d_{x2-y2}$ density wave state that it allows a charge transport avoiding strong Coulomb repulsion. This energy reduction is most effective for half filling whereas no gain of correlation energy can be expected for the motion of single electrons or holes across the empty lattice.

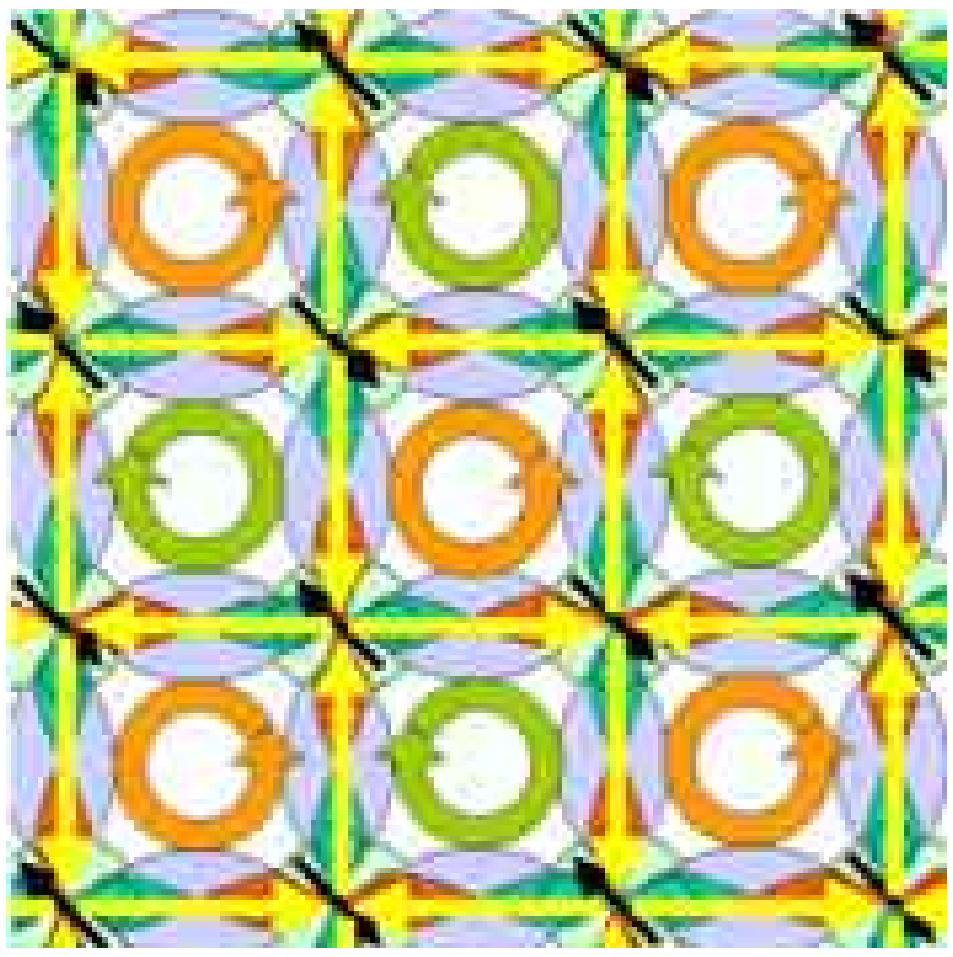

**Fig. 14** $d_{x2-y2}$ density wave on a $CuO_2$ lattice based on bond currents (yellow) summing up to circulating currents of alternating orientation (orange/green).

---

[46] Here it was called “Staggered Flux“ state due to the magnetic flux generated by the circulating currents (see Fig. 14).

The neighborhood of the pseudogap and superconductor phases to the AF phase in the temperature-doping phase diagram points to a spin-related superconductivity mechanism. Starting out from static AF order for the undoped case, it is plausible that on increasing the doping strong AF correlations will persist in the underdoped pseudogap regime. Actually, neutron scattering experiments show that the Bragg peaks which result from the doubling of the real space unit cell due to antiferromagnetism survive in broadened versions, at least in the form of spin fluctuation patterns which can be probed by inelastic neutron scattering (INS) [256,257]. Surprisingly, in the underdoped regime for $T < T^*$ the single peaks in reciprocal space found in the insulator split into four, each displaced in orthogonal directions by a small amount (see Fig. 15 a) [256,257,258,259,260,261,262,263,264].

This scattering had at first been interpreted in terms of spin fluctuations described by linear response theory, which corresponds to sinusoidal spin density waves (SDW) fluctuating slowly in space and time[47] [265]. Spin waves are the Goldstone modes with respect to the AF order parameter and are therefore expected as low energy excitations. However, the measured energy dispersion is not compatible with conventional SDWs. Nevertheless, the INS k-space pattern points to modified spin waves [257]. A prominent suggestion was that the charges introduced by doping into the AF spin lattice may align in equidistant rows to form parallel "stripes" in transversal orientation to the spin wave, thus forming a charge density wave running in the same direction but with half the spatial period of the corresponding spin wave[48] [266,267,268;261,262]. This construction results in perfect commensurate AF spin domains extending over half a spin wave period, enclosed by charge stripes. Here a phase jump occurs so that the spin pattern of the neighboring AF domains appears to be spatially shifted by one lattice position. Stripes have been suggested to constitute a fundamental structural element of the ground state of all doped AF insulators arising from a competition between phase separation (the tendency of an AF insulator to expel doped holes) and the long-range part of the Coulomb interaction[49] [272,273]. The incommensurability ε[50] increases monotonically with doping up to optimal doping where it saturates at $\varepsilon \sim \pi/4$. A quantitative description of this experimental incommensurability-doping relation [50] was given by the concept of "metallic stripes" [275,276,277,278] where only every other “spin”, i. e., hole along with its spin, is removed from the stripe positions: The remaining spins delocalize along the stripe forming metallic zone separating the AF domains.

---

[47] An effective correlation length of ~ 20 nm has been derived from the width of the scattering peaks [262]. This corresponds to an extension of the spin fluctuations over ~ 50 unit cells.

[48] INS observations of a broadening of the linewidth of phonons stemming from the motion of planar oxygen atoms had been interpreted as evidence for the existence of charge stripe fluctuations [269,270]. However, this interpretation has been disproved by later INS measurements [271].

[49] This interpretation was inspired by measurements on $La_{2-x-y}R_ySr_xCuO_4$ where partial substitution of La by R = Nd [261] or R = Eu [274] condenses the spin fluctuation into a static spin wave.

[50] measured as the splitting of the AF Bragg peak $(\pi,\pi) \rightarrow (\pi \pm \varepsilon, \pi \pm \varepsilon)$

The original stripe proposal [261,262] assumed one-dimensional stripes following a much earlier theoretical suggestion [266,267]. The four peaks (π ± ε, π ± ε) were believed to stem from a superposition of the signals (π ± ε, π) and (π, π ± ε), e. g., from the two twin domains of the usually twinned single crystal samples. INS measurements on partially detwinned samples seemed to back up this argument [279]. However, recent INS experiments on fully detwinned YBCO single crystals [280] as well as LSCO [281] showed that this decomposition is not appropriate. The correct Fourier transformation of the full INS k-space pattern leads to the 2-dimensional real space pattern as shown in Fig. 15 where the stripes run in "diagonal" spatial directions, in stark contrast to the reported naive interpretation of the INS Fourier peaks. Assuming again half-filling of the metallic stripes, the 2-dimensional stripe pattern reproduces equally well the experimental incommensurability-doping relation. Surprisingly, this stripe pattern was already presented in the original "stripe" paper as an alternate suggestion [267].

What has not yet been noticed is that it fits perfectly the "checkerboard" pattern which is observed in atomic resolution Scanning Tunneling Microscopy (STM) on Bi-2212 samples[51] [284] (Fig. 15 (e)) ! The caveat is that the STM pictures reproduce a static phenomenon whereas INS measures dynamic excitations[52] [285]. Nevertheless, pinning forces acting on these non-conventional spin-and-charge density waves may explain the actually observed tiny static spatially periodic charge accumulations[53]. The "checkerboard" pattern was first seen in STM only in the low-temperature SC regime. It has been interpreted in terms of elastic scattering between characteristic regions (gap nodes) of the Fermi surface as observed in photoelectron spectroscopy [286,287]. The observed spatial periodicity was attributed to the difference of the k-space wave vectors around the gap nodes on the Fermi surface to where the scattering is energetically restricted by the SC gap (see Fig. 16). Recent STM measurements on slightly underdoped Bi-2212 found such modulations above $T_c$ as well and attributed it to the energetic restrictions introduced by the pseudogap [288].

The topical question is whether this scattering is only an artifact of the STM observation or if it is already present in the HTS material as a particular spin-charge density wave [289]. The indicated compatibility of the collective spin excitation "stripe" pattern derived from INS with the STM "checkerboard" charge pattern speaks in favor of the latter one as the basic physical entity and against the idea that this real space pattern is only a mirage generated by the STM observation due to interference of single-particle excitations.

---

[51] Erroneous indications of the real space direction of the experimental stripes are widely encountered in literature. This confusion is mainly due to the counterintuitive result of the Fourier transformation (see caption of Fig. 15). The differerent crystallographic notations used for orthorhombic and tetragonal LSCO where the first stripe formation was observed by INS may be additionaly confusing: The so-called orthorhombic *a*- and *b*- directions of LSCO are rotated by 45° with respect to their tetragonal counterparts which are commonly denoted in literature as "HTS *a* and *b* axes" [282,283].

[52] Only in the case of $La_{2-x-y}R_ySr_xCuO_4$ with partial substitution of La by R = Nd [261] or R = Eu [274] for the doping x = 1/8 the spin fluctuation is condensed to a static spin wave, however, with the consequence of the disappearence of superconductivity [260].

[53] The dislocations in the STM picture Fig. 15 e indicate the presence of such pinning centers.

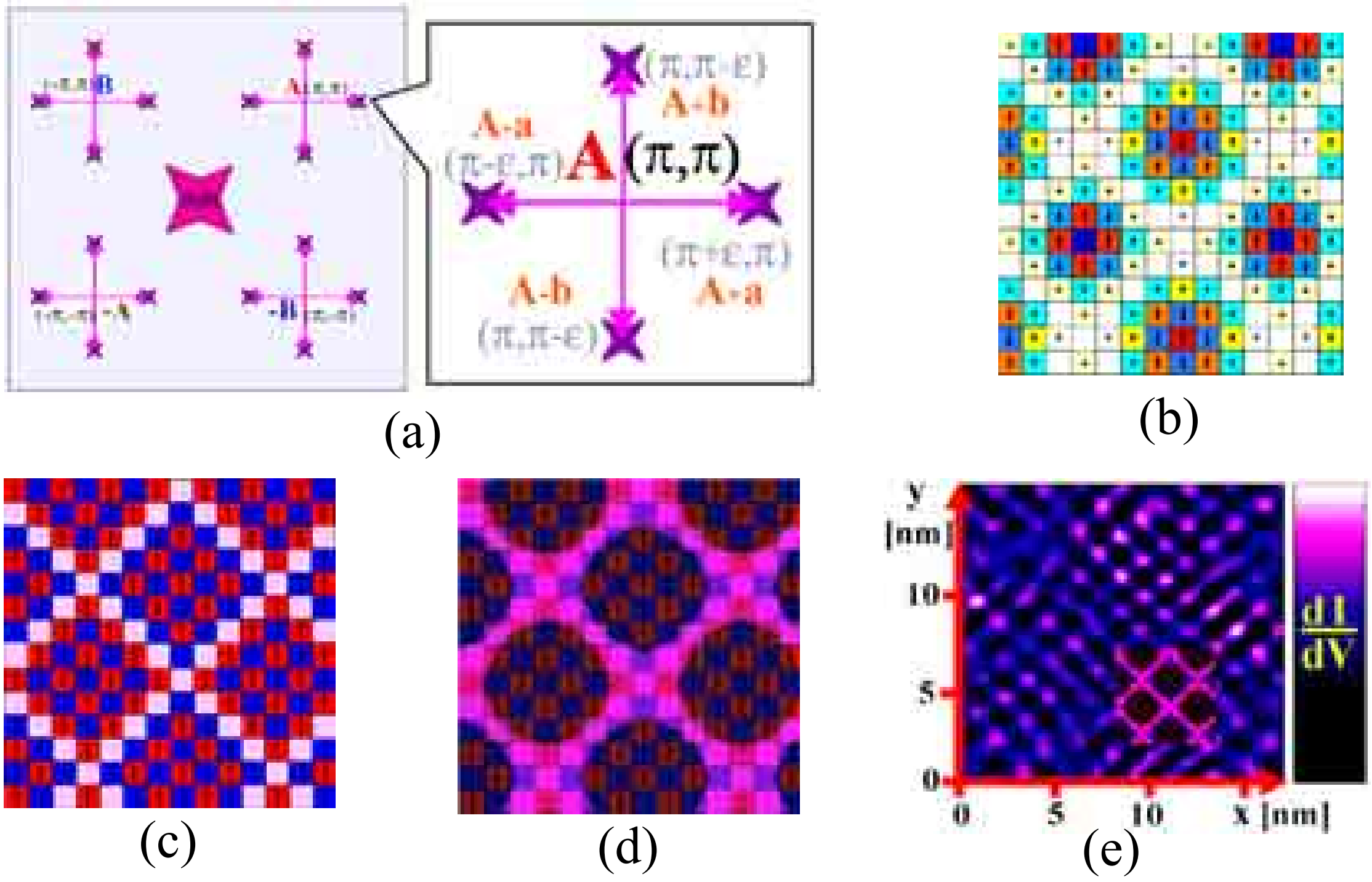

**Fig. 15** (a) Schematics of the incommensurate low-energy INS k-space spectrum [263].

(b) Schematics of the corresponding harmonic spin wave real space pattern for $\varepsilon = \pi/4$, the incommensurability observed for optimally and overdoped HTS [259]:

$$\begin{aligned} f(x,y) &= \cos(\pi x) \times \cos(\pi y) \times \cos(\{\pi/8\}\{x+y\}) \times \cos(\{\pi/8\}\{x-y\}) \\ &= [\cos(\mathbf{AR}) + \cos(\mathbf{BR})] \times [\cos(\mathbf{aR}) + \cos(\mathbf{bR})] / 4 \\ &= [\cos(\{\mathbf{A+a}\}\mathbf{R}) + \cos(\{\mathbf{A-a}\}\mathbf{R}) + \cos(\{\mathbf{A+b}\}\mathbf{R}) + \cos(\{\mathbf{A-b}\}\mathbf{R}) \\ &\quad + \cos(\{\mathbf{B+a}\}\mathbf{R}) + \cos(\{\mathbf{B-a}\}\mathbf{R}) + \cos(\{\mathbf{B+b}\}\mathbf{R}) + \cos(\{\mathbf{B-b}\}\mathbf{R})\ ] / 8 \\ &= [\exp(i\{\mathbf{A+a}\}\mathbf{R}) + \exp(i\{\mathbf{A-a}\}\mathbf{R}) + \exp(i\{\mathbf{A+b}\}\mathbf{R}) + \exp(i\{\mathbf{A-b}\}\mathbf{R}) \\ &\quad + \exp(i\{\mathbf{B+a}\}\mathbf{R}) + \exp(i\{\mathbf{B-a}\}\mathbf{R}) + \exp(i\{\mathbf{B+b}\}\mathbf{R}) + \exp(i\{\mathbf{B-b}\}\mathbf{R}) \\ &\quad + \exp(-i\{\mathbf{A+a}\}\mathbf{R}) + \exp(-i\{\mathbf{A-a}\}\mathbf{R}) + \exp(-i\{\mathbf{A+b}\}\mathbf{R}) + \exp(-i\{\mathbf{A-b}\}\mathbf{R}) \\ &\quad + \exp(-i\{\mathbf{B+a}\}\mathbf{R}) + \exp(-i\{\mathbf{B-a}\}\mathbf{R}) + \exp(-i\{\mathbf{B+b}\}\mathbf{R}) + \exp(-i\{\mathbf{B-b}\}\mathbf{R})] / 8 \end{aligned}$$

with $\mathbf{R} = (x,y)$, $\mathbf{A} = (\pi,\pi)$, $\mathbf{B} = (\pi,-\pi)$, $\mathbf{a} = (\pi/4,0)$, $\mathbf{b} = (0,\pi/4)$. [54]

(c) 2d-stripe distortion of the harmonic spin wave pattern (b).

(d) Schematical representation of pattern (c) with “diagonal” metallic stripes separating insulating AF spin domains.

(e) Fourier-filtered dI/dV real space pattern observed in STM over a 16 nm x 16 nm scan area of a slightly overdoped Bi-2212 single crystal (adapted from [284]). Figure (d) is integrated as appropriately scaled inset.

---

[54] $\cos(\alpha + \beta) + \cos(\alpha - \beta) = 2\cos(\alpha)\cos(\beta)$. The high-energy part of the incommensurate INS k-space spectrum has recently been determined to be described alike [264], but by $\mathbf{a} = (\varepsilon,\varepsilon)$ and $\mathbf{b} = (\varepsilon,-\varepsilon)$. This leads thus to $f(x,y) = \cos(\pi x) \times \cos(\pi y) \times \cos(\varepsilon x) \times \cos(\varepsilon y)$, i. e., a “parallel“ 2d-stripe pattern.

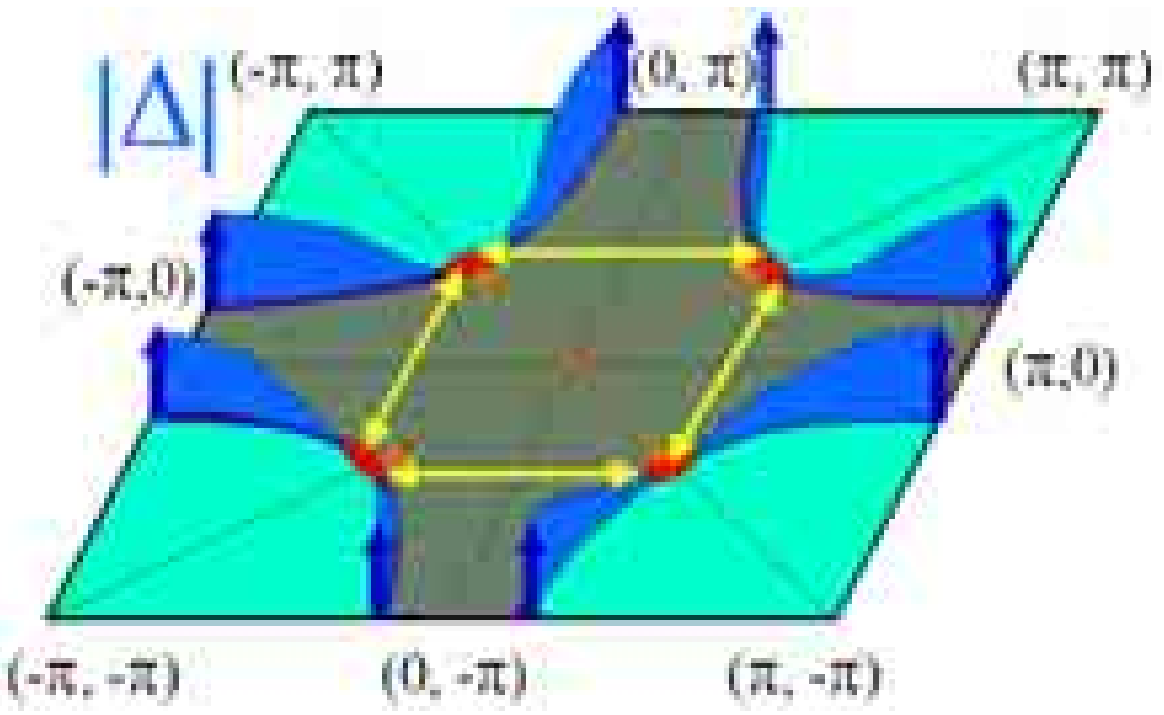

**Fig. 16** Schematics of the HTS Fermi surface with a d-wave-like gap and indicated scattering between the nodal points of the energy gap.

This discussion has interesting aspects for general physics. For a $d_{x2-y2}$–gapped Fermi surface, quasiparticles in the nodal regions make up the low-energy excitations (see Fig. 16). Time reversal symmetry connects the four nodal regions into two nonequivalent, but symmetry related pairs. Quasiparticles of either of these two pairs can be regarded as identicalparticles except for different "flavor" in the language of elementary particle physics. If the discussed "stripe-checkerboard" spin-charge agglomerations represent the basic physical entities these two pairs of quasiparticle excitations would also be intrinsically connected as a consequence of the energetic preference of the 2-d compared to the 1-d stripe configuration. As a nice toy model for quarks, this amounts to a "confinement" of two such single-flavor 2-d particles. It is charming to speculate that in our 3-d real world this argument would require a respective "confinement" of three 3-d particles as it is in fact observed for quarks. Moreover, this reasoning puts a question mark on the very idea of "elementary particles" which may turn out as artificial constructions for a single-particle description of many-body phenomena.

As a speculative summary of the presented HTS physics concepts, the stripe idea of an electronic phase separation in doped AF insulators may be extended into the following picture: The AF insulator bulk likes to get rid of the holes in order to maximize the exchange energy. The d- density wave state with its energetic preference for a "half-filling" environment provides a suitable mechanism for the formation and coherent connection of pure AF islands. The holes like to aggregate in a spatially extended metal state in order to reduce their kinetic exchange energy. An electronic phase separation into AF insulator islands surrounded by metallic 2-d network of stripes as shown in Fig. 15 c + d may be the energetically most favorable compromise[55]. As long as this spin-charge agglomeration is subject only to minor pinning forces it remains a mobile phenomenon, with the d-density wave as collective transport mechanism based on coherent particle-hole currents. The occurrence of superconductivity in the metallic stripe channels would hence be only a parasitic effect.

These recent findings and speculations show that the investigations on cuprate HTS gave and still give rise to novel theoretical concepts in solid state physics which may provide again new input for elementary particle physics.

---

[55] 1-d stripes were shown to face the "spaghetti" stability problem of getting in too close touch with each other [290]. Model simulations seem to result usually in electronic 2d-foam with metallic membranes around AF islands [268].

## 3H. TECHNICAL APPLICATIONS

The development of technically applicable HTS materials has progressed on several routes. Epitaxial YBCO [291,292] and Tl-HTS (Tl-2212) [293] *thin films* achieve excellent SC properties ($T_c > 90$ K; critical current density $J_c$ (77 K, 0 T) > $10^6$ A/cm$^2$; microwave surface resistance $R_s$(77 K, 10 GHz) < 500 μΩ, $R_s(T,f) \propto f^2$) that are well suited for superconductive electronics. They are already in use in commercial and military microwave filter systems (see Fig. 17). Since the HTS deposition takes place at 650 – 900 °C the thermal expansion coefficients of the substrates and the HTS films have to match or else the different contraction of HTS film and substrate on cooling to room or even cryogenic temperature will lead to mechanical stress which can be tolerated by the film only up to a certain maximum thickness without crack formation [295]. This limits the thickness of YBCO films on the technically most interesting substrates silicon and sapphire to ~ 50 nm [296] and ~ 250 - 300 nm [297], respectively.

YBCO *Josephson junctions* based either on Josephson tunneling through ultra-thin artificial barriers [298,299,300,301] or on the Josephson junction behavior of HTS grain boundaries [302,303,304,305,306,307] have become available with a reproducibility of the critical currents of about ± 10%. They can be used for the construction of highly sensitive SQUID magnetic field sensors. For 77 K operation a field resolution of ~ 10 fT has already been demonstrated [35] which compares favorably with the record resolution of ~ 1 fT of Nb SQUIDs operated at 4 K. Recent success in the fabrication of YBCO-Au-Nb ramp-type junctions [308] enables now the use of the d-wave symmetry related π phase shift as a new functional principle for the construction of novel electronic devices.

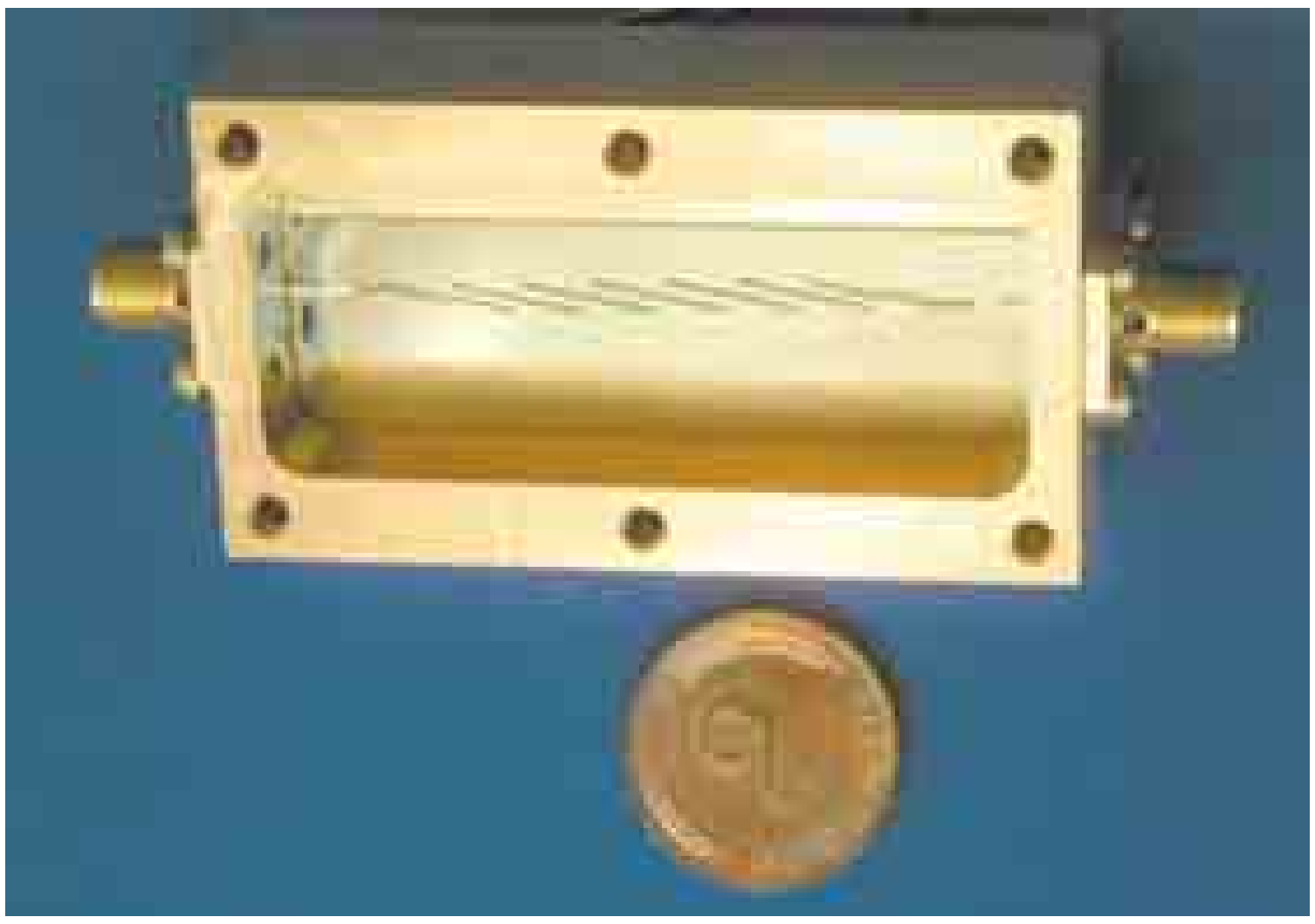

**Fig. 17** Microwave filter based on seven YBCO resonator stripes [294].

Among other types of HTS contacts, intrinsic Josephson junctions (IJJs) in Bi-HTS [309,310,311,312] are of technical interest. These junctions are formed by the layered crystal structure itself, with adjacent $CuO_2$ double layers acting as the superconducting electrodes and the BiO and SrO layers acting as the nonsuperconducting barrier layers. Stacks of IJJs can be realized by patterning single crystals or thin films into mesa structures consisting of between one and up to some hundreds of IJJs. Even two-dimensional arrays containing many thousands of IJJs can be formed using a two-dimensional fabrication technique. Besides from Bi-HTS stacks of IJJs have been realized from Tl-HTS [313] and also from oxygen deficient YBCO [314].

HTS single crystals are not suitable for applications due to their small size and their low $J_c$ values as a consequence of a low density of pinning centers. However, the "quick and dirty" version *melt-textured* 123-HTS *bulk material* can be grown reproducibly in sizes up to a diameter d = 6 cm even in complex shapes[56] [315], it shows superb magnetic pinning properties and is already tested as high-field permanent magnets, e. g., in magnetic bearings and motors [316]. Strong pinning allows the "freezing" of high magnetic fields, e. g., in single domain cylinders with a diameter of 30 mm an induction of 1.3 T can thus be fixed at 77 K [315], 17 T at 29 K has been demonstrated in-between two such samples [317]. This exceeds the potential of conventional permanent magnets based on ferromagnetic spin polarization by an order of magnitude and can be used for high-field "cryomagnets". The present field limit does not stem from the pinning, but from the involved mechanical forces, which exceed at high fields the fracture toughness of the YBCO ceramic [318]. The combination of melt-textured YBCO with permanent magnets leads to levitation properties that are attractive for bearing applications. In contrast to classic magnetic bearings, SC magnetic bearings are self-stabilizing due to flux pinning. The levitation forces with densities $> 10\ N/cm^2$ and the stiffness are already limited by the magnetic field strength of the permanent magnet [319,320,321].

In spite of the ceramic nature of the cuprate oxides, flexible HTS *wire* or *tape conductor* material can be obtained either by embedding thin HTS filaments in a silver matrix or by HTS coating of metal carrier tapes.

Bi-HTS /Ag tapes with cross sectional areas of ~ 1 $mm^2$ (Bi-HTS fill factor < 40 % [322]) are commercially available and can transport critical currents $I_c$ (77 K, 0 T) up to 130 A [323] over km lengths. In the "Powder-in-Tube" fabrication, Ag tubes are filled with Bi-HTS precursor powder and are subject to various thermomechanical processing steps where the tubes are flattened out to tapes that include the Bi-HTS material as thin filaments. The fragility of the Bi-22(n-1)n grains along neighboring Bi-oxide planes[57] is the physical origin for the successful mechanical alignability of the Bi-HTS grains by means of rolling or pressing which orients the grains preferably with the ab-planes parallel to the tape surface. Silver is the only matrix material that does not chemically react with HTS compounds and allows in addition sufficient oxygen diffusion. The common melting of Ag and Bi-HTS around ~ 850°C leads to a close mechanical and electrical contact that helps to bridge non-SC regions by means of low-resistivity shorts. Critical issues for technical applications of Bi-HTS/Ag conductors are the cost and the softness of Ag.

---

[56] comparable to what has been achieved in NdFeB permanent magnets.

[57] The two $BiO_x$ layers in the charge reservoir layers $SrO/(BiO_x)_2/SrO$ of the Bi-HTS are attached to each other only by weak van-der-Waals-like bonding.

Bi-2223/Ag tapes with $T_c$ > 100 K allow operation at $LN_2$ temperature [324,325,326]. However, the upper field limit of ~ 1 T at 77 K due to the pancake vortex disintegration of magnetic flux lines (see Fig. 7) [327] restricts the $LN_2$ operation of Bi-2223 tapes to low-field applications such as cables [328,329] or transformers. For extremely high magnetic field applications at low temperature operation, Bi-2212 conductors are economically more attractive than Bi-2223 tapes due to simpler processing and thus lower fabrication cost [330]. The fabrication of a complex standard conductor for accelerator magnets has already been demonstrated [331]. Another development aims at insertion magnets for 30 T class magnets where the Bi-HTS coils have to generate an additional field of several Tesla in a background field of ~ 20 T [58] [332]. The targeted high-field NMR application requires the persistent mode operation of the SC magnet system in order to fulfill the extreme temporal stability requirements based on an extremely slow decay of the supercurrents [333].

Cost arguments have always been in favor of HTS coating of simple robust metal carrier tapes. However, the major advantage would be the high-current operation in magnetic fields up to ~ 10 T at liquid nitrogen temperature. Meanwhile, $J_c$(77 K, 0 T) ~ 1 MA/cm$^2$ is achieved in YBCO-coated tapes over a length of ~ 10 meters[59]. If this performance can be extended within the next few years to a commercial status comparable to the present situation for Bi-2223/Ag conductors, the US military would be highly interested in respective small-size/high-power transformers, motors, generators, cables etc. for the next generation of "purely electric" tactical aircrafts without hydraulics [334] and "stealth" battleships [335].

Comparing the 18 years since HTS discovery with the time frame of about 50 years that classical superconductors and semiconductors required to establish as reliable high-tech materials these achievements of HTS applications are truly amazing. Nevertheless, there is still no "killer application" where HTS could prove its ability to provide a unique technical solution required in order to realize a broad-impact technology as it had been the case with MRI magnets for classical superconductors. The lesson from this example is that this breaktrough will probably happen in an application field which noone has even thought of today [336]. Until then, again in analogy with classical superconductors, HTS materials will probably find their main use in some niche applications, in particular in scientific research, where they have the chance to finally reach the status of a readily available mature technology.

---

[58] Present state of the art is the generation of an extra 5 T field by a Bi-HTS coil (at 4.2 K) on top of a 20 T background field, adding up to a total field of the SC coil system of > 25 T [26].

[59] For HTS conductor materials, the current density criterion $J_c$(77 K, 0 T) ≥ 1 MA/cm$^2$ has turned out to be good practical indicator for a sufficiently weak-link free microstructure. In principal, a lower $J_c$ level would be satisfactory as long as a total current capability of the conductor of ≥ 100 A can be achieved. However, with the only exception of Bi-HTS/Ag tapes $J_c$(77 K, 0 T) < 1 MA/cm$^2$ have been found to show a large reduction of the current capability in external magnetic fields due to weak links. In addition, it indicates that the reproducibility is not yet sufficient with respect to the fabrication of longer length conductor material.

## 4. OTHER OXIDE SUPERCONDUCTORS

The discovery of superconductivity in the *bismuthate* $BaPb_{1-x}Bi_xO_3$ in 1975 with a (in those days) rather high $T_c \sim 13$ K for x ~ 0.25 [337,338] raised great interest in the mechanism of superconductivity in this (at that time) quite exotic oxide compound with a low density of states at the Fermi level. The HTS cuprates soon chased away that exotic touch in spite of the rise of $T_c$ to > 30 K in $Ba_{1-x}K_xBiO_3$ ("BKBO"; x ~ 0.35) in the middle of the HTS bonanza days [339]: Tunneling showed clean gap structures consistent with weak-to-moderate coupling BCS theory [340]. The simple pseudocubic $ABO_3$ solid solution structure of a nonlayered nature derived for BKBO [341,342], with barium and potassium randomly occupying the A position, seemed to exclude for the SC phase the ordered, 3D charge-density wave ("CDW") arrangement of $Bi^{(4-\delta)+}O_6$ and $Bi^{(4+\delta)+}O_6$ octahedra ($0 < \delta << 1$) that renders the parent compound $BaBiO_3$ insulating.

In contrast to the HTS cuprates, it is not Coulomb correlation but this structural instability – in combination with charge disporportionation and in a 3D analog of the 1D Peierls gap formation – that causes the gap at the Fermi level for the undoped parent compounds. Just like in HTS cuprates, superconductivity is effected by introducing hole carriers into the stoichiometric parent compound, in this case by K or Pb doping. However, the doping dependence of $T_c$ is very different from the HTS doping scenario: For BKBO, the maximum occurs right at the metal-insulator transition at $x = x_c \sim 0.35$ when metallicity sets in; for $x > x_c$, $T_c$ rapidly decreases and finally disappears at the K solubility limit x ~ 0.65.

Actually, recent investigations indicate that CDWs do survive, in part, for all K or Pb doping levels and coexist with the doped hole states, which, in the case of BKBO, turn out to be bipolaronic in nature and originate from $Bi^{5+}O_6$ octahedra. In fact, a complex interplay of the remaining CDWs with the different electronic states introduced by the two doping routes is responsible for metallicity and superconductivity [338]. Recently, evidence has been reported for a non-cubic, non-centrosymmetric crystal structure of SC BKBO. This would point to a more layered electronic character and a closer analogy to the cuprate HTS compounds [343], but without their restrictions of the spin degrees of freedom by d-shell Coulomb correlation effects.

The extensive search for other SC transition metal oxides following the discovery of the cuprate HTS came in 1994 across strontium *ruthenate* ($Sr_2RuO_4$), a layered perovskite with an almost identical crystal structure as the cuprate HTS $La_{2-x}Sr_xCuO_4$ ("LSCO"), albeit only with a $T_c \sim 1.5$ K [344]. In both materials the conduction electrons stem from partially filled d-bands (of the Ru or Cu ions, respectively) that are strongly hybridized with oxygen p-orbitals. In contrast to the nearly filled Cu 3d-shell in HTS with only one hole state, in $Sr_2RuO_4$, in the formal oxidation state of the ruthenium ion $Ru^{4+}$ four electrons are left in the 4d-shell. The closely related ferromagnetic material $SrRuO_3$ shows the inherent tendency of $Ru^{4+}$ towards ferromagnetism. Hence, in analogy with the HTS cuprates, where on doping the AF ground state of the parent compounds seems to "dissolve" in spin-singlet Cooper pairs in a d-wave orbital channel, it was suggested that the superconductivity in $Sr_2RuO_4$ is brought about by spin-triplet pairing where the Ru ions "release" parallel-spin, i. e., triplet Cooper pairs in p-wave or even higher odd order angular orbital channels.

$RuSr_2GdCu_2O_8$ (Ru-1212) is a *ruthenate-cuprate hybrid* containing both $CuO_2$ and $RuO_2$ layers. It fits into the elucidated cuprate HTS layer structure scheme (see Fig. 3) substituting the Ca of the canonical 1212-HTS structure (or the Y in YBCO, or the RE in RE-123, respectively) by Gd to render $CuO_2$/Gd/$CuO_2$ stacks, separated by a SrO "wrapping layer" from the $RuO_2$ layers as "charge reservoir layers". Like rare-earth borocarbides (see chapt. 7), Ru-1212 and some other closely related rutheno-cuprate compounds display ferromagnetism and superconductivity coexisting on a microscopic scale [345], with $T_{Curie}$ ~ 135 K and $T_c$ up to 72 K for Ru-1212. The $CuO_2$/Gd/$CuO_2$ stacks are believed to be responsible for the superconductivity, whereas the (ferro)magnetic ordering arises from the $RuO_2$ layers. A clear intrinsic Josephson effect shows that the material acts as a natural superconductor-insulator-ferromagnet-insulator-superconductor superlattice [346]. Electronic phase separation as discussed for underdoped cuprates resulting in domains that are both SC and AF separated by ferromagnetic boundaries has been suggested as a possible explanation [347].

*Cobaltates* are a very recent entry in the SC oxide zoo. $T_c$ = 4.5 K has been achieved in hydrated sodium cobaltate $Na_{0.3}CoO_2$•1.4 $H_2O$ [348]. Na provides here the doping charge. The intercalation of water which increases the separation between the $CoO_2$ layers seems to be essential for the onset of superconductivity: $Na_{0.3}CoO_2$•0.6 $H_2O$, with the same Co formal oxidation state but substantially less separation between the $CoO_2$ layers is not SC [349]. This triggered theoretical interest in this compound, since it could hence provide an experimental test field for investigating the transition from a 3-dimensional to a 2-dimensional nature of the charge carriers which is believed to play a major role for the high $T_c$ of cuprate HTS. A major difference is the triangular lattice geometry of the $CoO_2$ layers (see Fig. 18) which introduces magnetic frustration into the Co spin lattice, in contrast to the square lattice geometry of Cu ions in cuprate HTS which favors unfrustrated AF spin orientation.

A recent report on superconductivity in *pyrochlore oxides* ($T_c$ = 9.6 K in $KOs_2O_6$, $T_c$ = 6.3 K in $RbOs_2O_6$, $T_c$ = 3.3 K in $CsOs_2O_6$ ) [351] with a triangle-based crystal structure which is even more subject to magnetic frustration is a further indication that magnetic frustration is apparently not a no-go criterion for superconductivity [352].

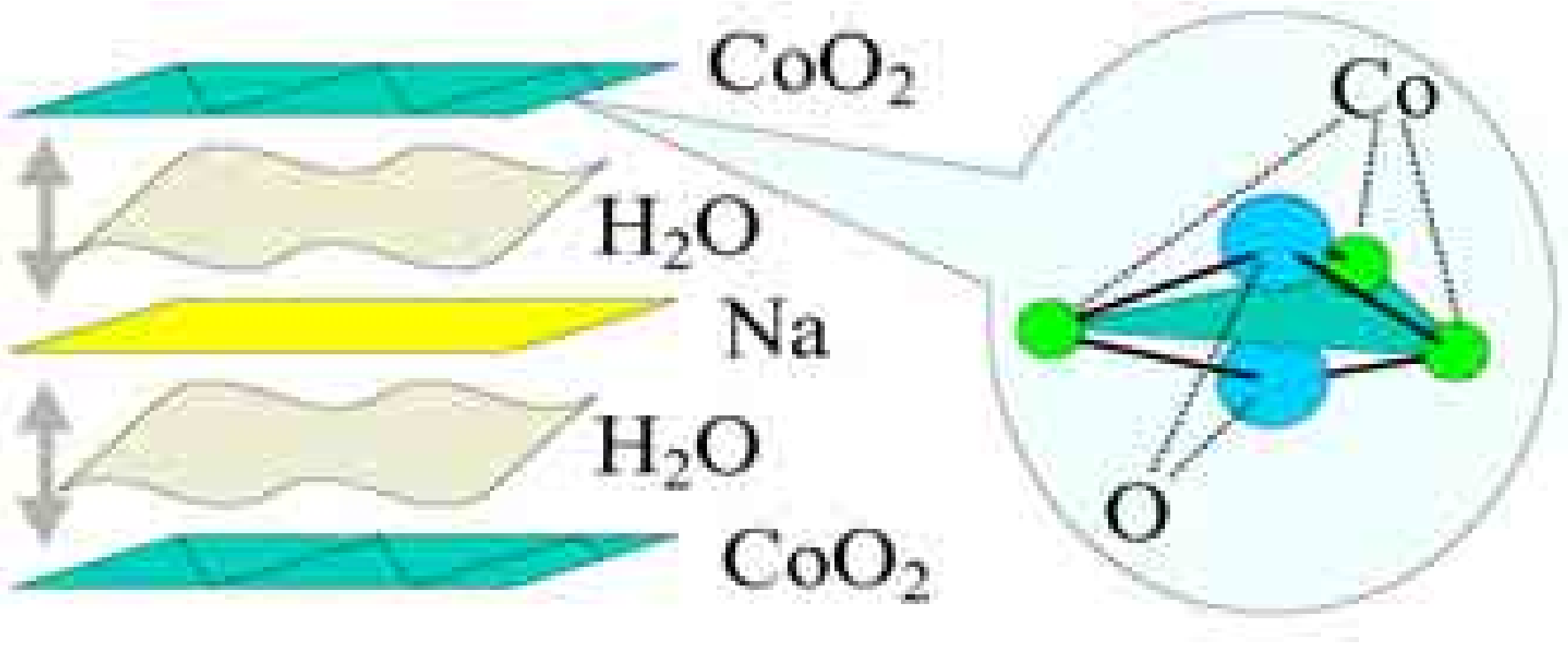

**Fig. 18** Schematics of the layer structure of hydrated sodium cobaltate $Na_{0.3}CoO_2$•1.4 $H_2O$ and atomic microstructure of the basic cobalt oxide triangular plaquets [350].

## 5. HEAVY FERMION SUPERCONDUCTORS

Heavy-Fermion (HF) systems are stoichiometric lanthanide or actinide compounds whose qualitative low-temperature behavior in the normal state closely parallels the one well-known from simple metals. The key features are the specific heat which varies approximately linearly $C \sim \gamma T$, the magnetic susceptibility which approaches a temperature independent constant $\chi(0)$ and the electrical resistivity which increases quadratically with temperature $\rho(T) = \rho_0 + AT^2$. The coefficient $\gamma \sim 1$ J / mole K$^2$ as well as $\chi(0)$ are enhanced by a factor of $100-1000$ as compared to the values encountered in ordinary metals while the Sommerfeld-Wilson ratio $[\pi (k_B)^2 \chi(0)] / [3(\mu_B)^2 \gamma]$ is of order unity. The large enhancement of the specific heat is also reflected in the quadratic temperature coefficient $A$ of the resistivity $A \sim \gamma^2$. These features indicate that the normal state can be described in terms of a Fermi liquid. The excitations determining the low-temperature behavior correspond to heavy quasiparticles whose effective mass $m^*$ is strongly enhanced over the free electron mass m. The characteristic temperature $T^*$ which can be considered as a ficticious Fermi temperature or, alternatively, as an effective band width for the quasiparticles is of the order 10 – 100 K. Residual interactions among the heavy quasiparticles lead to instabilities of the normal Fermi liquid state. A hallmark of these systems is the competition or coexistence of various different cooperative phenomena which results in highly complex phase diagrams. Of particular interest are the SC phases which typically form at a critical temperature $T_c \leq 2$ K[60].

The discovery of superconductivity in $CeCu_2Si_2$ [11] forced condensed-matter physicists to revise the generally accepted picture of the electrons occupying the inner shells of the atoms. Traditionally, the corresponding states were viewed as localized atomic-like orbitals which are populated according to Hund's rules in order to minimize the mutual Coulomb repulsion. This leads to the formation of local magnetic moments which tend to align and which are weakly coupled to the delocalized conduction electrons. The latter were viewed as "free" fermions which occupy coherent Bloch states formed by the valence orbitals of the atoms. An attractive residual interaction among the conduction electrons causes the normal metallic state to become unstable with respect to superconductivity. The Cooper pairs which characterize a superconducting phase are broken by magnetic centers. The damaging effect of 4f- and 5f-ions was well established by systematic studies of dilute alloys.

The key to the understanding of the origin and the nature of the unanticipated superconductivity in $CeCu_2Si_2$ [11] lies in a better appreciation of the physics of the highly unusual normal state out of which it forms. The characteristic features summarized above show that the magnetic degrees of freedom of the partially filled f-shells form a strongly correlated paramagnetic Fermi liquid with an effective Fermi energy of the order of 1-10 meV. The existence of a Fermi liquid state with heavy quasiparticles involving the f-degrees of freedom has been confirmed experimentally.

---

[60] The exceptionally high transition temperature of ~ 18.5 K was recently reported for the actinide compound $PuCoGa_5$ [12].

The superconducting phases of HF materials are characterized by BCS-type pair-correlations among the heavy quasiparticles of the normal state [353]. This picture is inferred from the fact that the discontinuities observed in various thermodynamic properties at the superconducting transition scale with the large effective masses of the normal state. The unusual energy and length scales in HF superconductors, however, lead to novel phenomena. First, the effective Fermi temperature typically exceeds the superconducting transition temperature only by one order of magnitude. This fact implies that standard weak-coupling theory which keeps only the leading contributions in an expansion with respect to the ratio $T_c / T^*$ can provide only qualitative results. Pronounced strong-coupling corrections are reflected in deviations from the universal behavior predicted for weak-coupling systems. Second, the small characteristic energy $k_B T^*$, or, equivalently, the narrow width of the quasiparticle bands implies a small value of the Fermi velocity $v_F$. As a result, the size of the Cooper pairs in the HF systems, i. e., the coherence length $\xi_0 = \frac{\hbar v_F}{k_B T_c}$ is much less than in a typical "ordinary" superconductor. Many of the systems can be considered as "clean" with the mean-free paths exceeding the coherence length. As a result, anisotropic pair states can form which are strongly suppressed by impurity scattering in "ordinary" superconductors with large coherence lengths.

Although we know the correlations which characterize the superconducting state we are still far from a complete theory of superconductivity in these materials. Major questions are still open. Among the prominent is the question of which order parameters characterize the superconducting states and what is the origin of the attraction between the quasiparticles. The correct microscopic description of the interaction would yield, of course, the superconducting transition temperatures as well as the detailed form of the order parameter.

To describe ordered phases, the determination of the type and the symmetry of the order parameter is of central importance. The latter restricts the possible excitations in the ordered phases and hence determines the low-temperature properties. Order parameters given in terms of expectation values of physical observables like spin- and charge densities can be directly measured by x-ray or neutron diffraction. The magnetic phases of the lanthanide and actinide compounds are therefore rather well characterized. Superconductivity, however, corresponds to an off-diagonal long-range order parameter which is not directy observable. The usual procedure to determine the symmetry of the superconducting order parameter is to select plausible candidate states corresponding to irreducible representations of the symmetry group, calculate expected behavior of physical quantities and compare the predictions with experiment.

The occurrence of long-range order described by an order parameter is most frequently associated with spontaneous symmetry breaking. The simplest superconductors where only gauge invariance is broken are called conventional. In this case the superconducting state has the same symmetry as the underlying crystal. It should be noted that conventional is not a synonym for isotropic. A superconductor who has additional broken symmetries besides gauge symmetry, i. e., whose symmetry is lower than that of the underlying crystal is called unconventional. Considerable progress has been made recently in detecting unconventional order parameter symmetries. Angle-resolved studies of thermodynamic and transport properties in the vortex phase determine the position of order parameter nodes relative to the crystal axes.

In several systems, the order parameter has been shown to be unconventional. This was inferred mainly from anisotropies displayed by thermal transport in the presence of an external magnetic field. The most spectacular examples of unconventional superconductors are $UPt_3$ [354] and $PrOs_4Sb_{12}$ [355] where the split transition points to a multicomponent order parameter. In all the other HF superconductors, the order parameter seems to be a complex scalar function.

The fact that HF superconductors can be considered as "clean" systems with the mean-free path exceeding the coherence length has rekindled speculations so as to find inhomogeneous superconductivity phases (Fulde-Ferrel-Larkin-Ovchinnikov states; “FFLO” / “LOFF”) at sufficiently low temperatures in sufficiently high magnetic magnetic fields [356]. In fact, the order of the phase transition in an applied magnetic field has been found to change from second to first order with decreasing temperature – in agreement with long-standing theoretical predictions [357].

An indispensable prerequisite to a microscopic theory of superconductivity is a microscopic theory of the normal state, of the quasiparticles and their interactions. The Landau theory does not make assumptions or predictions concerning the microscopic nature of the ground state and the low-lying excitations. It does not address the question how the latter emerge in an interacting electron system. During the past decade it became clear that there are different routes to heavy fermion behaviour.

In Ce-based compounds, the heavy quasiparticles with predominantly 4f-character arise through the Kondo effect in the periodic lattice. The Kondo picture for the Ce-based heavy-fermion compounds is supported by the fact that the thermodynamic properties at low temperatures (e. g., the specific heat, the magnetic susceptibility) as well as the temperature-dependence of the spectroscopic data can be reproduced by an Anderson model. This picture has been confirmed in detail by deHaas-vanAlphen and photoemission studies [358]. The strongly renormalized quasiparticles can be described by a semiphenomenological ansatz, the Renormalized Band Method which yields realistic quasiparticle bands [359]. The nesting features of the calculated Fermi surfaces can serve as a useful guideline in the search for instabilities with respect to modulated structures [360]. In particular, the calculations provide realistic models for the study of the competition/coexistence phase diagrams. Despite the efforts to implement modern many-body methods for strong correlations into realistic electronic structure calculations there is still no general concept for quantitative microscopic correlations. In particular, the subtle interplay between local and intersite effects continues to challenge theorists. The latter may lead to long-range order while the former favor the formation of a Fermi liquid state at low temperatures.

A microscopic picture for the heavy quasiparticles has finally emerged for the actinide compounds [361]. Increasing experimental evidence points towards a dual character of the 5f-electrons with some of them being delocalized forming coherent bands while others stay localized reducing the Coulomb repulsion by forming multiplets. The hypothesis of the dual character is translated into a calculational scheme which reproduces both the Fermi surfaces and the effective masses determined by deHaas-vanAlphen experiments without adjustable parameter. The method yields a model for the residual interaction leading to various instabilities of the normal phase. The dual model should also provide insight into the mysterious hidden order phases of U compounds.

In both cases, i. e., in Ce and in U HF compounds, the coherent heavy quasiparticles are derived from the partially filled f-shells whose degrees of freedom have to be (partially) included into the Fermi surface. The situation is different for the Pr skutterudites where the quasiparticles are derived from the conduction states whose effective masses are strongly renormalized by low-energy excitations of the Pr 4f-shells. From a microscopic point of view we are dealing with three different classes of HF superconducting materials.

It is generally agreed that the pairing interaction in HF superconductors is of electronic origin. Theoretical models usually involve the exchange of a boson. It is therefore convenient to classify the various mechanisms according to the boson which is exchanged [362]. The majority of models construct effective interactions based on the exchange of spin-fluctuations. Neither these models nor their refined variants which account for the internal orbital f-electron structure are able to properly predict the symmetry of the order parameter, e. g., in $UPt_3$. In particular, the calculations do not reproduce a multicomponent order parameter as stable solution . Recently, a model based on the exchange of weakly damped propagating magnetic excitons was suggested for U-based HF compounds [363,364]. First estimates of the transition temperature yield a value of the correct order of magnitude. Pairing due to intra-atomic excitations may also occur in the Pr-skutterudite HF superconductor. In this case, however, quadrupolar instead of magnetic excitons should be involved [365].

Superconductivity in HF compounds is usually found to coexist or to compete with various cooperative phenomena. Of particular importance in this context is itinerant antiferromagnetism as realized in a spin density wave (SDW). Since both order parameters appear in the itinerant quasiparticle system the observed behavior results from a subtle interplay between Fermi surface geometry and gap structures.

Many HF systems are on the verge of magnetic instability. By application of pressure these materials may be tuned in their normal states through a quantum critical point (QCP) from an antiferromagnet to a paramagnetic metal. The theoretical picture, however, is at present still rather controversial and contentious.

## 5A. Ce-BASED HF COMPOUNDS

The discovery of superconductivity in $CeCu_2Si_2$ ($T_c$ = 1.5 K; see Fig. 19) [11] initiated the rapid development of HF physics. For nearly two decades, this material was the only Ce-based heavy fermion superconductor at ambient pressure. Only recently, superconductivity at ambient pressure was found in the new class of heavy fermion materials $CeM_mIn_{3+2m}$ (M = Ir or Co; m = 0, 1) [366,367]. The most prominent member of this family is $CeCoIn_5$ which has a relatively high $T_c$ = 2.3 K (see Fig. 12) [368]. Of fundamental interest ist the discovery of HF superconductivity in $CePt_3Si$ which crystallizes in a lattice without inversion symmetry.

The superconducting phases in the Ce-based HF compounds are characterized by anisotropic order parameters which reflect the on-site repulsion introduced by the strong correlations of the partially filled Ce 4f shells. The current experimental and theoretical research focuses on the question which factors determine the character of the low-temperature phases. The subtle interplay between local singlet-formation via Kondo effect and long-range magnetic order can be monitored experimentally in pressure studies where isostructural relatives of the HF superconductors are tuned from magnetic phases at ambient pressure to SC states, e. g., $CeCu_2Ge_2$ [369], $CePd_2Si_2$ [370,371] $CeNi_2Ge_2$ [372] and $CeRh_2Si_2$ [373], $CeSn_3$ [374] and $CeIn_3$ (see Fig. 12 [370]. Similar behavior is found in doping experiments where constituents of the metallic host are successively replaced. In $CeCu_2Si_2$ a highly interesting type of competition between (anisotropic) superconductivity and magnetic order is encountered. In this material, the heavy Fermi liquid is unstable with respect to both superconductivity and a spin-density wave. The actual ground state realized in a sample depends sensitively on the composition of the sample [360].

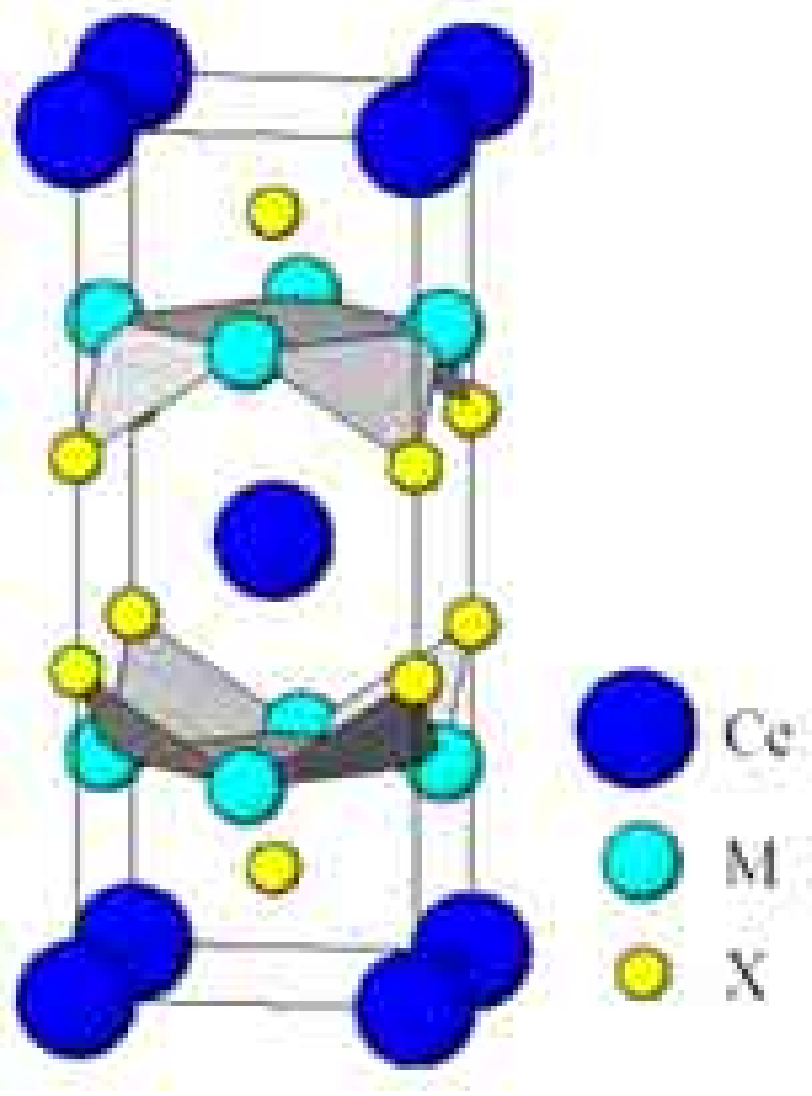

**Fig. 19** Conventional unit cell of $CeM_2X_2$ (M = Cu, Ni, Ru, Rh, Pd, Au, ...; X = Si, Ge) and $URu_2Si_2$.

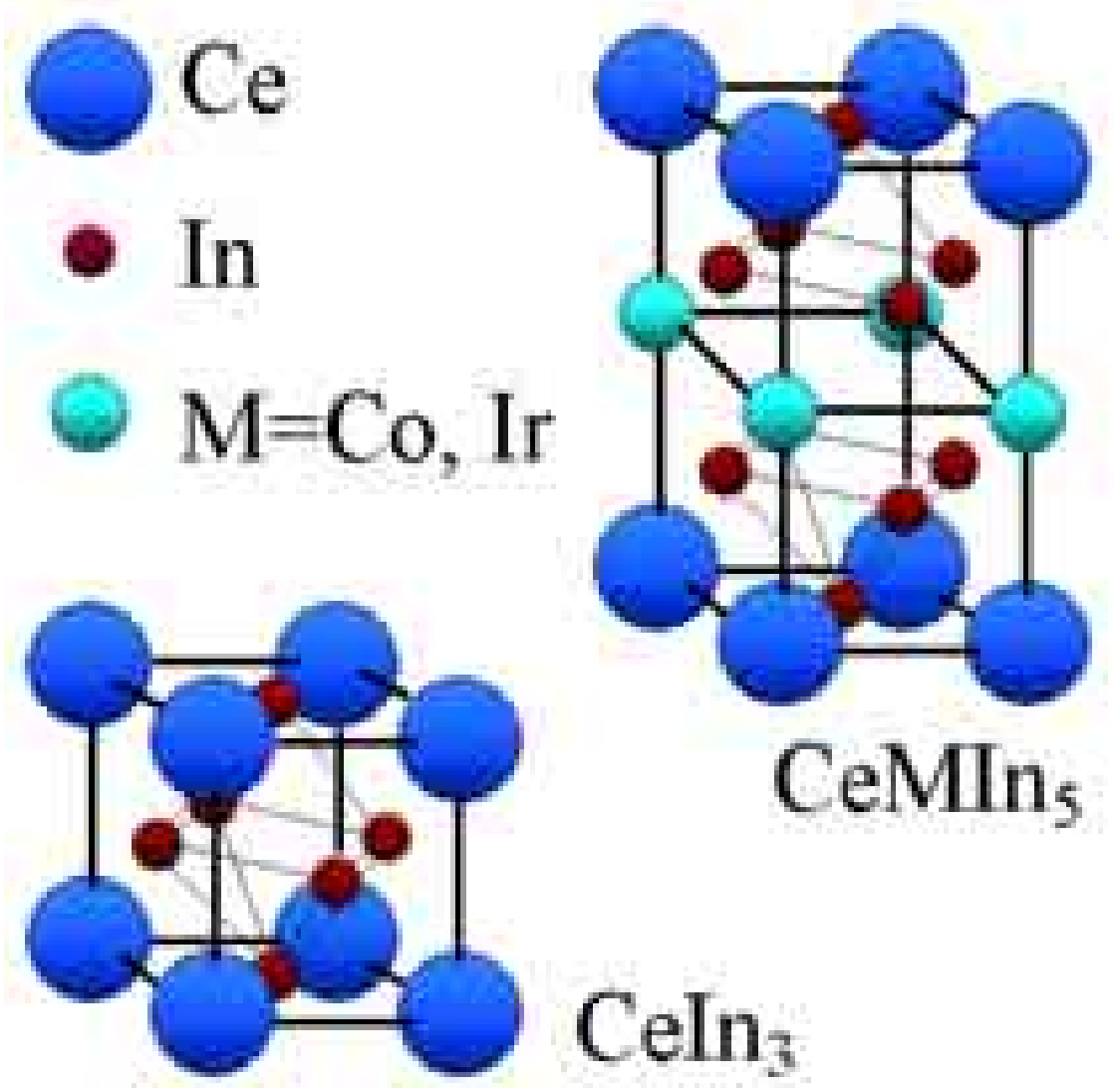

**Fig. 20** Unit cell of $CeIn_3$ and $CeMIn_5$ (M = Co, Ir).

## 5B. U-BASED HF COMPOUNDS

Heavy Fermion superconductivity is found more frequently in intermetallic U-compounds than in Ce-compounds. This may be related to the different nature of heavy quasiparticles in U-compounds where the 5f-electrons have a considerable, though orbitally dependent, degree of delocalization. The genuine Kondo mechanism is not appropriate for heavy quasiparticle formation as in Ce-compounds. This may lead to more pronounced delocalized spin fluctuations in U-compounds which mediate unconventional Cooper pair formation. Antiferromagnetic ("AF") order, mostly with small moments of the order $10^{-2}$ $\mu_B$ is frequently found to envelop and coexist with the SC phase.

The hexagonal compound $UPt_3$ (see Fig. 21) [375] exhibits triplet pairing. It sticks out as the most interesting case of unconventional superconductivity with a multicomponent order parameter whose degeneracy is lifted by a symmetry-breaking field due to a small moment AF order. In contrast, in $UPd_2Al_3$ (see Fig. 22) [376] superconductivity coexists with large moment antiferromagnetism. Probably spin singlet pairing is realized. There is experimental evidence for a new kind of magnetic pairing mechanism mediated by propagating magnetic exciton modes. The sister compound $UNi_2Al_3$ [377] is an example of coexistence of large moment antiferromagnetism with a SC triplet order parameter. In $URu_2Si_2$ [378] the SC order parameter symmetry is still undetermined. The interest in this compound is focused more on the enveloping phase with a "hidden" order parameter presumably of quadrupolar type or an "unconventional" spin density wave (SDW). The oldest cubic U-HF superconductor $UBe_{13}$ [379] and its thorium alloy $U_{1-x}Th_xBe_{13}$ is also the most mysterious one. While for the pure system there is a single SC phase of yet unknown symmetry, in the small Th concentration range two distinct phases exist which may either correspond to two different SC order parameters or may be related to a coexistence of superconductivity with a SDW phase. In addition, in $UBe_{13}$ SC order appears in a state with clear non-Fermi liquid type anomalies. More recently the coexistence of ferromagnetism and superconductivity in $UGe_2$ [380] has been found. Due to the ferromagnetic polarization the triplet gap function contains only equal spin pairing.

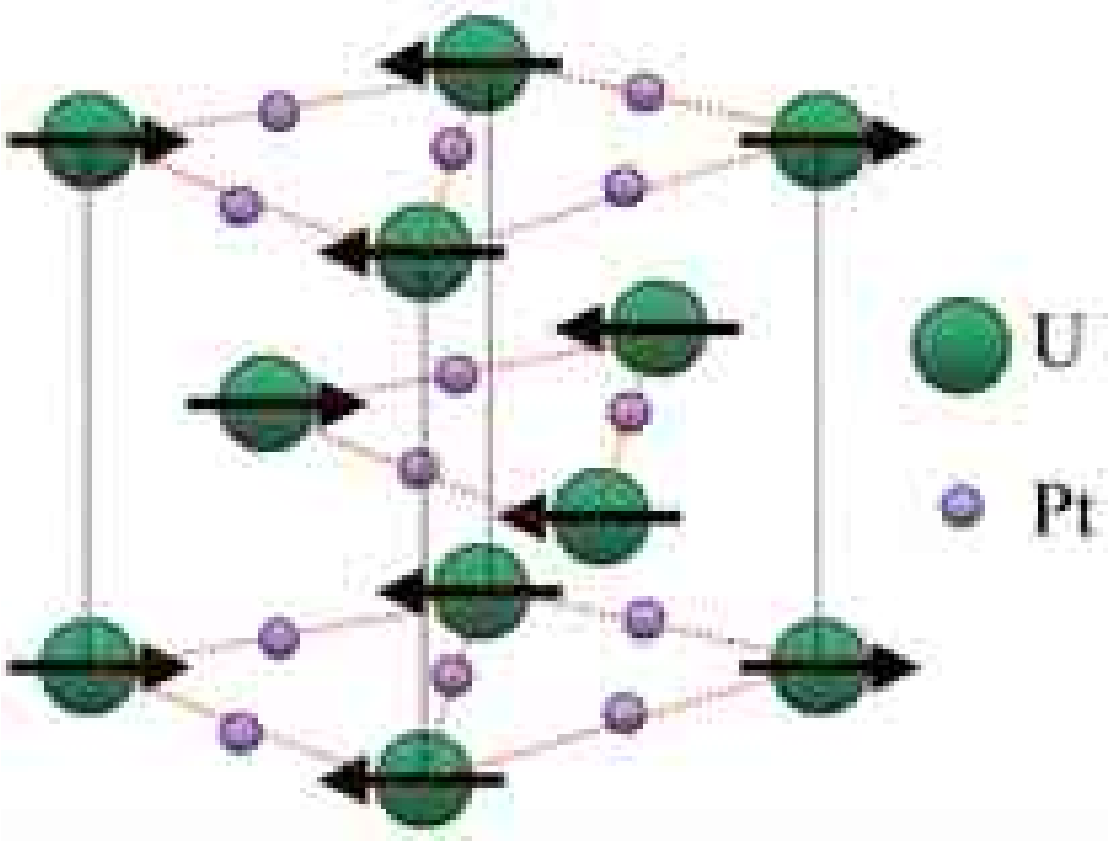

**Fig. 21** Crystal structure of $UPt_3$ ($a = 0.5764$ nm, $c = 0.4884$ nm) and AF magnetic structure ($T < T_N = 5.8$ K).

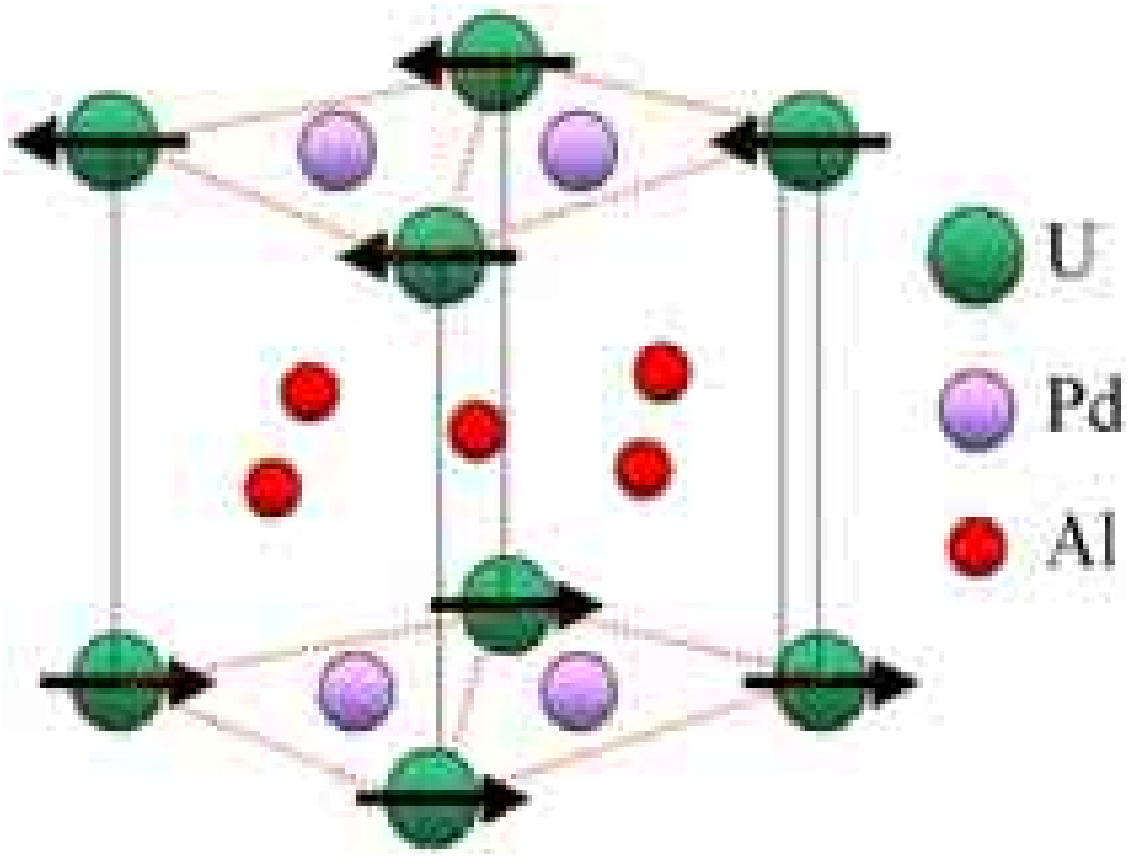

**Fig. 22** Conventional unit cell of $UPd_2Al_3$ ($a = 0.5350$ nm, $c = 0.4185$ nm) and simple AF magnetic structure.

The hexagonal heavy fermion compound $UPt_3$ (see Fig. 21) is the “flagship” of unconventional superconductivity, despite its rather low critical temperature of slightly less than 1 K. It is set aside from all other unconventional superconductors insofar as it exhibits two SC phase transitions. The exciting discovery of the split superconductive transitions in $UPt_3$ at $T_{c1}$ = 530 mK and $T_{c2}$ = 480 mK in specific heat measurements [354] has led to an enormous amount of experimental and theoretical work on $UPt_3$ [381]. The additional small moment antiferromagnetism observed in $UPt_3$ ($T_N$ = 5.8 K, $\mu$ = 0.035 $\mu_B$) plays a key role in the identification of the SC order parameter since the in-plane staggered magnetization acts as a symmetry breaking field (SBF) to the SC multicomponent order parameter. The SBF is believed to be responsible for the appearance of two SC transitions which otherwise would merge into one. Therefore one can identify three distinct SC phases. Despite the wealth of experimental results on $UPt_3$ there is no unequivocal consensus on the symmetry and node structure of the SC gap.

$UPd_2Al_3$ (see Fig. 22) [376] is a rather special case among the U-based HF superconductors. There is also AF order below $T_N$ = 14.3 K with almost atomic size local moments ($\mu$ = 0.85 $\mu_B$) in contrast to the small moments in other U-compounds. The AF order coexists with superconductivity below $T_c$ = 1.8 K. This suggests that in addition to the heavy itinerant quasiparticles nearly localized 5f-electrons should be present. They result from the dominating $5f^2$ configuration of the $U^{4+}$ ion [382]. This dual nature of 5f-electrons is even more obvious than in $UPt_3$ as is seen in various experiments. A direct confirmation of this dual nature of 5f-electrons in $UPd_2Al_3$ was obtained from inelastic neutron scattering (INS) [383] which found excitations that originate from local CEF transitions and disperse into bands of “magnetic excitons” due to intersite exchange. Complementary tunneling experiments with epitaxial $UPd_2Al_3$-$AlO_x$-Pb heterostructures probed the response of the itinerant quasiparticles and their SC gap. Strong coupling features in the tunneling density of states were reported [384] suggesting that the magnetic excitons identified in INS are the bosonic “glue” which binds the electrons to Cooper pairs [363]. This new mechanism for superconductivity is distinctly different from both the electron-phonon and spin fluctuation mechanism known so far. The pairing potential is mediated by a propagating boson (the magnetic exciton) as in the former case but depends on the spin state of conduction electrons as in the latter case.

The possibility of coexisting ferromagnetism and superconductivity was first considered by Ginzburg [385] who noted that this is only possible when the internal ferromagnetic field is smaller than the thermodynamic critical field of the superconductor. Such a condition is hardly ever fulfilled except immediately below the Curie temperature $T_C$ where coexistence has been found in a few superconductors with local moment ferromagnetism and $T_C < T_c$ such as $ErRh_4B_4$ and $HoMo_6S_8$. If the temperature drops further below $T_C$ the internal ferromagnetism molecular field rapidly becomes larger than $H_{c2}$ and superconductivity is destroyed. The reentrance of the normal state below $T_C$ has indeed been observed in the above compounds. The only compound known so far where local moment ferromagnetism coexists homogeneously with superconductivity for all temperatures below $T_c$ is the borocarbide compound $ErNi_2B_2C$ [386] (see chapter 6). The competition between ferromagnetism and superconductivity becomes more interesting if ferromagnetic order is due to itinerant electrons which also form the SC state. If the interaction slightly exceeds a critical value one has weak ferromagnetic order such as in $ZrZn_2$ with large longitudinal ferromagnetic spin fluctuations. In this case p-wave superconductivity may actually be mediated by the exchange of ferromagnetic spin fluctuations and coexist with the small ferromagnetic moments [387]. p-wave superconductivity was predicted to exist in this case both on the ferromagnetic as well as on the paramagnetic side of this critical parameter region. The discovery of unconventional superconductivity under pressure in the itinerant ferromagnets $UGe_2$ [380] and later for URhGe [388] and the 3d ferromagnet $ZrZn_2$ [389] under ambient pressure has opened this theoretical scenario to experimental investigation.

$URu_2Si_2$ (see Fig. 19), a "moderately heavy" fermion compound has mystified experimentalists and theorists alike with the discovery of AF order and another still unidentified "hidden order" phase. In addition, the compound becomes a nodal superconductor below $T_c$ = 1.2 K [378, 390, 391]. The simple tetragonal AF order with wave vector in c-direction has tiny moments $\mu \sim 0.02\ \mu_B$ along c-axis [392, 393] which are of the same order as in $UPt_3$. However, unlike in $UPt_3$, very large thermodynamic effects, e. g., in specific heat, thermal expansion etc., occur which are hard to reconcile with the small ordered moments. These pronounced anomalies were interpreted as evidence for the presence of a second "hidden order" parameter, which cannot be seen in neutron or x-ray diffraction. In the itinerant models the order parameter is due to an unconventional pairing in the particle-hole channel leading to an unconventional SDW which has vanishing moment in the clean limit and also does not break time reversal invariance. The small staggered moments may then be induced in the condensate due to impurity scattering. Finally, in the dual models one assumes a localized singlet-singlet system in interaction with the itinerant electrons to cause induced moment magnetism with small moments but large anomalies. In all models it was previously taken for granted that both the primary "hidden" order parameter and AF order coexist homogeneously within the sample. However, hydrostatic and uniaxial pressure experiments [394, 395] have radically changed this view, showing that the order parameters exist in different parts of the sample volume; the tiny AF moment is not intrinsic but due to the small AF volume fraction under ambient pressure. Applying hydrostatic pressure or lowering the temperature increases the AF volume fraction and hence the ordered moment until it saturates at an atomic size moment of 0.4 $\mu_B$ per U atom. This means that the evolution of antiferromagnetism arises from the increase of AF volume with pressure rather than the increase of the ordered moment $\mu$ per U-atom.

The cubic compound $UBe_{13}$ and $U_{1-x}Th_xBe_{13}$ was discovered rather early [379, 396] as a SC HF system. The U atom in this structure is embedded in an icosahedral cage of 12 Be-atoms. A global understanding of the normal state and symmetry breaking in both SC and magnetic state is still elusive. Firstly, $UBe_{13}$ single crystals do not yet achieve the quality as, e. g., $UPt_3$ single crystals so that the symmetry of the anisotropic SC gap functions could not be identified yet. Furthermore, the Th-doped crystals $U_{1-x}Th_xBe_{13}$ show a perplexing variety of SC and possibly also magnetic phases whose microscopic origin and order parameter symmetries are not understood. Already in the normal state $UBe_{13}$ is a rather anomalous metal, e. g. non-Fermi-liquid behavior has been observed and attributed to a multichannel Kondo effect. The $T_c$ values for superconductivity, which occurs in the non-Fermi-liquid state, depend considerably on the type of sample. There are two classes with 'high' $T_c \sim 0.9$ K and 'low' $T_c \sim 0.75$ K which, however, are not much different in their impurity content.

## 5C. RARE-EARTH SKUTTERUDITES

The recently discovered heavy fermion superconductor $PrOs_4Sb_{12}$ [355] is potentially of similar interest as $UPt_3$ because it represents the second example of multiphase superconductivity [397] with a critical temperature $T_c = 1.85$ K. The skutterudites $RT_4X_{12}$ (R = alkaline earth, rare earth or actinide; T = Fe, Ru or Os and X=P, As or Sb) show a cage structure where large voids formed by tilted $T_4X_{12}$ octahedrons can be filled with R atoms (see Fig. 23). They are however rather loosely bound and are therefore subject to large anharmonic oscillations ("rattling") in the cage. In addition, the presence of several equivalent equilibrium positions may give rise to tunneling split states. Both effects may lead to interesting low temperature elastic and transport phenomena, such as thermoelectric effects [398,399]. Depending on the cage-filling atom this large class of compounds displays also a great variety of interesting effects of strong electron correlation. Mixed valent and HF behavior, magnetic and quadrupolar order, non-Fermi liquid and Kondo insulating behavior has been found [355,399]. Recently the superconductivity in non-stoichiometric skutterudites $Pr(Os_{1-x}Ru_x)Sb_{12}$ has been investigated throughout the whole concentration range of $0 \leq x \leq 1$ [400]. While for x = 0 one has an unconventional HF superconductor, the x = 1 compound $PrRu_4Os_{12}$ on the other hand is a conventional superconductor with $T_c \sim 1$ K. The type of superconductivity changes at x ~ 0.6 where the transition temperature $T_c(x)$ has a minimum value of 0.75 K.

In the HF multiphase superconductor $PrOs_4Sb_{12}$ (x = 0) the specific heat jump due to superconductivity is superposed on a Schottky anomaly due to the lowest CEF excitation. Nevertheless, its detailed analysis provides clear evidence for a split superconductivity transition [355,401] at $T_{c1} = 1.85$ K and $T_{c2} = 1.75$ K. The total jump of both transitions amounts to $\Delta_{SC}$ $C/\gamma T_c$ ~3 which considerably exceeds the BCS value 1.43 for a single transition. It also proves that the SC state is formed from the heavy quasiparticles that cause the enhanced γ-value of the electronic low-temperature specific heat contribution. A $T_c$-splitting of similar size also was clearly seen in thermal expansion measurements [402]. The two SC transitions are reminiscent of the split transition in $UPt_3$. There, a twofold orbitally degenerate SC state is split by weak AF order that reduces the hexagonal symmetry to an orthorhombic one. This also leads to two critical field curves in the B-T phase diagram. In $PrOs_4Sb_{12}$ no such symmetry breaking field exists.

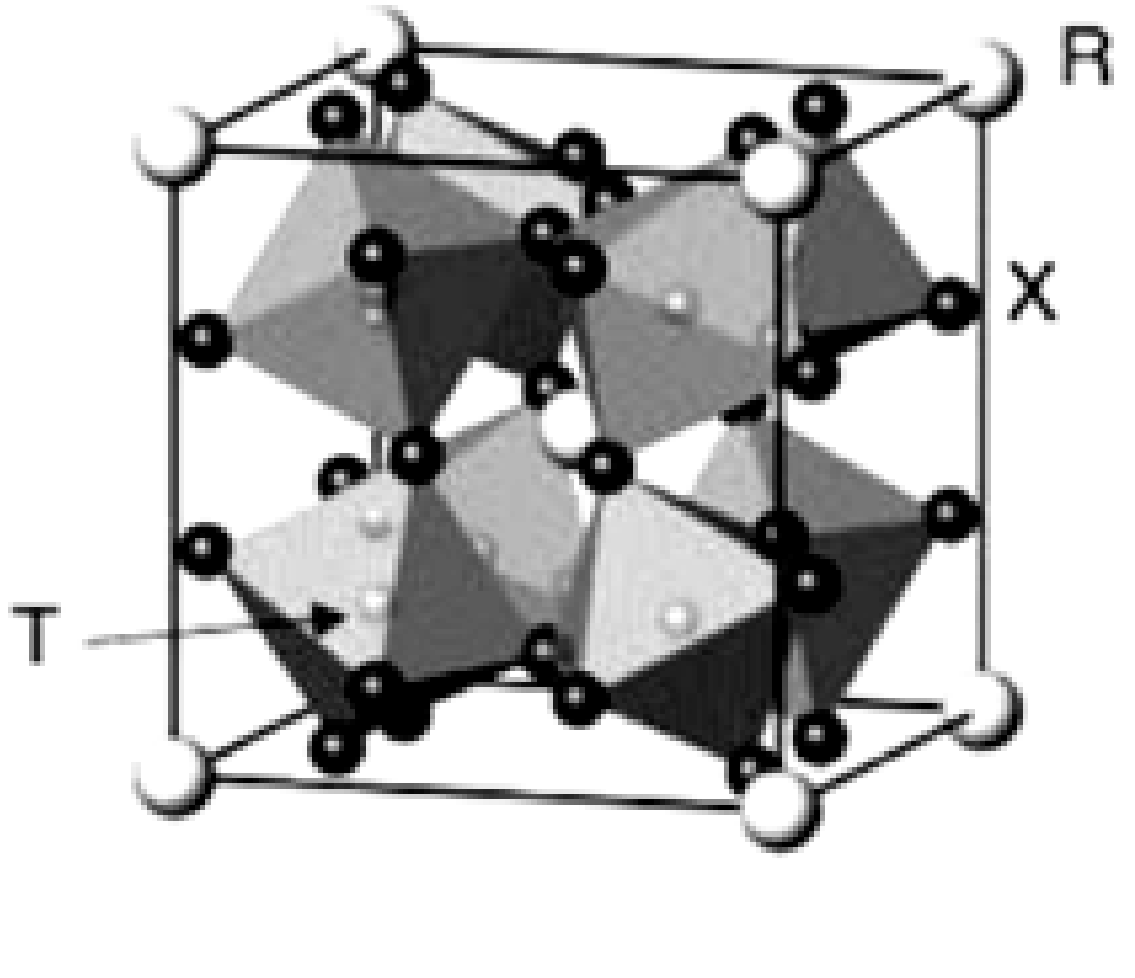

**Fig. 23** Cubic crystal structure of the filled skutterudite $RT_4X_{12}$. The T atoms are located in the center of the X octahedra. For $PrOs_4Sb_{12}$ the lattice constant is a = 0.93017 nm.

In this still rather early stage of investigation, various experiments gave inconclusive results on the question of the nature of gap anisotropy. As a preliminary conclusion, it seems clear that $PrOs_4Sb_{12}$ is a very unconventional multiphase HF superconductor of potentially the same interest as $UPt_3$. Since that heavy quasiparticles are presumably caused by coupling with virtual quadrupolar excitations from the nonmagnetic 5f ground state one is lead to speculate that superconductivity in $PrOs_4Sb_{12}$ might also result from an unprecedented pairing mechanism based on the exchange of quadrupolar fluctuations. At the moment, this quadrupolar superconductivity mechanism in $PrOs_4Sb_{12}$ as a third possibility for Cooper pair formation in heavy fermion compounds in addition to the spin-fluctuation and magnetic-exciton exchange mechanisms is still a conjecture.

## 6. ORGANIC AND OTHER CARBON BASED SUPERCONDUCTORS

Carbon, being perhaps the most important element for life, has not played a big role in superconductivity for a long time. Things changed in 1980 when superconductivity was discovered below 0.9 K in the compound $(TMTSF)PF_6$, with the organic molecule TMTSF (tetra-methyl-tetra-selenium-fulvalene; see Fig. 24) [14]. To suppress a metal-insulator transition the material had to be kept under a hydrostatic pressure of 12 kbar. In 1991 it was found that compounds based on the "soccer ball" $C_{60}$ (see Fig. 25) become superconductors when doped with alkali atoms [403]. In 1994 superconductivity in Boron carbides was reported [404,405] (see the chapter 7), in 2001 there was evidence that carbon nanotubes when imbedded in a zeolite matrix can become superconductors with a $T_c$ of about 15 K [406]. Most recent are reports on superconductivity in diamond with $T_c$ up to 4 K when doped with boron [3], and in yttrium carbide compounds with $T_c$ as high as 18 K [407,408].

Regarding superconductors based on the TMTSF molecule a variety of compounds are known by now, exhibiting transition temperatures around 1 K. An example is $(TMTSF)_2ClO_4$, which also under normal pressure remains metallic down to the lowest temperatures and becomes superconducting at 1 K [409]. The general formula is $(TMTSF)_2X$ where X denotes an electron acceptor such as $PF_6$, $ClO_4$, $AsF_6$ or $TaF_6$. In the TMTSF compounds, the organic molecules are stacked upon each other (see Fig. 24).

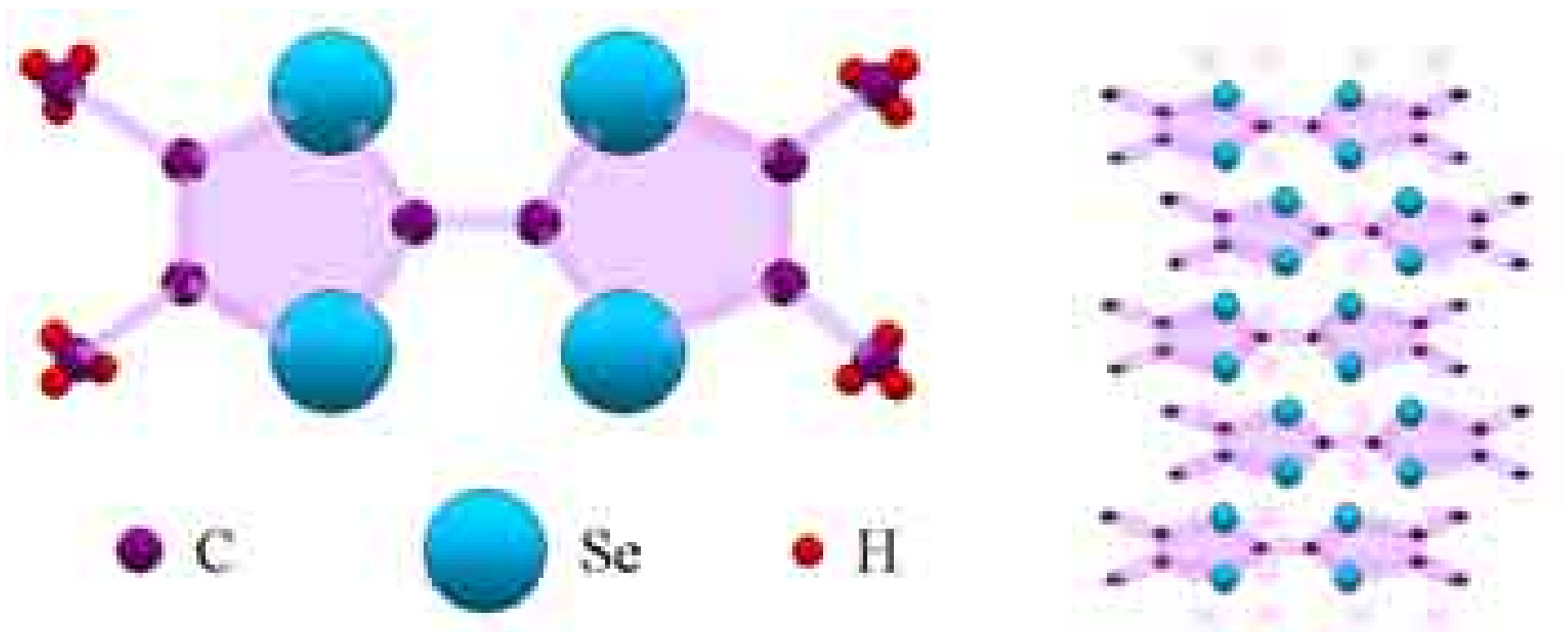

**Fig. 24** Structure of the organic molecule tetra-methyl-tetra-selenium-fulvalene (TMTSF) and stack arrangement of the molecules forming one-dimensional conduction channels.

In the normal state, the TMTSF compounds have a relatively large electric conductivity along the stacks, but only a small conductivity perpendicular to the stacks, thus forming *nearly-one-dimensional conductors*. The TMTSF compounds are type-II superconductors with highly anisotropic properties. For example, in $(TMTSF)_2ClO_4$ along the stacks the Ginzburg-Landau coherence length is about 80 nm, whereas along the two perpendicular directions of the crystal axes it is about 35 nm and 2 nm, respectively. The latter value is of the same order of magnitude as the lattice constant along the c axis. Hence, the compound nearly represents a *two-dimensional superconductor* [410].

Another important class of organic superconductors, often exhibiting $T_c$ well above 1 K is based on the bis-ethylene-dithia-tetra-thiafulvalene molecule, abbreviated as "BEDT-TTF" or "ET". For example, the compound $(BEDT\text{-}TTF)_2Cu[N(CN)_2]Br$ becomes superconducting at 11.2 K [411]. The transition temperature of $(BEDT\text{-}TTF)_2Cu(NCS)_2$ is 10.4 K. The ET-compounds are also highly anisotropic. However, in contrast to the TMTSF-compounds, in the normal state they form two-dimensional layered structures with a large electric conductivity in two dimensions. Like the TMTSF based materials the ET-compounds are type-II superconductors as well, with very short out-of-plane coherence lengths. These compounds thus also represent superconducting layered structures making them in many respects similar to HTS. Like for HTS, the pairing mechanism of the organic superconductors is at present still unclear. At least some compounds appear to be d-wave superconductors. However, in the compound $(TMTSF)_2PF_6$ one may even deal with a spin-triplet superconductor [412].

Fullerides are compounds of the form $A_3C_{60}$. They can become superconducting at surprizingly high temperatures [18,413,414]. By now, a number of SC fullerides based on the admixture of alkali atoms or of alkaline earth atoms are known. $Rb_3C_{60}$ has a value of $T_C$ of 29.5 K, and the present record under pressure is held by $Cs_3C_{60}$ with $T_C$ = 40 K. The crystal structure of the fullerides is face centered cubic, with the alkali atoms occupying interstitial sites between the large $C_{60}$ molecules. Fullerides are BCS-like s-wave superconductors. Intramolecular $C_{60}$ phonons (see Fig. 25) seem to contribute the most important part of the pairing interactions [18].

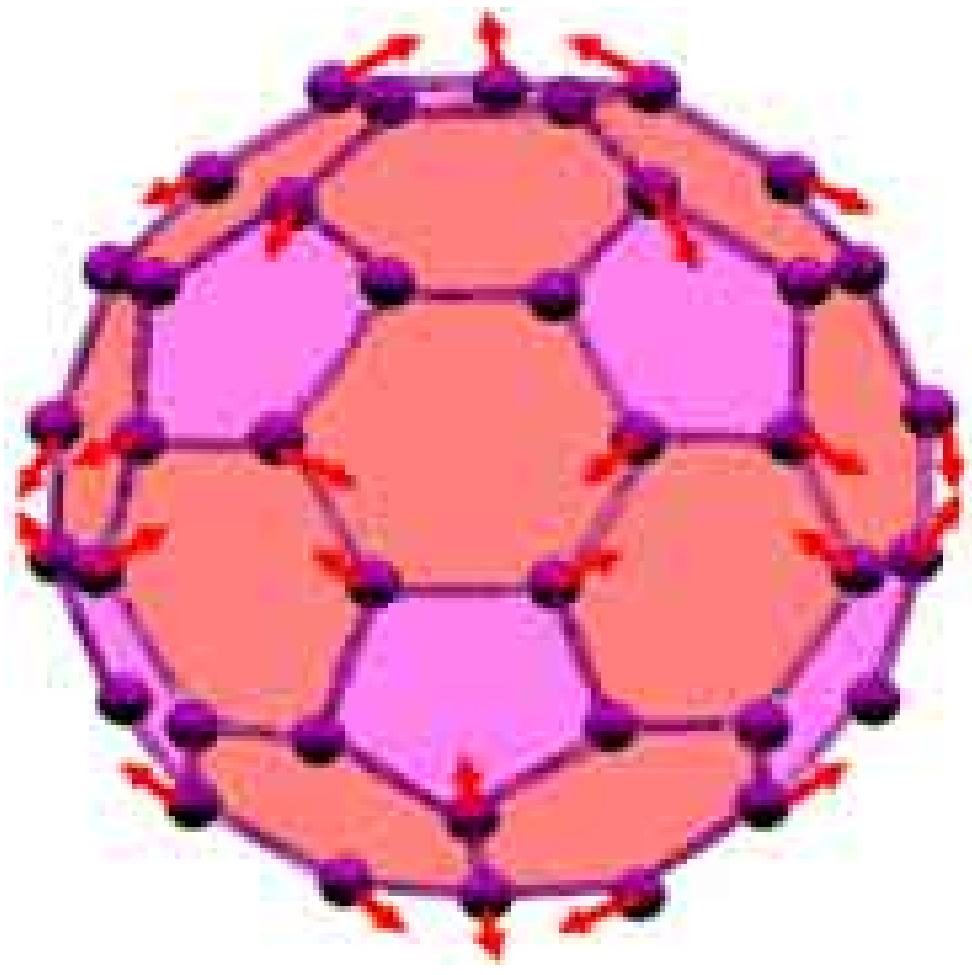

**Fig. 25** Structure of the $C_{60}$ molecule. The red arrows indicate one of the intramolecular $H_g$ phonon modes which are believed to be mainly responsible for the SC pairing [18].

## 7. BORIDES AND BOROCARBIDES

Rare-earth borocarbide superconductors have provided the first example of a homogeneous coexistence of superconductivity and ferromagnetism for all temperatures below $T_c$: The two antagonistic long-range orders are carried by different species of electrons that interact only weakly through contact exchange interaction leading to a small effect of the local moment molecular field on the SC conduction electrons. This allows a much better understanding of coexistence behavior as compared to the HF systems. Moreover, the nonmagnetic rare earth borocarbides have extremely large gap anisotropy ratios $\Delta_{max}/\Delta_{min} \geq 100$. Surely the standard electron-phonon mechanism has to be supplemented by something else, perhaps anisotropic Coulomb interactions to achieve this “quasi-unconventional” behavior in borocarbides.

The SC class of layered transition metal borocarbides $RNi_2B_2C$ (nonmagnetic R = Y, Lu, Sc; magnetic R = lanthanide elements in a $R^{3+}$ state; see Fig. 26) was discovered in 1994 [404,415,416,417]. The crystal structure consists of RC rock salt type planes separated by $Ni_2B_2$ layers built from $NiB_4$ tetrahedra and stacked along the c-axis. More general structures with more than one RC layer are possible [416]. The nonmagnetic borocarbides have relatively high $T_c$ values around 15 K. There is evidence that the SC mechanism is primarily of the electron-phonon type although this cannot explain the large anisotropy of the SC gap. At first sight the layered structure is similar to the HTS cuprates. However, unlike the copper oxide planes the $NiB_2$ planes show buckling. As a consequence, the electronic states at the Fermi level in the borocarbides do not have quasi-2-dimensional $d_{x^2-y^2}$ character and, therefore, have much weaker correlations excluding the possibility of AF spin-fluctuation mediated superconductivity.

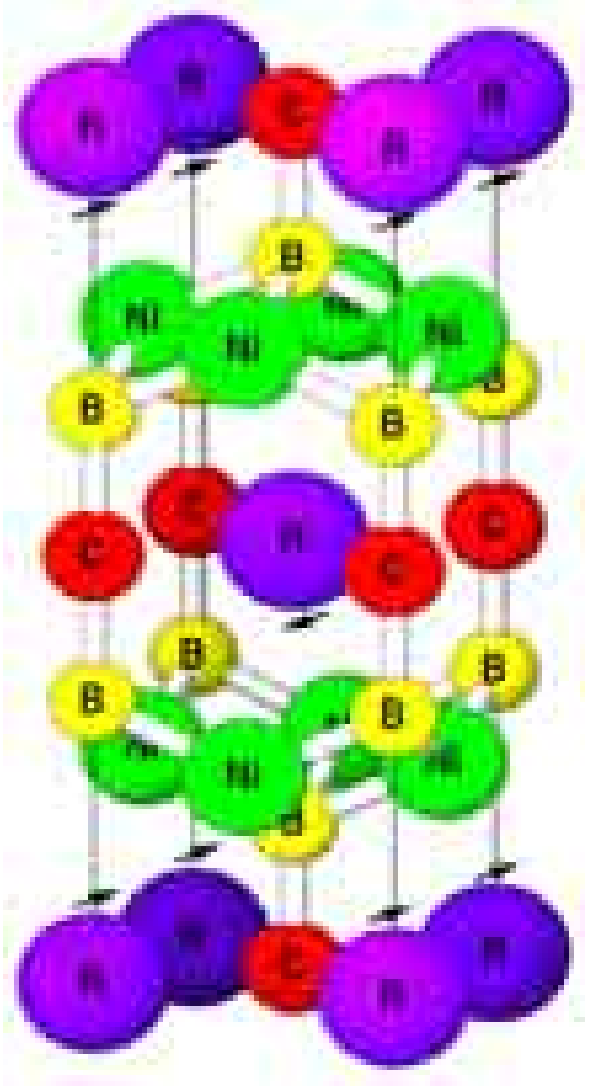

**Fig. 26** Tetragonal crystal structure of $RNi_2B_2C$ (a = 0.352 nm, c = 1.053 nm for R = Ho) and low temperature AF magnetic structure with [110] easy axis as indicated by the arrows.

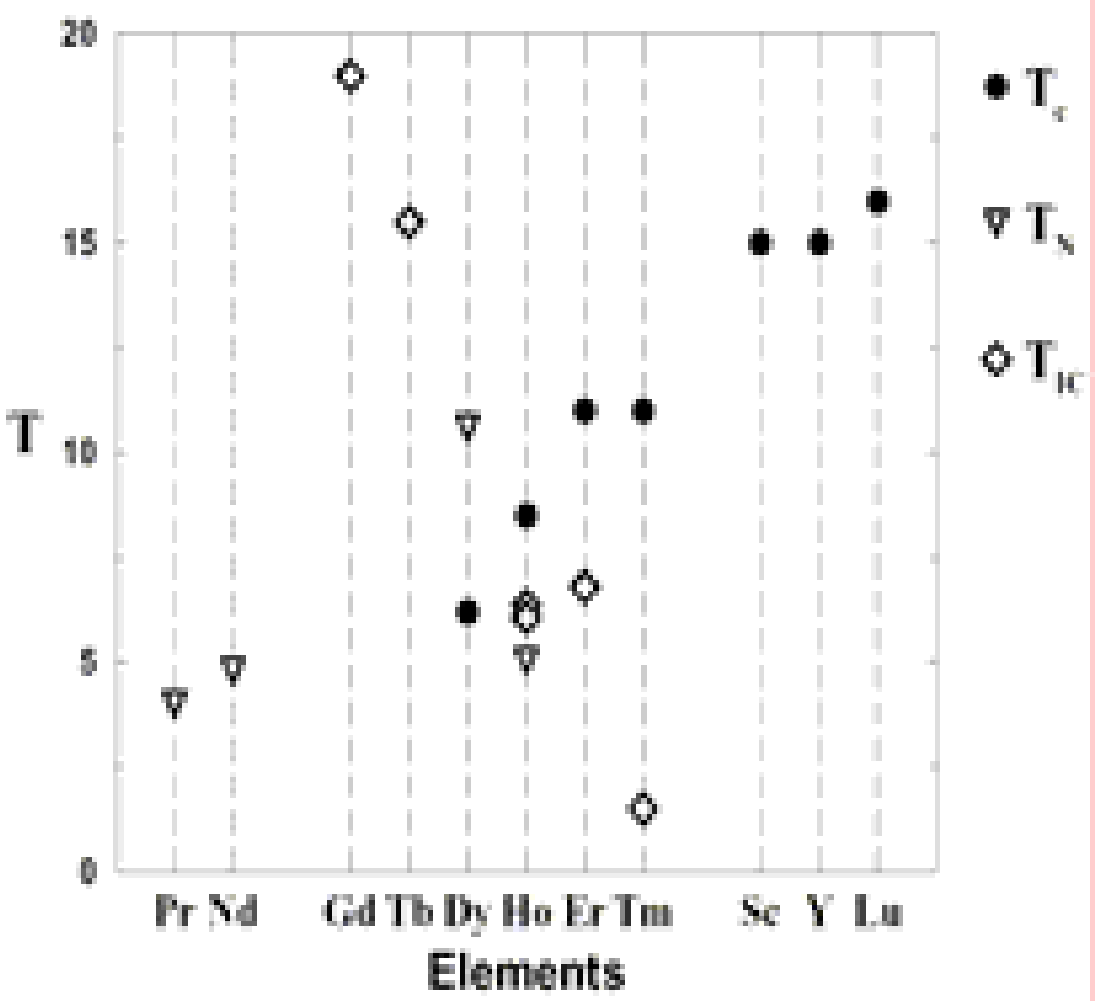

**Fig. 27** Magnetic ($T_{IC}$: incommensurate magnetic structure, $T_N$: simple AF structure) and SC ($T_c$) transition temperatures in Kelvin for the $RNi_2B_2C$ series.

The nonmagnetic borocarbides serve as a kind of reference point to separate the fascinating effects of AF and SC order parameter coupling in the magnetic $RNi_2B_2C$. However, the former have their own peculiarities, which are not yet completely understood. Foremost, despite their alleged electron-phonon nature, $LuNi_2B_2C$ and $YNi_2B_2C$ have strongly anisotropic gap functions and low energy quasiparticle states as is evident from specific heat and thermal conductivity. Furthermore, an anomalous upturn in $H_{c2}$ has been observed.

The magnetic $RNi_2B_2C$ are an excellent class of materials to study the effects of competition of magnetic order and superconductivity for the following reasons: The $T_c$ values are relatively high, and the $T_c / T_N$ ratio varies systematically across the R-series (see Fig. 27). Especially interesting are the cases of $RNi_2B_2C$ with R = Dy, Ho and Er where $T_c$ and $T_N$ (or $T_C$) are not too different, leading to strong competition of the magnetic and superconductivity order parameters. Furthermore, the SC condensate and magnetic moments are carried by different types of electrons, namely itinerant 3d-electrons for the $Ni_2B_2$ layers and localized $R^{3+}$ 4f-electrons for the R C layers, respectively. Finally, they are well separated and their coupling which is of the local exchange type can be treated in a controlled perturbative way, somewhat akin to the situation in the well known classes of Chevrel phase [418] and ternary compound [419] magnetic superconductors.

The AF molecular field establishes a periodic perturbation characterized by a length scale of the order of the Fermi wavelength. This implies that the spatial extent of the Cooper pairs extends over many periods of the alternating molecular field. The latter is therefore effectively averaged to zero and does not suppress superconductivity via an orbital effect. The system is invariant under the combined operation of time inversion followed by a translation with a lattice vector which allows to form Cooper pairs in a spin singlet state with vanishing (crystal) momentum in the AF lattice. This pair-state can be considered as a natural generalization of the pairing in time-reversed states encountered in usual non-magnetic superconductors.

The nonmagnetic $YNi_2B_2C$ and $LuNi_2B_2C$ compounds with comparatively high $T_c$ values of 16.5 K and 15.5 K serve as reference systems for the more difficult systems $RNi_2B_2C$ with both magnetic and SC phases. The electron-phonon nature of superconductivity in $YNi_2B_2C$ and $LuNi_2B_2C$ is inferred from a substantial s-wave character of the order parameter as witnessed by the appearance of a moderate Hebel-Slichter peak in the $^{13}C$ NMR relaxation rate [420].

On the other hand, the gap function is strongly anisotropic as can be seen both from temperature and field dependence of thermodynamic and transport quantities [416,421] indicating the presence of gap nodes [422]. More precisely, within experimental accuracy, there must be at least a gap anisotropy $\Delta_{max}/\Delta_{min} \geq 100$ [422]. For an electron-phonon superconductor this would be the largest anisotropy ever observed. This conjecture is also supported by the field dependence of the low temperature specific heat [423] and of the thermal conductivity along (001) [421]. Since in the latter case the heat current is perpendicular to the vortices this proves that quasiparticles must be present in the inter-vortex region. This is also required to explain the observation of dHvA oscillations far in the vortex phase. Experimental evidence therefore demands a nodal gap function for borocarbides. A

s+g wave model [424] fulfills the requirements. In addition, it explains recent results on ultrasonic attenuation, which also confirmed the existence of gap nodes in the cubic plane [425].

The classical argument in favor of the electron-phonon mechanism is the observation of an isotope effect characterized by the isotope exponent for a specific atom with mass M as given by $\alpha_X = d [\ln T_c] / d [\ln M_X]$. However, this cannot be applied easily to complex layered superconductors, as is evident from the existence of an isotope effect in the probably nonphononic cuprate superconductors with nonoptimal doping. The boron isotope effect found in $YNi_2B_2C$ ($\alpha_B = 0.2$) and $LuNi_2B_2C$ ($\alpha_B = 0.11$) is much smaller than the BCS value $\alpha_X = 0.5$. A nonphononic origin stemming from the influence of boron on the charge density in the $B_2Ni_2$ layers has therefore been suggested [426]. The most direct support for phonon-mediated Cooper pairing comes from scanning tunneling spectroscopy ("STS")[427] which has shown the existence of a strong coupling signature in the tunneling DOS due to a soft optical phonon close to the Fermi surface nesting wave vector Q. The STS-derived electron-phonon coupling constant $\lambda = 0.5 - 0.8$ is compatible with the value $\lambda = 0.53$ obtained from resistivity data.

The discovery of superconductivity in $MgB_2$ (see Fig. 28) in early 2001 with $T_c \sim 40$ K [61], almost twice the record value before HTS [20] brought back a bit of the HTS bonanza spirit of the late 1980s [429,430,431,432,433]. The fact that such a simple material which was known since the early 1950s had been missed in the systematic research for superconductivity on this class of compounds was as amazing as the "materials preparation" that was usually practiced immediately after the discovery by most of the research groups for reestablishing this high $T_c$ value: They simply ordered $MgB_2$ powder from chemicals wholesale where it was commercially available in quantities of metric tons already for years!

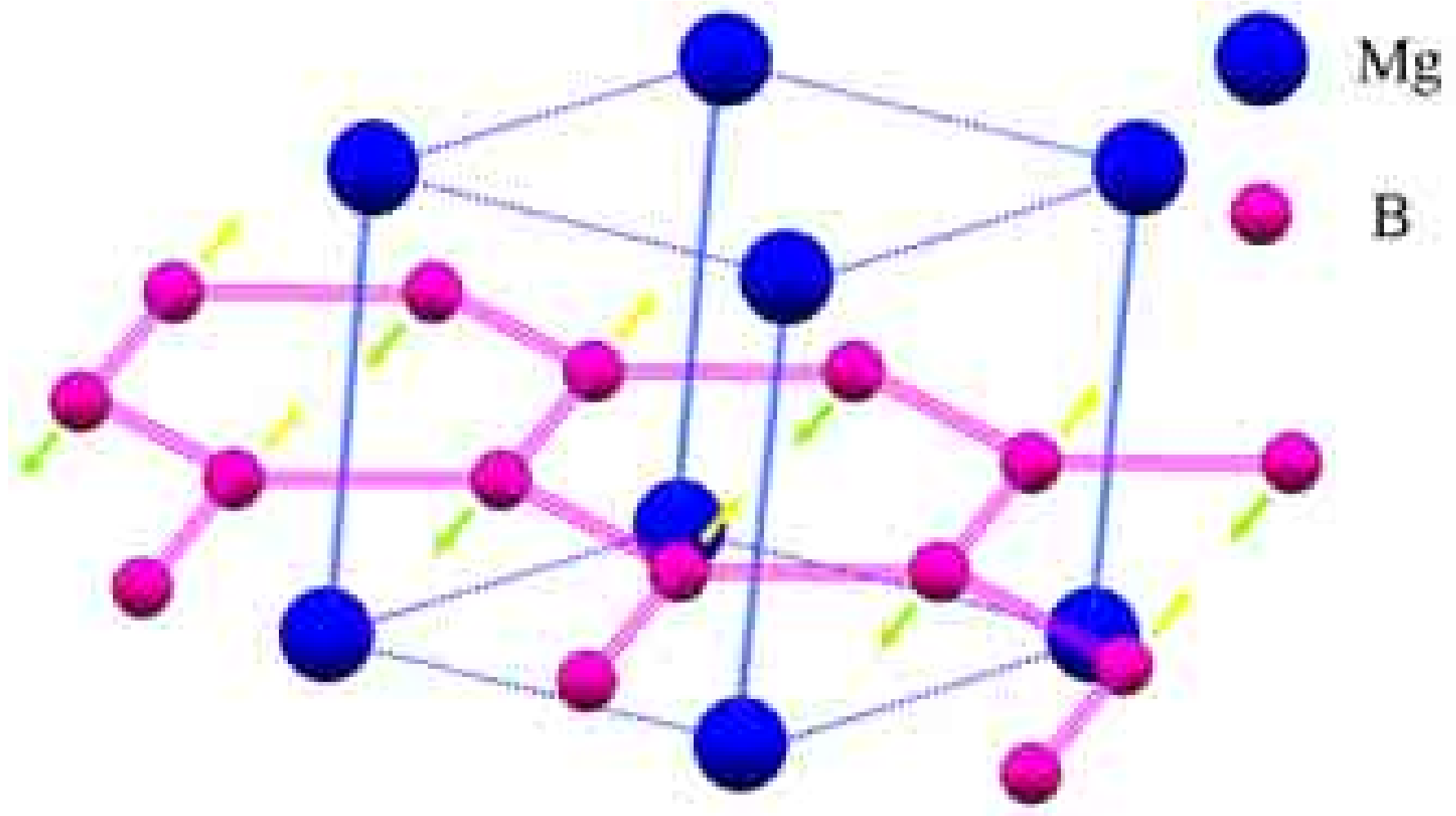

**Fig. 28** Hexagonal crystal structure of $MgB_2$. The arrows indicate the B-phonon mode, which presumably introduces the strongest SC coupling.

[61] Isotopically pure $Mg^{11}B_2$ and $Mg^{10}B_2$ show sharp superconducting transitions in resistivity with $T_c(Mg^{11}B_2) = 39.2$ K and $T_c(Mg^{10}B_2) = 40.2$ K [428].

No strong Coulomb correlation effects can be expected in $MgB_2$: No atomic d- or f-shells are involved in the conduction electron system of this binary compound of light elements. In addition, the simple crystal structure consisting of graphite-like B-layers with intercalated Mg clearly favors conduction along these layers and a respective superconductive and normal state anisotropy, but it does not introduce a reduction of the effective dimensionality, as in the case of organic superconductors due to the stacking of isolated aromatic rings. The coupling of the conduction electrons to a particular boron phonon mode (see Fig. 28) was hence identified right from the start as basic origin of superconductivity in $MgB_2$ [434, 435]. The observation of two energy gaps (at 1.8 and 6.8 meV [436, 437]) and the considerable superconductive anisotropy as large as 6–9 [428] challenged a more thorough theoretical investigation which tries at present to explain these findings in terms of two-band superconductivity [439] on the basis of the large anharmonicity of the involved phonon mode and a refined treatment of its coupling with the different sheets of the electronic conduction band [436,438,440,441].

$MgB_2$ shows great promise as material for both large-scale and electronic *applications* [4]:

- low cost and abundant availability of the raw materials
- high $T_c$
- extreme type-II SC nature as a consequence of a large coherence lengths $\xi_0 \sim 4.4$ nm and an even larger penetration depth $\lambda \sim 132$ nm in combination with the occurrence of suitable pinning centers
- both SC gaps larger than that of Nb (1.5 meV)
- high critical current densities $J_c$ (10 K, 0 T) ~ 10 MA/cm$^2$
- high critical fields ($H_{c2}$(0 K) ~ 16 T in bulk, > 40 T achieved for thin films)
- transparency of grain boundaries to current
- remarkably low normal state resistivity $\rho$(42 K) = 0.38 μΩ cm [62] [428]

$MgB_2$ has already been fabricated in the form of bulk, single crystals, thin films [442], tapes and wires [443]. Serious difficulties in depositing $MgB_2$ films, such as the volatility of Mg, the phase stability of $MgB_2$, the low sticking coefficients of Mg at elevated temperatures, and the reactivity of Mg with oxygen have meanwhile been overcome. Epitaxial $MgB_2$ films with superior superconducting properties have been produced [444]. $MgB_2$ wires are already used in first real applications [443].

$MgB_2$ is hence a further excellent example for the progress and excitement of the present research on superconductors. Superconductivity with its wide range from the most fundamental aspects of physics to hands-on applications is, one century after its discovery, still one of the most fascinating topics of modern science.

---

[62] This value is comparable to that of copper wire and over a factor of 20 times lower than that of polycrystalline $Nb_3Sn$ [436]. It means that $MgB_2$ can carry in the normal state substantially higher current densities than other superconductors with less Joule heating. This combination of a high $T_c$ and a low normal state resistivity makes $MgB_2$ particularly appealing for wire applications [428].

## REFERENCES

[1] W. Buckel, R. Kleiner, *Supraleitung – Grundlagen und Anwendungen,* WILEY-VCH Verlag, Weinheim (2004);

W. Buckel, R. Kleiner, *Superconductivity - Fundamentals and Applications,* WILEY-VCH Verlag, Weinheim (2004)

[2] T. H. Geballe, Science **293 (**2001) 223

[3] E. A. Ekimov, V. A. Sidorov, E. D. Bauer, N. N. Mel'nik, N. J. Curro, J. D. Thompson, S. M. Stishov, Nature **428 (**2004) 542

[4] C. Buzea, T. Yamashita, Supercond. Sci. Techn. **14 (**2001) R115

[5] J. Bardeen, L. N. Cooper, J. R. Schrieffer, Phys. Rev **108** (1957) 1175

[6] J. G. Bednorz, K. A. Müller, Z. Phys. B **64** (1986) 189

[7] S. N. Putilin, E. V. Antipov, A. M. Abakumov, M. G. Rozova, K. A. Lokshin, D. A. Pavlov, A. M. Balagurov, D. V. Sheptyakov, M. Marezio, Physica C **338** (2001) 52

[8] W. L. McMillan, J. M. Rowell, Phys. Rev. Lett. **14** (1965) 108

[9] R. Zeyer, G. Zwicknagl, Z. Phys. B **78** (1990) 175

[10] J. R. Gavaler, Appl. Phys. Lett. **23** (1973) 480

[11] F. Steglich, J. Aarts, C. D. Bredl, W. Lieke, D. Meschede, W. Franz, H. Schäfer, Phys. Rev. Lett. **43** (1979) 1892

[12] J. L. Sarrao, L. A. Morales, J. D. Thompson, B. L. Scott, G. R. Stewart, F. Wastin, J. Rebizant, P. Boulet, E. Colineau, G. H. Lander, Nature **420** (2002) 297

[13] W. A. Little, Phys. Rev. A **134** (1964) 1416

[14] D. Jérome, A. Mazaud, M. Ribault, K. Bechgaard, J. Phys. (Paris), Lett. **41** (1980) L95;

D. Jerome, A. Mazaud, M. Ribault,. K. Bechgaard; C. R. Hebd. Seances Acad. Sci. Ser. **B 290** (1980)

[15] G. Saito, H. Yamochi, T. Nakamura, T. Komatsu, M. Nakashima, H. Mori, K. Oshima, Physica B **169** (1991) 372

[16] N. D. Mermin, H. Wagner, Phys. Rev. Lett. **17** (1966) 1133 & 1307

[17] K. Tanigaki, T. W. Ebbesen, S. Saito, J. Mizuki, J. S. Tsai, Y. Kubo, S. Kuroshima, Nature **352** (1991) 222

[18] O. Gunnarsson, Rev. Mod. Phys. **69** (1997) 575

[19] P. C. Canfield, P. L. Gammel, D. J. Bishop, Physics Today, October 1998, p. 40

[20] J. Nagamatsu, N. Nakagawa, T. Muranaka, Y. Zenitani, J. Akimitsu, Nature **410** (2001) 63

[21] T. Dahm, this volume

[22] R. Hott, *High Temperature Superconductivity 2 - Engineering Applications* (Ed. A. V. Narlikar), Springer Verlag, Berlin (2004) pp.35; cond-mat/0306444; http://wwwifp.fzk.de/ISAS/

*Handbook of Applied Superconductivity* (Ed. B. Seeber),
Institute of Physics Publishing, Bristol and Philadelphia (1998)

[23] P. Komarek, Supercond. Sci. Technol. **13** (2000) 456

[24] Y. Wang, S. Ono, Y. Onose, G. Gu, Y. Ando, Y. Tokura, S. Uchida, N. P. Ong, Science **299** (2003) 86

[25] G. Blatter, M. V. Feigel'man, V. B. Geshkenbein, A. I. Larkin, V. M. Vinokur, Rev. Mod. Phys. **66** (1994) 1125

[26] H. W. Weijers, U. P. Trociewitz, K. Marken, M. Meinesz, H. Miao, J. Schwartz, Supercond. Sci. Technol. **17** (2004) 636

[27] http://www.lanl.gov/superconductivity/wire_conductors.shtml

[28] CERN Weekly Bulletin No. 21/2004

[29] P. Renton, Nature **428** (2004) 141

[30] Physics Today **54,** May 2001, pp.27

[31] H. J. Chaloupka, T. Kässer, *High Temperature Superconductivity 2 - Engineering Applications* (Ed. A. V. Narlikar), Springer Verlag, Berlin (2004) pp.411

[32] K. K. Likharev, *Dynamics of Josephson Junctions and Circuits,* Gordon & Breach Science Publishers (1986) p.30

[33] *SQUID Handbook* (Ed. J. Clarke, A. Braginski), WILEY-VCH Verlag, Berlin (2003)

[34] J. Malmivuo, V. Suihko, H. Eskola, IEEE Trans. Biomed. Eng. **44** (1997) 196

[35] D. Koelle, R. Kleiner, F. Ludwig, E. Dantsker, J. Clarke, Rev. Mod. Phys. **71** (1999) 631;

H. J. Barthelmess, F. Ludwig, M. Schilling, D. Drung, T. Schurig, *High Temperature Superconductivity 2 -Engineering Applications* (Ed. A. V. Narlikar), Springer Verlag, Berlin (2004) pp.299

[36] L. A. Knauss, A. B. Cawthorne, N. Lettsome, S. Kelly, S. Chatraphorn, E. F. Fleet, F. C. Wellstood, W. E. Vanderlinde, Microelectronics Reliability **41** (2001)1211

[37] M.v.Kreutzbruck, *High Temperature Superconductivity 2 - Engineering Applications* (Ed. A. V. Narlikar), Springer Verlag, Berlin (2004) pp.223

[38] *WTEC Panel Report on Electronic Applications of Superconductivity in Japan*, International Technology Institute, Maryland (1998);
http://www.wtec.org/loyola/scel96/toc.htm

[39] A. Ekert, R. Jozsa, Rev. Mod. Phys. **68** (1996) 733;

http://www.cs.caltech.edu/~westside/quantum-intro.html;

http://www.qubit.org/library/intros/comp/comp.html

[40] I. Chiorescu, Y. Nakamura, C. J. P. M. Harmans, J. E. Mooij, Science **299** (2003) 1869

[41] H. Hilgenkamp, Ariando, H.-J. H. Smilde, D. H. A. Blank, G. Rijnders, H. Rogalla, J. R. Kirtley, C. C. Tsuei, Nature **422** (2003) 50

[42] R. Hott, *High Temperature Superconductivity,* Vol. II *Engineering Applications*, (Ed. A. V. Narlikar), Springer Verlag, Berlin (2004) pp.41;

H. J. M. ter Brake, G. F. M. Wiegerinck, Cryogenics **42** (2002) 705

[43] A. Damascelli, Z. Hussain, Z.-X. Shen, Rev. Mod. Phys. **75** (2003) 473

[44] J. C. Campuzano, M. R. Norman, M. Randeria, in *Physics of Superconductors*, Vol. II, (Ed. K. H. Bennemann & J. B. Ketterson), Springer Verlag, Berlin (2004), p. 167-273; cond-mat/0209476;

M. R. Norman, C. Pepin, Rep. Prog. Phys. **66** (2003) 1547

[45] *Neutron Scattering in Layered Copper-Oxide Superconductors* (Ed. A. Furrer), Kluwer Academic Publishers (1998), ISBN 0-7923-5226-2

[46] B. W. Hoogenboom, M. Kugler, B. Revaz, I. Maggio-Aprile, Ø. Fischer, Ch. Renner, Phys. Rev. B **62** (2000) 9179

[47] K. McElroy, D.-H. Lee, J. E. Hoffman, K. M. Lang, E. W. Hudson, H. Eisaki, S. Uchida, J. Lee, J. C. Davis, Nature **428** (2004) 542; cond-mat/0404005

[48] R. F. Service, Science **271** (1996) 1806

[49] A. Erb, E. Walker, R. Flükiger, Physica C **245** (1995) 245;

A. Erb, E. Walker, J.-Y. Genoud, R. Flükiger, Physica C **282-287** (1997) 89

[50] J. Orenstein, A. J. Millis, Science **288** (2000) 468

[51] P. W. Anderson, Science **235** (1987) 1196

[52] P. W. Anderson, *The Theory of Superconductivity in High-$T_c$ Cuprates Superconductors,* Princeton University Press, Princeton (1997)

[53] A. B. Fowler, Physics Today, October 1993, 59

[54] C. W. Chu, J. Supercond. **12** (1999) 85

[55] C. W. Chu, IEEE Trans. Appl. Supercond. **7** (1997) 80

[56] B. Raveau, in *High-$T_c$ Superconductivity 1996: Ten Years after the Discovery*, NATO ASI Series E Vol. **343,** Kluwer Academic Publishers (1997) p.109

[57] J. L. Tallon, G. V. M. Williams, J. W. Loram, Physica C **338** (2000) 9

[58] H. Lütgemeier, S. Schmenn, P. Meuffels, O. Storz, R. Schöllhorn, Ch. Niedermayer, I. Heimaa, Yu. Baikov, Physica C **267** (1996) 191

[59] R. Bormann, J. Nölting, Appl. Phys. Lett. **54** (1989) 2148

[60] A. Mawdsley, J. L. Tallon, M. R. Presland, Physica C **190** (1992) 437

[61] E. L. Brosha, P. K. Davis, F. H. Garzon, I. D. Raistrick, Science **260** (1993) 196

[62] H. Yamauchi, M. Karppinen, S. Tanaka, Physica C **263** (1996) 146

[63] H. Yamauchi, M. Karppinen, Supercond. Sci. Technol. **13** (2000) R33

[64] P. Bordet, S. LeFloch, C. Chaillout, F. Duc, M. F. Gorius, M. Perroux, J. J. Capponi, P. Toulemonde, J. L. Tholence, Physica C **276** (1997) 237

[65] N. L. Wu, Z. L. Du, Y. Y. Xue, I. Rusakova, D. K. Ross, L. Gao, Y. Cao, Y. Y. Sun, C. W. Chu, M. Hervieu, B. Raveau, Physica C **315** (1999) 227

[66] D. R. Harshman, A. P. Mills, Jr., Phys. Rev. B **45** (1992) 10684

[67] M. K. Wu, J. R. Ashburn, C. J. Torng, P. H. Hor, R. L. Meng, L. Gao, Z. J. Huang, Y. Q. Wang, C. W. Chu, Phys. Rev. Lett. **58** (1987) 908

[68] J. G. Lin, C. Y. Huang, Y. Y. Xue, C. W. Chu, X. W. Cao, J. C. Ho, Phys. Rev. B **51** (1995) 12900

[69] J. W. Chu, H. H. Feng, Y. Y. Sun, K. Matsuishi, Q. Xiong, C. W. Chu, *HTS Materials, Bulk Processing and Bulk Applications* - Proceedings of the 1992 TCSUH Workshop (Ed. C. W. Chu, W. K. Chu, P. H. Hor and K. Salama), World Scientific, Singapore (1992), p. 53; TCSUH preprint 92:043

[70] G. V. M. Williams, J. L. Tallon, Physica C **258** (1996) 41

[71] M. Guillaume, P. Allenspach, W. Henggeler, J. Mesot, B. Roessli, U. Staub, P. Fischer, A. Furrer, V. Trounov, J. Phys.: Condens. Matter **6** (1994) 7963

[72] M. Guillaume, P. Allenspach, J. Mesot, B. Roessli, U. Staub, P. Fischer, A. Furrer, Z. Phys. B **90** (1993) 13

[73] M. Al-Mamouri, P. P. Edwards, C. Greaves, M. Slaski, Nature **369** (1994) 382

[74] T. Kawashima, Y. Matsui, E. Takayama-Muromachi, Physica C **257** (1996) 313

[75] Z. Hiroi, N. Kobayashi, M. Takano, Nature **371** (1994) 139

[76] L. Gao, Y. Y. Xue, F. Chen, Q. Xiong, R. L. Meng, D. Ramirez, C. W. Chu, J. H. Eggert, H. K. Mao, Phys. Rev. B **50** (1994) 4260

[77] B. W. Roberts, J. Phys. Chem. Ref. Data **5** (1976) 581

[78] B. W. Roberts, NBS Technical Note **983** (1978), *Properties of Selected Superconductive Materials* 1978 Supplement

[79] Landolt-Börnstein, New Series, Group III, Vol. **21**, *Superconductors* (1990-2002)

[80] R. K. Williams; D. M. Kroeger, P. M. Martin, J. R. Mayotte, E. D. Specht, J. Brynestad, J. Appl. Phys. **76** (1994) 3673;

W. Zhang, K. Osamura, Physica C **190** (1992) 396;

G. F. Voronin, S. A. Degterov, Physica C **176** (1991) 387;

J. Karpinski, S. Rusiecki, B. Bucher, E. Kaldis, E. Jilek, Physica C **161** (1989) 618;

J. Karpinski, S. Rusiecki, B. Bucher, E. Kaldis, E. Jilek, Physica C **160** (1989) 449;

J. Karpinski, E. Kaldis, E. Jilek, S. Rusiecki, B. Bucher, Nature **336** (1988) 660

[81] R. Liang, D. A. Bonn, W. N. Hardy, J. C. Wynn, K. A. Moler, L. Lu, S. Larochelle, L. Zhou, M. Greven, L. Lurio, S. G. J. Mochrie, Physica C **383** (2002) 1

[82] Th. Wolf, A.-C. Bornarel, H. Küpfer, R. Meier-Hirmer, B. Obst, Phys. Rev. B **56** (1997) 6308

[83] P. Zoller, J. Glaser, A. Ehmann, C. Schultz, W. Wischert, S. Kemmler-Sack, T. Nissel, R. P. Huebener, Z. Phys. B **96** (1995) 505

[84] D. L. Feng, A. Damascelli, K. M. Shen, N. Motoyama, D. H. Lu, H. Eisaki, K. Shimizu, J.-i. Shimoyama, K. Kishio, N. Kaneko, M. Greven, G. D. Gu, X. J. Zhou, C. Kim, F. Ronning,N. P. Armitage, Z.-X. Shen, Phys. Rev. Lett. **88** (2002) 107001

[85] H. Eisaki, N. Kaneko, D. L. Feng, A. Damascelli, P. K. Mang, K. M. Shen, Z.-X. Shen, M. Greven, Phys. Rev. B **69** (2004) 064512

[86] S. Lösch, H. Budin, O. Eibl, M. Hartmann, T. Rentschler, M. Rygula, S. Kemmler-Sack, R. P. Huebener, Physica C **177** (1991) 271

[87] H. Yamauchi, T. Tamura, X.-J. Wu, S. Adachi, S. Tanaka, Jpn. J. Appl. Phys. **34** (1995) L349

[88] T. Tamura, S. Adachi, X.-J. Wu, T. Tatsuki, K. Tanabe, Physica C **277** (1997) 1

[89] A. Iyo, Y. Tanaka, Y. Ishiura, M. Tokumoto, K. Tokiwa, T. Watanabe, H. Ihara, Supercond. Sci. Technol. **14** (2001) 504

[90] A. Iyo, Y. Aizawa, Y. Tanaka, M. Tokumoto, K. Tokiwa, T. Watanabe, H. Ihara, Physica C **357-360** (2001) 324

[91] D. Tristan Jover, R. J. Wijngaarden, R. Griessen, E. M. Haines, J. L. Tallon, R. S. Liu, Phys. Rev. B **54** (1996) 10175

[92] Z. Y. Chen, Z. Z. Sheng, Y. Q. Tang,Y. F. Li, L. M. Wang, D. O. Pederson, Supercond. Sci. Technol. **6** (1993) 261

[93] E. V. Antipov, A. M. Abakumov, S. N. Putilin, Supercond. Sci. Technol. **15** (2002) R31

[94] C. Acha, S. M. Loureiro, C. Chaillout, J. L. Tholence, J. J. Capponi, M. Marezio, M. Nunez-Regueiro, Solid State Comm. **102** (1997) 1

[95] T. Tatsuki, A. Tokiwa-Yamamoto, A. Fukuoka, T. Tamura, X.-J. Wu, Y. Moriwaki, R. Usami, S. Adachi, K. Tanabe, S. Tanaka, Jpn. J. Appl. Phys. **35** (1996) L205

[96] E. Stangl, S. Proyer, M. Borz, B. Hellebrand, D. Bäuerle, Physica C **256** (1996) 245

[97] H. Ihara, Physica C **364-365** (2001) 289

[98] P. W. Klamut, B. Dabrowski, S. M. Mini, M. Maxwell, S. Kolesnik, J. Mais, A. Shengelaya, R. Khasanov I. Savic, H. Keller, T. Graber, J. Gebhardt, P. J. Viccaro, Y. Xiao, Physica C **364-365** (2001) 313;

B. Lorenz, R. L. Meng, J. Cmaidalka, Y. S. Wang, J. Lenzi, Y. Y. Xue, C. W. Chu, Physica C **363** (2001) 251

[99] T. Kawashima, Y. Matsui, E. Takayama-Muromachi, Physica C **254** (1995) 131

[100] I. Bozovic, G. Logvenov, I. Belca, B. Narimbetov, I. Sveklo, Phys. Rev. Lett. **89** (2002) 107001

[101] S. Karimoto, H. Yamamoto, T. Greibe, M. Naito, Jpn. J. Appl. Phys. **40** (2001) L127

[102] M. Naito, M. Hepp, Jpn. J. Appl. Phys. **39** (2000) L485

[103] L. Alff, S. Meyer, S. Kleefisch, U. Schoop, A. Marx, H. Sato, M. Naito, R. Gross, Phys. Rev. Lett. **83** (1999) 2644

[104] C. Q. Jin, Y. S. Yao, S. C. Liu, W. L. Zhou, W. K. Wang, Appl. Phys. Lett. **62** (1993) 3037

[105] A. Iyo, Y. Tanaka, M. Tokumoto, H. Ihara, Physica C **366** (2001) 43

[106] C. U. Jung, J. Y. Kim, S. M. Lee, M.-S. Kim, Y. Yao, S. Y. Lee, S.-I. Lee, D. H. Ha Physica C **364-365** (2001) 225

[107] P. Majewski, J. Mater. Res. **15** (2000) 854

[108] M. P. Siegal, E. L. Venturini, B. Morosin, T. L. Aselage, J. Mater. Res. **12** (1997) 2825

[109] J. J. Capponi, J. L. Tholence, C. Chaillout, M. Marezio, P. Bordet, J. Chenavas, S. M. Loureiro, E. V. Antipov, E. Kopnine, M. F. Gorius, M. Nunez-Regueiro, B. Souletie, P. Radaelli, F. Gerhards, Physica C **235-240** (1994) 146

[110] T. Ito, H. Suematsu, M. Karppinen, H. Yamauchi, Physica C **308** (1998) 198

[111] D. Monster, P.-A. Lindgard, N. H. Andersen, Phys. Rev. B **60** (1999) 110

[112] C. L. Jia, K. Urban, Science **303** (2004) 2001

[113] Y. Yan, M. G. Blanchin, Phys. Rev. B **43** (1991) 13717

[114] S. H. Pan, J. P. O'Neal, R. L. Badzey, C. Chamon, H. Ding, J. R. Engelbrecht, Z. Wang, H. Eisaki, S. Uchida, A. K. Gupta, K.-W. Ng, E. W. Hudson, K. M. Lang, J. C. Davis, Nature **413** (2001) 282

[115] M. Karppinen, A. Fukuoka, L. Niinistö, H. Yamauchi, Supercond. Sci. Technol. **9** (1996) 121

[116] S. Sadewasser, J. S. Schilling, A. M. Hermann, Phys. Rev. B **62** (2000) 9155

[117] Y. Ohta, T. Tohyama, S. Maekawa, Phys. Rev. B **43** (1991)2968

[118] H. Kotegawa, Y. Tokunaga, K. Ishida, G. -q. Zheng, Y. Kitaoka, K. Asayama, H. Kito, A. Iyo, H. Ihara, K. Tanaka, K. Tokiwa, T. Watanabe, J. Phys. Chem. Solids **62** (2001) 171

[119] H. Kotegawa, Y. Tokunaga, K. Ishida, G.-q. Zheng, Y. Kitaoka, H. Kito, A. Iyo, K. Tokiwa, T. Watanabe, H. IharaPhys. Rev. B **64** (2001) 064515

[120] S. Chakravarty, H.-Y. Kee, K. Völker, Nature **428** (2004) 53;
P. Coleman, Nature **428** (2004) 26

[121] J. Leggett, Phys. Rev. Lett. **83** (1999) 392

[122] C. Meingast, O. Kraut, T. Wolf, H. Wühl, A. Erb, G. Müller–Vogt, Phys. Rev. Lett. **67** (1991) 1634

[123] K. Kuroda, K. Abe, S. Segawa, J.-G. Wen, H. Unoki, N. Koshizuka, Physica C **275** (1997) 311

[124] J. P. Attfield, A. L. Kharlanov, J. A. McAllister, Nature **394** (1998) 157

[125] M. W. Pieper, F. Wiekhorst, T. Wolf, Phys. Rev. B **62** (2000) 1392;
A. J. Markwardsen, A. T. Boothroyd, B. Buck, G. J. McIntyre, Th. Wolf, J. Magn. Magn. Mater. **177-181** (1998) 502

[126] K. Oka, Z. Zou, J. Ye, Physica C **300** (1998) 200;
Z. Zou, J. Ye, K. Oka, Y. Nishihara, Phys. Rev. Lett. **80** (1998) 1074

[127] J. Ye, Z. Zou, A. Matsushita, K. Oka, Y. Nishihara, T. Matsumoto, Phys. Rev. B **58** (1998) 619

[128] Z. Hu, R. Meier, C. Schüßler-Langeheine, E. Weschke, G. Kaindl, I. Felner, M. Merz, N. Nücker, S. Schuppler, A. Erb, Phys. Rev. B **60** (1999) 1460

[129] M. Merz, S. Gerhold, N. Nücker, C. A. Kuntscher, B. Burbulla, P. Schweiss, S. Schuppler, V. Chakarian, J. Freeland, Y. U. Idzerda, M. Kläser, G. Müller-Vogt, Th. Wolf, Phys. Rev. B **60** (1999) 9317

[130] Y. S. Lee, F. C. Chou, A. Tewary, M. A. Kastner, S. H. Lee, R. J. Birgeneau, Phys. Rev. B **69** (2004) 020502

[131] P. K. Mang, S. Larochelle, A. Mehta, O. P. Vajk, A. S. Erickson, L. Lu, W. J. L. Buyers, A. F. Marshall, K. Prokes, M. Greven, cond-mat/0403258, submitted to Phys. Rev. **B**

[132] O. Chmaissem, J. D. Jorgensen, S. Short, A. Knizhnik, Y. Eckstein, H. Shaked, Nature **397** (1999) 45

[133] F. Izumi, J. D. Jorgensen, Y. Shimakawa, Y. Kubo, T. Manako, S. Pei, T. Matsumoto, R. L. Hitterman, Y. Kanke, Physica C **193** (1992) 426

[134] Z. F. Ren, J. H. Wang, D. J. Miller, Appl. Phys. Lett. **69** (1996) 1798

[135] A. Yamato, W. Z. Hu, F. Izumi, S. Tajima, Physica C **351** (2001) 329

[136] K. A. Moler, J. R. Kirtley, D. G. Hinks, T. W. Li, M. Xu, Science **279** (1998) 1193

[137] A. A. Tsvetkov, D. Van der Marel, K. A. Moler, J. R. Kirtley, J. L. De Boer, A. Meetsma, Z. F. Ren, N. Koleshnikov, D. Dulic, A. Damascelli, M. Grüninger, J. Schützmann, J. W. van der Eb, H. S. Somal, J. H. Wang, Nature **395** (1998) 360

[138] T. H. Geballe, B. Y. Moyzhes, Physica C **341-348** (2000) 1821

[139] D. H. Lu, D. L. Feng, N. P. Armitage, K. M. Shen, A. Damascelli, C. Kim, F. Ronning, Z.-X. Shen, D. A. Bonn, R. Liang, W. N. Hardy, A. I. Rykov, S. Tajima, Phys. Rev. Lett. **86** (2001) 4370

[140] D. A. Wollman, D. J. Van Harlingen, J. Giapintzakis, D. M. Ginsberg, Phys. Rev. Lett. **74** (1995) 797

[141] D. J. Van Harlingen, J. E. Hilliard, B. L. T. Blourde, B. D. Yanoff, Physica C **317-318** (1999) 410

[142] C. C. Tsuei, J. R. Kirtley, Rev. Mod. Phys. **72** (2000) 969;
C. C. Tsuei, J. R. Kirtley, Physica C **367** (2002) 1

[143] J. R. Kirtley, C. C. Tsuei, H. Raffy, Z. Z. Li, A. Gupta, J. Z. Sun, S. Megert, Europhys. Lett. **36** (1996) 707

[144] L. N. Bulaevskii, V. V. Kuzii, A. A. Sobyanin, JETP Lett. **25** (1977) 290

[145] V. B. Geshkenbein, A. I. Larkin, A. Barone, Phys. Rev. B **36** (1987) 235

[146] V. G. Kogan, J. R. Clem, J. R. Kirtley, Phys. Rev. B **61** (2000) 9122

[147] J. R. Kirtley,C. C. Tsuei, K. A. Moler, Science **285** (1999) 1373

[148] B. Chesca , Ann. Phys. (Leipzig) **8** (1999) 511

[149] B. Chesca, Physica B **284–288** (2000) 2124

[150] R. R. Schulz, B. Chesca, B. Goetz, C. W. Schneider, A. Schmehl, H. Bielefeldt, H. Hilgenkamp, J. Mannhart, C. C. Tsuei, Appl. Phys. Lett. **76** (2000) 912

[151] B. Chesca, R. R. Schulz, B. Goetz, C. W. Schneider, H. Hilgenkamp, J. Mannhart, Phys. Rev. Lett. **88** (2002) 177003

[152] C. C. Tsuei, J. R. Kirtley, Phys. Rev. Lett. **85** (2000) 182

[153] B. Chesca, K. Ehrhardt, M. Mößle, R. Straub, D. Koelle, R. Kleiner, A. Tsukada Phys. Rev. Lett. **90** (2003) 057004

[154] T. Sato, T. Kamiyama, T. Takahashi, K. Kurahashi, K. Yamada, Science **291** (2001) 1517

[155] M.-S. Kim, J. A. Skinta, T. R. Lemberger, A. Tsukada, M. Naito, Phys. Rev. Lett. **91** (2003) 087001

[156] J. A. Skinta, T. R. Lemberger, T. Greibe, M. Naito, Phys. Rev. Lett. **88** (2002) 207003

[157] S. Kashiwaya, Y. Tanaka, M. Koyanagi, K. Kajimura, Phys. Rev. B **53** (1996) 2667

[158] S. Kleefisch, B. Welter, A. Marx, L. Alff, R. Gross, M. Naito, Phys. Rev. B **63** (2001) 100507

[159] I. Iguchi, W. Wang, M. Yamazaki, Y. Tanaka, S. Kashiwaya, Phys. Rev. B **62** (2000) R6131

[160] C.-R. Hu, Phys. Rev. Lett. **72** (1994) 1526

[161] L. Alff, A. Beck, R. Gross, A. Marx, S. Kleefisch, Th. Bauch, H. Sato, M. G. Koren, Phys. Rev. B **58** (1998) 11197

[162] R. Krupke, G. Deutscher, Phys. Rev. Lett. **83** (1999) 4634

[163] E. Maiser, W. Mexner, R. Schäfer, T. Schreiner, P. Adelmann, G. Czjzek, J. L. Peng, R. L. Greene, Phys. Rev. B **56** (1997) 12961

[164] P. Fulde, Ann. Phys. (Leipzig) **9** (2000) 871

[165] D. Dunlap, M. Slaski, Z. Sungaila, D. G. Hinks, K. Zhang, C. Segre, S. K. Malik, E. E. Alp, Phys. Rev. B **37** (1988) 592

[166] R. J. Ormeno, C. E. Gough, G. Yang, Phys. Rev. B **63** (2001) 104517

[167] J. C. Wynn, D. A. Bonn, B. W. Gardner, Yu-Ju Lin, R. Liang, W. N. Hardy, J. R. Kirtley, K. A. Moler, Phys. Rev. Lett. **87** (2001) 197002

[168] N. P. Armitage, D. H. Lu, D. L. Feng, C. Kim, A. Damascelli, K. M. Shen, F. Ronning, Z.-X. Shen, Y. Onose, Y. Taguchi, Y. Tokura, Phys. Rev. Lett. **86** (2001) 1126

[169] J. A. Skinta, M.-S. Kim, T. R. Lemberger, T. Greibe, M. Naito, Phys. Rev. Lett. **88** (2002) 207005

[170] Y. Dagan, M. M. Qazilbash, C. P. Hill, V. N. Kulkarni, R. L. Greene, Phys. Rev. Lett. **92** (2004) 167001

[171] R. C. Budhani, M. C. Sullivan, C. J. Lobb, R. L. Greene, Phys. Rev. B **66** (2002) 052506

[172] E. H. Brandt, cond-mat/0104518

[173] G. Grasso, R. Flükiger, Supercond. Sci. Technol. **10** (1997) 223

[174] J. Sato, K. Ohata, M. Okada, K. Tanaka, H. Kitaguchi, H. Kumakura, T. Kiyoshi, H. Wada, K. Togano, Physica C **357-360** (2001) 1111;
M. Okada, Supercond. Sci. Technol. **13** (2000) 29

[175] N. P. Plakida, *High-Temperature Superconductivity,* Springer-Verlag Berlin-Heidelberg (1995)

[176] E. H. Brandt, Rep. Prog. Phys. **58** (1995) 1465; condmat/9506003

[177] K. Tanaka, A. Iyo, Y. Tanaka, K. Tokiwa, M. Tokumoto, M. Ariyama, T. Tsukamoto, T. Watanabe, H. Ihara, Physica B **284-288** (2000) 1081;

T. Watanabe, S. Miyashita, N. Ichioka, K. Tokiwa, K. Tanaka, A. Iyo, Y. Tanaka, H. Ihara,Physica B **284-288** (2000) 1075

[178] B. Götz, "*Untersuchung der Transporteigenschaften von Korngrenzen in Hochtemperatur-Supraleitern*", PhD-thesis, University of Augsburg, 2000

[179] H. Hilgenkamp, J. Mannhart, Rev. Mod. Phys. **74** (2002) 485

[180] H. Hilgenkamp, J. Mannhart, B. Mayer, Phys. Rev. B **53** (1996) 14586

[181] P. A. Nilsson, Z. G. Ivanov, H. K. Olsson, D. Winkler, T. Claeson, E. A. Stepantsov, A. Ya. Tzalenchuk, J. Appl. Phys. **75** (1994) 7972

[182] J. Betouras, R. Joynt, Physica C **250** (1995) 256

[183] J. Alarco, E. Olsson, Phys. Rev. B **52** (1995) 13625

[184] D. Agassi, D. K. Christen, S. J. Pennycook, Appl. Phys. Lett. **81** (2002) 2803

[185] H. Hilgenkamp, J. Mannhart, Appl. Phys. Lett. **73** (1998) 265

[186] G. Hammerl, A. Schmehl, R. R. Schulz, B. Goetz, H. Bielefeldt, C. W. Schneider, H. Hilgenkamp, J. Mannhart, Nature **162** (2000) 162;

C. W. Schneider, R. R. Schulz, B. Goetz, A. Schmehl, H. Bielefeldt, H. Hilgenkamp, J. Mannhart, Appl. Phys. Lett. **75** (1999) 850;

H. Hilgenkamp, C. W. Schneider, R. R. Schulz, B. Goetz, A. Schmehl, H. Bielefeldt, J. Mannhart, Physica C **326-327** (1999) 7

[187] W. M. Temmermann, H. Winter, Z. Szotek, A. Svane, Phys. Rev. Lett. **86** (2001) 2435

[188] O. K. Andersen, A. I. Liechtenstein, O. Jepsen, F. Paulsen, J. Phys. Chem. Solids **56** (1995) 1567

[189] J. Zaanen, G. Sawatzky, J. W. Allen, Phys. Rev. Lett. **55** (1985) 418;

A. Fujimori, F. Minami, Phys. Rev. B **30** (1984) 957

[190] R. J. Cava, J. Am. Ceram. Soc. **83** (2000) 5

[191] E. Dagotto, Rev. Mod. Phys. **66** (1994) 763

[192] R. M. White and T. H. Geballe: *Long Range Order in Solids*, Academic Press, New York - San Francisco – London (1979)

[193] S. Sachdev, Rev. Mod. Phys. 75, 913-932 (2003)

[194] A. Paramekanti, M. Randeria, N. Trivedi, Phys. Rev. Lett. **87** (2001) 217002

[195] P. W. Anderson, P. A. Lee, M. Randeria, T. M. Rice, N. Trivedi, F. C. Zhang, J. Phys. Condens. Matter **16** (2004) R755

[196] C. M. Varma, cond-mat/0312385

[197] B. A. Bernevig, G. Chapline, R. B. Laughlin, Z. Nazario, D. I. Santiago, cond-mat/0312573

[198] P. Coleman, Nature **424** (2003) 625

[199] F. C. Zhang, T. M. Rice, Phys. Rev. B **37** (1988) 3759

[200] T. M. Rice, Physica B **241-243** (1997) 5

[201] Y. Onose, Y. Taguchi, K. Ishizaka, Y. Tokura, Phys. Rev. Lett. **87** (2001) 217001

[202] Y. Ando, A. N. Lavrov, S. Komiya, K. Segawa, X. F. Sun, Phys. Rev. Lett. **87** (2001) 017001

[203] A. N. Lavrov, S. Komiya, Y. Ando, Nature **418** (2002) 385

[204] M. A. Kastner, R. J. Birgenau, G. Shirane, Y. Endoh, Rev. Mod. Phys. **70** (1998) 897

[205] J. M. Tranquada, D. E. Cox, W. Kunnmann, H. Moudden, G. Shirane, M. Suenaga, P. Zolliker, D. Vaknin, S. K. Sinha, M. S. Alvarez, A. J. Jacobson, D. C. Johnston, Phys. Rev. Lett. **60** (1988) 156

[206] J. L. Tallon, J. W. Loram, Physica C **349** (2001) 53

[207] C. M. Varma, Phys. Rev. B **55** (1997) 14554

[208] L. Alff, Y. Krockenberger, B. Welter, M. Schonecke, R. Gross, D. Manske, M. Naito, Nature **422** (2003) 698

[209] B. W. Veal, A. P. Paulikas, Physica C **184** (1991) 321

[210] K. Segawa, Y. Ando, J. Low Temp. Phys. **131** (2003) 821

[211] U. Tutsch, P. Schweiss, H. Wühl, B. Obst, Th. Wolf, to be published

[212] J. W. Loram, K. A. Mirza, J. M. Wade, J. R. Cooper, W. Y. Liang, Physica C **235-240** (1994) 134

[213] T. Schneider, H. Keller, cond-mat/0404702

[214] T. Timusk, B. Statt, Rep. Prog. Phys. **62** (1999) 61

[215] M. Buchanan, Nature **409** (2001) 11

[216] C. Meingast, V. Pasler, P. Nagel, A. Rykov, S. Tajima, P. Olsson, Phys. Rev. Lett. **86** (2001) 1606

[217] B. Batlogg, Solid State Comm. **107** (1998) 639

[218] J. Ruvalds, Supercond. Sci. Technol. **9** (1996) 905

[219] H. Ding, M. R. Norman, J. C. Campuzano, M. Randeria, A. F. Bellman, T. Yokoya, T. Takahashi, T. Mochiku, K. Kadowaki, Phys. Rev. B **54** (1996) 9678

[220] P. V. Bogdanov, A. Lanzara, X. J. Zhou, S. A. Kellar, D. L. Feng, E. D. Lu, H. Eisaki, J.-I. Shimoyama, K. Kishio, Z. Hussain, Z. X. Shen, Phys. Rev. B **64** (2001) 180505

[221] T. Valla, A. V. Fedorov, P. D. Johnson, Q. Li, G. D. Gu, N. Koshizuka, Phys. Rev. Lett. **85** (2000) 828

[222] A. A. Kordyuk, S. V. Borisenko, M. S. Golden, S. Legner, K. A. Nenkov, M. Knupfer, J. Fink, H. Berger, L. Forro, Phys. Rev. B **66** (2002) 014502

[223] C. M. Varma, P. B. Littlewood, S. Schmitt–Rink, E. Abrahams, A. E. Ruckenstein, Phys. Rev. Lett. **63** (1989) 1996

[224] R. W. Hill, C. Proust, L. Taillefer, P. Fournier, R. L. Greene, Nature **414** (2001) 711

[225] R. Lortz, C. Meingast, A. I. Rykov, S. Tajima, Phys. Rev. Lett. **91** (2003) 207001

[226] N. P. Ong, Y. Wang, S. Ono, Y. Ando, S. Uchida, Annalen der Physik 13 (2004) 9

[227] C. Proust, E. Boaknin, R. W. Hill, L. Taillefer, A. P. Mackenzie, Phys. Rev. Lett. **89** (2002)147003

[228] S. Nakamae, K. Behnia, N. Mangkorntong, M. Nohara, H. Takagi, S. J. C. Yates, N. E. Hussey, Phys. Rev. B **68** (2003) 100502

[229] R. Bel, K. Behnia, C. Proust, P. van der Linden, D. Maude, S. I. Vedeneev, Phys. Rev. Lett. **92** (2004) 177003

[230] T. Tohyama, S. Maekawa, Supercond. Sci. Technol. **13** (2000) R17

[231] B. Keimer, Science **292** (2001) 1498

[232] G. Yang, J. S. Abell, C. E. Gough, Appl. Phys. Lett. **75** (1999) 1955

[233] R. Joynt, Science **284** (1999) 777

[234] S. V. Borisenko, M. S. Golden, S. Legner, T. Pichler, C. Dürr, M. Knupfer, J. Fink, G. Yang, S. Abell, H. Berger, Phys. Rev. Lett. **84** (2000) 4453

[235] D. L. Feng, D. H. Lu, K. M. Shen, C. Kim, H. Eisaki, A. Damascelli, R. Yoshizaki, J.-I. Shimoyama, K. Kishio, G. D. Gu, S. Oh, A. Andrus, J. O'Donnell, J. N. Eckstein, Z.-X. Shen, Science **289** (2000) 277

[236] A. Junod, A. Erb, C. Renner, Physica C **317-318** (1999) 333

[237] A. V. Fedorov, T. Valla, P. D. Johnson, Q. Li, G. D. Gu, N. Koshizuka, Phys. Rev. Lett. **82** (1999) 2179

[238] J. R. Schrieffer, M. Tinkham, Rev. Mod. Phys. **71** (1999) S313

[239] P. A. Lee, Physica C **317-318** (1999) 194

[240] Z. A. Xu, N. P. Ong, Y. Wang, T. Kakeshita, S. Uchida, Nature **406** (2000) 486

[241] P. A. Lee, Nature **406** (2000) 467

[242] I. Iguchi, T. Yamaguchi, A. Sugimoto, Nature **412** (2001) 420

[243] Y. Wang,. Z. A. Xu, T. Kakeshita, S. Uchida, S. Ono, Y. Ando, N. P. Ong, Phys. Rev. B **64** (2001) 224519

[244] V. Pasler, P. Schweiss, C. Meingast, B. Obst, H. Wühl, A. I. Rykov, S. Tajima, Phys. Rev. Lett. **81** (1998) 1094

[245] A. S. Alexandrov, N. F. Mott, Rep. Prog. Phys. **57** (1994) 1197

[246] V. J. Emery, S. A. Kivelson, Nature **374** (1995) 434

[247] M. Randeria, in *Bose-Einstein condensation* (Ed. A. Griffin, D. W. Snoke, S. Stringari), Cambridge University Rress 1995, pp.355

[248] J. Corson, R. Mallozzi, J. Orenstein, J. N. Eckstein, I. Bozovic, Nature **398** (1999) 221

[249] A. J. Millis, Nature **398** (1999) 193

[250] S. Chakravarty, R. B. Laughlin, D. K. Morr, C. Nayak, Phys. Rev. B **63** (2001) 094503

[251] I. Affleck, J. B. Marston, Phys. Rev. B **37** (1988) 3774

[252] J. B. Marston, I. Affleck, Phys. Rev. B **39** (1989) 11538

[253] A. G. Loesser, Z.-X. Shen, D. S. Dessau, D. S. Marshall, C. H. Park, P. Fournier, A. Kapitulnik, Science **273** (1996) 325

[254] H. Ding, T. Yokoya, J. C. Campuzano, T. Takahashi, M. Randeria, M. R. Norman, T. Mochiku, K. Kadowaki, J. Giapintzakis, Nature **382** (1996) 51

[255] C. Nayak, Phys. Rev. B **62** (2000) 4880

[256] G. Aeppli, D. J. Bishop, C. Broholm, E. Bucher, S.-W. Cheong, P. Dai, Z. Fisk, S. M. Hayden, R. Kleiman, T. E. Mason, H. A. Mook, T. G. Perring, A. Schroeder, Physica C **317-318** (1999) 9

[257] P. Bourges, *The Gap Symmetry and Fluctuations in High Temperature Superconductors* (Eds. J. Bok, G. Deutscher, D. Pavuna, S. A. Wolf), Plenum Press (1998); Proc. NATO ASI summer school, September 1-13, 1997 in Cargese, France; cond-mat/9901333

[258] M. Arai, T. Nishijima, Y. Endoh, T. Egami, S. Tajima, K. Tomimoto, Y. Shiohara, M. Takahashi,A. Garrett, S. M. Bennington, Phys. Rev. Lett. **83** (1999) 608

[259] K. Yamada, C. H. Lee, K. Kurahashi, J. Wada, S. Wakimoto, S. Ueki, H. Kimura, Y. Endoh, S. Hosoya, G. Shirane, R. J. Birgeneau, M. Greven, M. A. Kastner, Y. J. Kim, Phys. Rev. B **57** (1998) 6165

[260] J. M. Tranquada, J. D. Axe, N. Ichikawa, A. R. Moodenbaugh, Y. Nakamura, S. Uchida, Phys. Rev. Lett. **78** (1997) 338

[261] J. M. Tranquada, Physica B **241-243** (1998) 745

[262] J. M. Tranquada, J. D. Axe, N. Ichikawa, Y. Nakamura, S. Uchida, B. Nachumi, Phys. Rev. B **54** (1996) 7489

[263] J. M. Tranquada, H. Woo, T. G. Perring, H. Goka, G. D. Gu, G. Xu, M. Fujita, K. Yamada, Nature **429** (2004) 534

[264] S. M. Hayden, H. A. Mook, Pengcheng Dai, T. G. Perring, F. Dogan, Nature **429** (2004) 531

[265] G. Grüner, Rev. Mod. Phys. **66** (1994) 1

[266] K. Machida, Physica C **158** (1989) 192

[267] J. Zaanen, O. Gunnarsson, Phys. Rev. **B** 40 (1989) 7391

[268] J. Zaanen, Nature **404** (2000) 714

[269] H. A. Mook, F. Dogan, Physica C **364-365** (2001) 553

[270] H. A. Mook, F. Dogan, Nature **401** (1999) 145

[271] L. Pintschovius, W. Reichardt, M. Kläser, T. Wolf, H. v. Löhneysen, Phys. Rev. Lett. **89** (2000) 037001

[272] M. I. Salkola, V. J. Emery, S. A. Kivelson, Phys. Rev. Lett. **77** (1996) 156

[273] S. A. Kivelson, E. Fradkin, V. J. Emery, Nature **393** (1998) 550

[274] H.-H. Klauss, W. Wagener, M. Hillberg, W. Kopmann, H. Walf, F. J. Litterst, M. Hücker, B. Büchner, Phys. Rev. Lett. **85** (2000) 4590

[275] T. Tohyama, C. Gazza, C. T. Shih, Y. C. Chen, T. K. Lee, S. Maekawa, E. Dagotto, Phys. Rev. B **59** (1999) R11649

[276] M. G. Zacher, R. Eder, E. Arrigoni, W. Hanke, Phys. Rev. Lett. **85** (2000) 2585

[277] G. B. Martins, C. Gazza, J. C. Xavier, A. Feiguin, E. Dagotto, Phys. Rev. Lett. **84** (2000) 5844

[278] A. L. Chernyshev, A. H. Castro Neto, A. R. Bishop, Phys. Rev. Lett. **84** (2000) 4922

[279] H. A. Mook, P. Dai, F. Dogan, R. D. Hunt, Nature **404** (2000) 729

[280] V. Hinkov, S. Pailhès, P. Bourges, Y. Sidis, A. Ivanov, A. Kulakov, C. T. Lin, D. P. Chen, C. Bernhard, B. Keimer, Nature **430** (2004) 650

[281] N. B. Christensen, D. F. McMorrow, H. M. Ronnow, B. Lake, S. M. Hayden, G. Aeppli, T. G. Perring, M. Mangkorntong, M. Nohara, H. Tagaki, cond-mat/0403439

[282] Y. S. Lee, R. J. Birgeneau, M. A. Kastner, Y. Endoh, S. Wakimoto, K. Yamada, R. W. Erwin, S.-H. Lee, G. Shirane, Phys. Rev. B **60** (1999) 3643

[283] M. Fujita, K. Yamada, H. Hiraka, P. M. Gehring, S. H. Lee, S. Wakimoto, G. Shirane, Phys. Rev. B **65** (2002) 064505

[284] C. Howald, H. Eisaki, N. Kaneko, M. Greven, A. Kapitulnik, Phys. Rev. B **67** (2003) 014533;

A. Fang, C. Howald, N. Kaneko, M. Greven, A. Kapitulnik, cond-mat/0404452

[285] S. A. Kivelson, I. P. Bindloss, E. Fradkin, V. Oganesyan, J. M. Tranquada, A. Kapitulnik, C. Howald, Rev. Mod. Phys. **75** (2003) 1201

[286] J. E. Hoffman, K. McElroy, D.-H. Lee, K. M Lang, H. Eisaki, S. Uchida, J. C. Davis, Science **297** (2002) 1148

[287] K. McElroy, R. W. Simmonds, J. E. Hoffman, D.-H. Lee, J. Orenstein, H. Eisaki, S. Uchida, J. C. Davis, Nature **422** (2003) 592

[288] M. Vershinin, S. Misra, S. Ono,Y. Abe, Y. Ando, A. Yazdani, Science **303** (2004) 1995;

K. McElroy, D.-H. Lee, J. E. Hoffman, K. M Lang, J. Lee, E. W. Hudson, H. Eisaki, S. Uchida, J. C. Davis, cond-mat/0406491

[289] M. Norman, Science **303** (2004) 1985

[290] L. P. Pryadko, S. A. Kivelson, V. J. Emery, Y. B. Bazaliy, E. A. Demler, Phys. Rev. B **60** (1999) 7541

[291] W. Prusseit, S. Furtner, R. Nemetschek, Supercond. Sci. Technol. **13** (2000) 519

[292] V. Matijasevic, Z. Lu, T. Kaplan, C. Huang, Inst. Phys. Conf. Ser. No. **158** (1997) 189

[293] D. W. Face, R. J. Small, M. S. Warrington, F. M. Pellicone, P. J. Martin, Physica C **357-360** (2001) 1488;

D. W. Face, C. Wilker, J. J. Kingston, Z.-Y. Shen, F. M. Pellicone, R. J. Small, S. P. McKenna, S. Sun, P. J. Martin, IEEE Trans. Appl. Supercond. **7** (1997) 1283

[294] R. Aidam, J. Geerk, G. Linker, F. Ratzel, J. Reiner, R. Schneider, R. Smithey, A. G. Zaitsev, E. Gaganidze, R. Schwab, IEEE Trans. Appl. Supercond. **11** (2001) 357

[295] Y. Yamada, J. Kawashima, J. G. Wen, Y. Niiori, I. Hirabayashi, Jpn. J. Appl. Phys. 39 (2000) 1111

[296] P. Seidel, S. Linzen G. Kaiser, F. Schmidl, Y. Tian, A. Matthes, S. Wunderlich, H. Schneidewind, Proc. SPIE **3481** (1998) 366

[297] A. G. Zaitsev, G. Ockenfuss, R. Wördenweber, Inst. Phys. Conf. Ser. No. **158** (1997) 25

[298] D. H. A. Blank, H. Rogalla, J. Mater. Res. **12** (1997) 2952;
K. Verbist, O. I. Lebedev, M. A. J. Verhoeven, R. Wichern, A. J. H. M. Rijnders, D. H. A. Blank, F. Tafuri, H. Bender, G. Van Tendeloo, Supercond. Sci. Technol. **11** (1998) 13;
K. Verbist, O. I. Lebedev, G. Van Tendeloo, M. A. J. Verhoeven, A. J. H. M. Rijnders, D. H. A. Blank, H. Rogalla, Appl. Phys. Lett. **70** (1997) 1167

[299] J. Talvacchio, M. G. Forrester, B. D. Hunt, J. D. McCambridge, R. M. Young, X. F. Zhang, D. J. Miller, IEEE Trans. Appl. Supercond. **7** (1997) 2051

[300] Y. Soutome, T. Fukazawa, A. Tsukamoto, K. Saitoh, K. Takagi, Physica C **372-376** (2002) 143

[301] T. Satoh, M. Hidaka, S. Tahara, J. G. Wen, N. Koshizuka, S. Tanaka, IEEE Trans. Appl. Supercond. **11** (2001) 770

[302] J. Mannhart, P. Chaudhari, Physics Today, November 2001, p.48

[303] F. Herbstritt, T. Kemen, A. Marx, R. Gross, J. Appl. Phys. **91** (2002) 5411

[304] Z. G. Ivanov, E. A. Stepantsov, T. Claeson, F. Wenger, S. Y. Lin, N. Khare, P. Chaudhari, Phys. Rev. B **57** (1998) 602

[305] T. Minotani, S. Kawakami, T. Kiss, Y. Kuroki, K. Enpuku, Jpn. J. Appl. Phys. **36** (1997) L 1092

[306] H. Q. Li, R. H. Ono, L. R. Vale, D. A. Rudman, S. H. Liou, Appl. Phys. Lett. **71** (1997) 1121

[307] F. Schmidl, S. Linzen, S. Wunderlich, P. Seidel, Appl. Phys. Lett. **72** (1998) 602

[308] H. J. H. Smilde, H. Hilgenkamp, G. Rijnders, H. Rogalla, D. H. A. Blank, Appl. Phys. Lett. **80** (2002) 4579;
H. J. H. Smilde, H. Hilgenkamp, G. J. Gerritsma, D. H. A. Blank, H. Rogalla, Physica C **350** (2001) 269;
H. J. H. Smilde, Ariando, D. H. A. Blank, G. J. Gerritsma, H. Hilgenkamp, H. Rogalla, Phys. Rev. Lett. **88** (2002) 057004

[309] R. Kleiner, F. Steinmeyer, G. Kunkel, P. Müller, Phys. Rev. Lett. **68** (1992) 2394; R. Kleiner, P. Müller, Phys. Rev. B **49** (1994) 1327

[310] D. Winkler, N. Mros, E. J. Tarte, A. Yurgens, V. M. Krasnov, D. T. Foord, W. E. Booij, M. G. Blamire, Supercond. Sci. Technol. **12** (1999) 1013;
for a review, see A. Yurgens, Supercond. Sci. Technol. **13** (2000) R85

[311] H. B. Wang, L. X. You, J. Chen, P. H. Wu, T. Yamashita, Supercond. Sci. Technol. **15** (2002) 90;
H. B. Wang, J. Chen J, P. H. Wu, T. Yamashita, D. Vasyukov, P. Müller, Supercond. Sci. Technol. **16** (2003) 1375

[312] T. Clauss, V. Oehmichen, M. Mößle, A. Müller, A. Weber, D. Koelle, R. Kleiner, Supercond. Sci. Technol. **15** (2002) 1651

[313] K. Schlenga, R. Kleiner, G. Hechtfischer, M. Mößle, S. Schmitt, P. Müller, Ch. Helm, Ch. Preis, F. Forsthofer, J. Keller, H. L. Johnson, M. Veith, E. Steinbeiß, Phys. Rev. B **57** (1998) 14518

[314] M. Rapp, A. Murk, R. Semerad, W. Prusseit, Phys. Rev. Lett. **77** 1996) 928

[315] D. Litzkendorf, T. Habisreuther, R. Müller, S. Kracunovska, O. Surzhenko, M. Zeisberger, J. Riches, W. Gawalek, Physica C **372-376** (2002) 1163

[316] G. Desgardin, I. Monot. B. Raveau, Supercond. Sci. Technol. **12** (1999) R115

[317] M. Tomita, M. Murakami, Nature **421** (2003) 517

[318] M. Tsuchimoto, H. Takashima, M. Tomita, M. Murakami, Physica C **378-381** (2002) 718

[319] H. Teshima, M. Sawamura, M. Morito, M. Tsuchimoto, Cryogenics **37** (1997) 505

[320] P. Stoye, W. Gawalek, P. Görnert, A. Gladun, G. Fuchs, Inst. Phys. Conf. Ser. No. **158** (1997) 1535

[321] W. Hennig, D. Parks, R. Weinstein, R.-P. Sawh, Appl. Phys. Lett. **72** (1998) 3059

[322] K. Tanaka, M. Okada, K. Ohata, J. Sato, H. Kitaguchi, H. Kumakura, K. Togano, Physica C **357-360** (2001) 1102;

K. Ohata, J. Sato, M. Okada, K. Tanaka, H. Kitaguchi, H. Kumakura, K. Togano, T. Kiyoshi, H. Wada, Physica C **357-360** (2001) 1107

[323] L. Masur, D. Parker, M. Tanner, E. Podtburg, D. Buczek, J. Scudiere, P. Caracino, S. Spreafico, P. Corsaro, M. Nassi, IEEE Trans. Appl. Supercond. **11** (2001) 3256

[324] J. Kellers, L. J. Masur, Physica C **372-376** (2002) 1040

[325] T. J. Arndt, B. Fischer, J. Gierl, H. Krauth, M. Munz, A. Szulczyk, M. Leghissa, H.-W. Neumueller, IEEE Trans. Appl. Supercond. **11** (2001) 3261

[326] J. Fujikami, T. Kaneko, N. Ayai, S. Kobayashi, K. Hayashi, H. Takei, K. Sato, Physica C **378-381** (2002) 1061

[327] Q. Li, M. Suenaga, T. Kaneko, K. Sato, Ch. Simmon, Appl. Phys. Lett. **71** (1997) 1561

[328] P. Corsaro, M. Bechis, P. Caracino, W. Castiglioni, G. Cavalleri, G. Coletta, G. Colombo, P. Ladiè, A. Mansoldo, R. Mele, S. Montagnero, C. Moro, M. Nassi, S. Spreafico, N Kelley, C. Wakefield, Physica C **378-381** (2002) 1168

[329] T. Masuda, T. Kato, H. Yumura, M. Watanabe, Y. Ashibe, K. Ohkura, C. Suzawa, M. Hirose, S. Isojima, K. Matsuo, S. Honjo, T. Mimura, T. Kuramochi, Y. Takahashi, H. Suzuki, T. Okamoto, Physica C **378-381** (2002) 1174

[330] H. Kitaguchi, K. Itoh, T. Takeuchi, H. Kumakura, H. Miao, H. Wada, K. Togano, T. Hasegawa, T. Kizumi, Physica C **320** (1999) 253;

T. Hasegawa, T. Koizumi, N. Ohtani, H. Kitaguchi, H. Kumakura, K . Togano, H. Miao, Supercond. Sci. Technol. **13** (2000) 23

[331] Y. Aoki, T. Koizumi, N. Ohtani, T. Hasegawa, L. Motowidlo, R. S. Sokolowski, R. M. Scanlan, S. Nagaya, Physica C **335** (2000) 1

[332] T. Wakuda, M. Okada, S. Awaji, K. Watanabe, Physica C **357-360** (2001) 1293

[333] T. Hase, Y. Murakami, S. Hayashi, Y. Kawate, T. Kiyoshi, H. Wada, S. Sairote, R. Ogawa, Physica C **335** (2000) 6;

K. Fukushima, K. Tanaka, T. Wakuda, M. Okada, K. Ohata, J. Sato, T. Kiyoshi, H. Wada, Physica C **357-360** (2001) 1297

[334] J. Cloyd, "2002 DOE Wire Workshop", January 22, 2002, St. Petersburg, FL, http://www.eren.doe.gov/superconductivity/docs/10mil_perspective_cloyd.ppt

[335] M. Gouge, "2002 DOE Wire Workshop", January 22, 2002, St. Petersburg, FL, http://www.eren.doe.gov/superconductivity/docs/09dod_perspective_gouge.ppt

[336] M. Wood, *Trials and Triumphs in Superconductivity: The Making of Oxford Instruments*, IEEE Trans. Appl. Supercond. **5** (1995) 152

[337] A. W. Sleight, J. L. Gillson, P. E. Bierstedt, Solid State Commun. **17** (1975) 27

[338] M. Merz, N. Nücker, S. Schuppler, D. Arena, J. Dvorak, Y. U. Idzerda, S. N. Ustinovich, A. G. Soldatov, S. V. Shiryaev, S. N. Barilo, to be published

[339] L. F. Mattheiss, E. M. Gyorgy, D. W. Johnson Jr., Phys. Rev. B **37** (1988) 3745

[340] E. S. Hellman, E. H. Hartford, Jr., Phys. Rev. B **52** (1995) 6822

[341] R. J. Cava, B. Batlogg, J. J. Krajewski, R. Farrow, L. W. Rupp, Jr., A. E. White, K. Short, W. F. Peck, T. Kometani, Nature **332** (1988) 814

[342] S. Pei, J. D. Jorgensen, B. Dabrowski, D. G. Hinks, D. R. Richards, A. W. Mitchell, J. M. Newsam, S. K. Sinha, D. Vaknin, A. J. Jacobson, Phys. Rev. B **41** (1990) 4126

[343] L. A. Klinkova, M. Uchida, Y. Matsui, V. I. Nikolaichik, N. V. Barkovskii, Phys. Rev. B **67** (2003) 140501

[344] Y. Maeno, T. M. Rice, M. Sigrist, Physics Today, January 2001, p. 42

[345] C. Bernhard, J. L. Tallon, Ch. Niedermayer, Th. Blasius, A. Golnik, E. Brücher, R. K. Kremer, D. R. Noakes, C. E. Stronack, E. J. Asnaldo, Phys. Rev. B **59** (1999) 14099

[346] T. Nachtrab, D. Koelle, R. Kleiner, C. Bernhard, C. T. Lin, Phys. Rev. Lett. **92** (2004)117001

[347] B. Lorenz, Y. Y. Xue, C. W. Chu, this volume

[348] K. Takada, H. Sakurai, E. Takayama-Muromachi, F. Izumi, R. A. Dilanian, T. Sasaki, Nature **422** (2003) 53

[349] M. L. Foo, R. E. Schaak, V. L. Miller, T. Klimczuk, N. S. Rogado, Y. Wang, G. C. Lau, C. Craley, H. W. Zandbergen, N. P. Ong, R. J. Cava, Solid State Commun. **127** (2003) 33

[350] J. W. Lynn, Q. Huang, C. M. Brown, V. L. Miller, M. L. Foo, R. E. Schaak, C. Y. Jones, E. A. Mackey, R. J. Cava, Phys. Rev. B **68** (2003) 214516

[351] S. Yonezawa, Z. Muraoka, Y. Matsushita, Z. Hiroi, J. Phys.: Condensed Matter **16** (2004) L9;
S. Yonezawa, Z. Muraoka, Z. Hiroi, cond-mat/0404220;
S. Yonezawa, Y. Muraoka, Y. Matsushita, Z. Hiroi, J. Phys. Soc. Jpn. **73** (2004) 819;
Z. Hiroi, S. Yonezawa, Y. Muraoka, cond-mat/0402006

[352] N. P. Ong, R. J. Cava, Science **305** (2004) 52

[353] L. Taillefer, G. G. Lonzarich, Phys. Rev. Lett. **60** (1988) 1570;
H. Aoki, S. Uji, A. K. Albessard, Y. Onuki, Phys. Rev. Lett. **71** (1993) 2110;
F. S. Tautz, S. R. Julian, G. J. McMullen, G. G. Lonzarich, Physica B **206-207** (1995) 29

[354] R. A. Fisher, S. Kim, B. F. Woodfield, N. E. Phillips, L. Taillefer, K. Hasselbach, J. Flouquet, A. L. Giorgi, J. L. Smith, Phys. Rev. Lett. **62** (1989) 1411

[355] E. D. Bauer, N. A. Frederick, P.-C. Ho, V. S. Zapf, M. B. Maple, Phys. Rev. B **65** (2002) 100506

[356] T. Watanabe, Y. Kasahara, K. Izawa, T. Sakakibara, C. vanderBeek, T. Hanaguri, H. Shishido, R. Settai, Y. Onuki, Y. Matsuda, Phys. Rev. B **70** (2004) 020506;
For a recent review on the LOFF / FFLO state see R. Casalbuoni, G. Nardulli, Rev. Mod. Phys. **76** (2004) 263

[357] A. Bianchi, R. Movshovich, C. Capan, A. Lacerda, P. G. Pagliuso, J. L. Sarrao, Phys. Rev. Lett. **91** (2003) 187004

[358] J. D. Denlinger, G.-H. Gweon, J. W. Allen, C. G. Olson, M. B. Maple, J. L. Sarrao, P. E. Armstrong, Z. Fisk, H. Yamagami, J. Electron. Spectrosc. Relat. Phenom**. 117&118** (2001) 347;
F. Reinert, D. Ehm, S. Schmidt, G. Nicolay, S. Hüfner, J. Kroha, O. Trovarelli, C. Geibel, Phys. Rev. Lett. **87** (2001) 106401

[359] G. Zwicknagl, Adv. Phys. **41** (1992) 203

[360] O. Stockert, E. Faulhaber, G. Zwicknagl, N. Stüßer, H. S. Jeevan, M. Deppe, R. Borth, R. Küchler, M. Loewenhaupt, C. Geibel, F. Steglich, Phys. Rev. Lett. **92** (2004) 136401

[361] G. Zwicknagl, A. N. Yaresko, P. Fulde, Phys. Rev. B **63** (2002) 081103
G. Zwicknagl, A. N. Yaresko, P. Fulde, Phys. Rev. B **68** (2003) 052508

[362] For a recent review see
P. Thalmeier, G. Zwicknagl, *Handbook on the Physics and Chemistry of Rare Earths*, Vol. 36, chapter "*Unconventional Superconductivity and Magnetism in Lanthanide and Actinide Intermetallic Compounds*"; cond-mat/0312540

[363] N. K. Sato, N. Aso, K. Miyake, R. Shiina, P. Thalmeier, G. Varelogiannis, C. Geibel, F. Steglich, P. Fulde, T. Komatsubara, Nature **410** (2001) 340

[364] P. McHale, P. Thalmeier, P. Fulde, cond-mat/0401520

[365] P. Thalmeier, G. Zwicknagl, G. Sparn, F. Steglich, "*Superconductivity in Heavy Fermion Compounds*", this volume

[366] J. D. Thompson, R. Movshovich, Z. Fisk, F. Bouquet, N. J. Curro, R. A. Fisher, P. C. Hammel, H. Hegger, M. F. Hundley, M. Jaime, P. G. Pagliuso, C. Petrovic, N. E. Phillips, J. L. Sarrao, J. Magn. Magn. Mat. **226-230** (2001) 5

[367] C. Petrovic, R. Movshovich, M. Jaime, P. G. Pagliuso, M. F. Hundley, J. L. Sarrao, Z. Fisk, J. D. Thompson, Europhys. Lett. **53** (2001) 354

[368] C. Petrovic, P. G. Pagliuso, M. F. Hundley, R. Movshovich, J. D. Sarrao, J. D. Thompson, Z. Fisk, P. Monthoux, J. Phys.: Condens. Matter **13** (2001) L337

[369] D. Jaccard, K. Behnia, J. Sierro, Phys. Lett. A **163** (1992) 475

[370] F. M. Grosche, S. R. Julian, N. D. Mathur, G. G. Lonzarich, Physica B **223&224** (1996) 50

[371] N. D. Mathur, F. M. Grosche, S. R. Julian, I. R. Walker, D. M. Freye, R. K. W. Haselwimmer, G. G. Lonzarich, Nature **394** (1998) 39

[372] F. M. Grosche, P. Agarwal S. R. Julian, N. J. Wilson, R. K. W. Haselwimmer, S. J. S. Lister, N. D. Mathur, F. V. Carter, S. S. Saxena, G. G. Lonzarich, J. Phys. Condens. Matter. **12** (2000) L533

[373] R. Movshovich, T. Graf, D. Mandrus, J. D. Thompson, J. L. Smith, Z. Fisk, Phys. Rev. B **53** (1996) 8241

[374] I. R. Walker, F. M. Grosche, D. M. Freye, G. G. Lonzarich, Physica C **282** (1997) 303

[375] G. R. Stewart, Z. Fisk, J. O. Willis, J. L. Smith, Phys. Rev. Lett. **52** (1984) 679

[376] C. Geibel, C. Schank, S. Thies, H. Kitazawa, C. D. Bredl, A. Böhm, M. Rau, A. Grauel,R. Caspary, R. Helfrich, U. Ahlheim, G. Weber, F. Steglich, Z. Phys. B **84** (1991) 1

[377] C. Geibel, S. Thies, D. Kczorowski, A. Mehner, A. Grauel, B. Seidel, U. Ahlheim, R. Helfrich, K. Petersen, C. D. Bredl, F. Steglich, Z. Phys. **83** (1991) 305

[378] T. T. M. Palstra, A. A. Menovsky, J. van den Berg, A. J. Dirkmaat, P. H. Res, G. J. Nieuwenhuys, J. A. Mydosh, Phys. Rev. Lett. **55** (1985) 2727

[379] H. R. Ott, H. Rudigier, Z. Fisk, J. L. Smith, Phys. Rev. Lett. **80** (1983) 1595

[380] S. S. Saxena, P. Agarwal, K. Ahllan, F. M. Grosche, R. W. K. Haselwimmer, M. J. Steiner, E. Pugh, I. R. Walker, S. R. Julian, P. Monthoux, G. G. Lonzarich, A. Huxley, I. Shelkin, D. Braithwaite, J Flouquet, Nature **406** (2000) 587

[381] R. Joynt, L. Taillefer, Rev. Mod. Phys. **74** (2002) 237

[382] A. Grauel, A. Böhm, H. Fischer, C. Geibel, R. Köhler, R. Modler, C. Schank, F. Steglich, G. Weber, Phys. Rev. B **46** (1992) 5818

[383] T. E. Mason, G. Aeppli, Matematisk-fysiske Meddelelser 45 (1997) 231

[384] M. Jourdan, M. Huth, H. Adrian, Nature **398** (1999) 47

[385] V. Ginzburg, JETP **4** (1957) 153

[386] P. C. Canfield, S. L. Bud'ko, Physica C **262** (1996) 249

[387] D. Fay, J. Appel, Phys. Rev. B **22** (1980) 3173

[388] D. Aoki, A. Huxley, E. Ressouche, D. Braithwaite, J. Flouquet, J.-P. Brison, E. Lhotel, Nature **413** (2001) 613

[389] C. Pfleiderer, M. Uhlarz, S. M. Hayden, R. Vollmer, H. V. Löhneysen, N. R. Bernhoeft, G. G. Lonzarich, Nature **412** (2001) 58

[390] W. Schlabitz, J. Bauman, B. Politt, U. Rauchschwalbe, H. M. Mayer, U. Ahlheim, C. D. Bredl, Z. Phys. B **52** (1986) 171

[391] A. de Visser, F. E. Kayzel, A. A. Menovsky, J. J. M. Franse, J. van der Berg, G. J. Nieuwenhuys, Phys. Rev. B **34** (1986) 8168

[392] C. Broholm, J. K. Kjems, W. J. L. Buyers, P. Matthews, T. T. M. Palstra, A. A. Menovsky, J. A. Mydosh, Phys. Rev. Lett. **58** (1987) 1467

[393] M. B. Walker, W. J. L. Buyers, Z. Tun, W. Que, A. A. Menovsky, J. D. Garrett, Phys. Rev. Lett. **71** (1993) 2630

[394] H. Amitsuka, M. Sato, N. Metoki, M. Yokohama, K. Kuwahara, T. Sakakibara, H. Morimoto, S. Kawarazaki, Y. Miyako, J. A. Mydosh, Phys. Rev. Lett. **83** (1999) 5114

[395] H. Amitsuka, M. Yokoyama, S. Miyazaki, K. Tenya, T. Sakakibara, W. Higemoto, K. Nagamine, K. Matsuda, Y. Kohori, T. Kohara, Physica B **312-313** (2002) 390

[396] H. R. Ott, H. Rudigier, T. M. Rice, K. Ueda, Z. Fisk, J. L. Smith, Phys. Rev. Lett. **82** (1984) 1915

[397] K. Izawa, Y. Nakajima, J. Goryo, Y. Matsuda, S. Osaki, H. Sugawara, H. Sato, P. Thalmeier, K. Maki, Phys. Rev. Lett. **90** (2003) 117001

[398] B. C. Sales, D. Mandrus, R. K. Williams, Science **272** (1996) 1325

[399] B. C. Sales, *Handbook on the Physics and Chemistry of the Rare Earths*, Vol. 33, Elsevier (2003) chapter 211, p. 1

[400] N. A. Frederick, T. D. Do, P.-C. Ho, N. P. Butch, V. S. Zapf, M. B. Maple, Phys. Rev. B **69** (2004) 024523

[401] R. Vollmer, A. Faißt, C. Pfleiderer, H. v. Löhneysen, E. D. Bauer, P.-C. Ho, V. Zapf, M. B. Maple, Phys. Rev. Lett. **90** (2003) 5700

[402] N. Oeschler, P. Gegenwart, F. Steglich, N. A. Frederick, E. D. Bauer, M. B. Maple, Acta Phys. Pol. B **34** (2003) 959

[403] A. Hebard, M. J. Rosseinsky, R. C. Haddon, D. W. Murphy, S. H. Glarum, T. T. M. Palstra, A. P. Ramirez, A. R. Kortam, Nature **350** (1991) 600;

R. M. Fleming, A. P. Ramirez, M. J. Rosseinsky, D. W. Murphy, R. C. Haddon, S. M. Zahurak, A. V. Markhija, Nature **352** (1991) 787

[404] R. J. Cava, H. Takagi, H. W. Zandbergen, J. J. Krajewski, W. F. Peck, Jr., T. Siegrist, B. Batlogg, R. B. van Dover, R. J. Felder, K. Mizuhashi, J. O. Lee, H. Eisaki, S. Uchida, Nature **367** (1994) 252;

[405] J. Cava, H. Takagi, B. Batlogg, H. W. Zandbergen, J. J. Krajewski, W. F. Peck Jr., R. B. van Dover, R. J. Felder, K. Mizuhashi, J. O. Lee, H. Eisaki, S. A. Carter, S. Uchida, Nature **367** (1994) 146

[406] Z. K. Tang, L. Zhang, N. Wang, X. X. Zhang, G. W. Wen, G. D. Li, J. N. Wang, C. T. Chan, P. Sheng, Science **292** (2001) 2462

[407] G. Amano, S. Akutagawa, T. Muranaka, Y. Zenitani, J. Akimitsu, J. Phys. Soc. Jpn. **73** (2004) 530

[408] T. Nakane, T. Mochiku, H. Kito, J. Itoh, M. Nagao, H. Kumakura, Y. Takano, Appl. Phys. Lett. **84** (2004) 2859

[409] K. Bechgaard, K. Carneiro, M. Olsen, R B. Rasmussen, C. B. Jacobsen, Phys. Rev. Lett. **46** (1981) 852

[410] P. A. Mansky, G. Danner, P. M. Chaikin, Phys. Rev. B **52** (1995) 7554

[411] G. Saito, H. Yamochi, T. Nakamura, T. Komatsu, M. Nakashima, H. Mori, K. Oshima, Physica B **169** (1991) 372

[412] I. J. Lee, S. E. Brown, W. G. Clark, M. J. Strouse, M. J. Naughton, W. Kang, P. M. Chaikin, Phys. Rev. Lett. **88** (2002) 017004

[413] C. H. Pennington, V. A. Stenger, Rev. Mod. Phys. **68** (1996) 855

[414] V. Buntar, H. W. Weber, Supercond. Sci. Technol. **9** (1996), 599

[415] R. Nagarajan, C. Mazumdar, Z. Hossain, S. K. Dhar, K. V. Gopalakrishnan, L. C. Gupta, C. Godart, B. D. Padalia, R. Vijayaraghavan, Phys. Rev. Lett. **72** (1994) 274

[416] G. Hilscher, H. Michor, *Studies in High Temperature Superconductors*, Vol. 28, Nova Science Publishers (1999) pp. 241–286

[417] K.-H. Müller, V. N. Narozhnyi, Rep. Prog. Phys. **64** (2001) 943

[418] O. Fischer, *Magnetic Superconductors*, Vol. 5, North Holland, Amsterdam (1990)

[419] *Topics in Current Physics* (Eds. O. Fischer & M. B. Maple), Vol. 32 and 34 Springer Verlag, Berlin (1982)

[420] K. Mizuno, T. Saito, H. Fudo, K. Koyama, K. Endo, H. Deguchi, Physica B **259-261** (1999) 594

[421] E. Boaknin, R. W. Hill, C. Proust, C. Lupien, L. Taillefer, P. C. Canfield, Phys. Rev. Lett. **87** (2001) 237001

[422] K. Izawa, K. Kamata, Y. Nakajima, Y. Matsuda, T. Watanabe, M. Nohara, H. Takagi, P. Thalmeier, K. Maki, Phys. Rev. Lett. **89** (2002) 137006

[423] M. Nohara, M., Isshiki, H., Takagi, R. Cava, J. Phys. Soc. Jpn. **66** (1997) 1888

[424] K. Maki, P. Thalmeier, H. Won, Phys. Rev. B **65** (2002) 140502

[425] T. Watanabe, M. Nohara, T. Hanaguri, H. Takagi, 2003, preprint

[426] S. -L. Drechsler, H. Rosner, S. V. Shulga, I. Opahle, H. Eschrig, J. Freudenberger, G. Fuchs, K. Nenkov, K. -H. Müller, H. Bitterlich, W. Löser, G. Behr, D. Lipp, A. Gladun, Physica C **364-365** (2001) 31

[427] H. Suderow, P. Martinez-Samper, N. Luchier, J. P. Brison, S. Vieira, P. C. Canfield, Phys. Rev. B. **64** (2003) 020503

[428] S. L. Bud'ko, P. C. Canfield, V. G. Kogan, Physica C **382** (2002) 85

[429] P. Grant, Nature **411** (2001) 532

[430] A. M. Campbell, Science **292** (2001) 65
[431] R. F. Service, Science **291** (2001) 1476
[432] R. F. Service, Science **291** (2001) 2295
[433] S. Tomlin, Nature **410** (2001) 407
[434] J. Kortus, I. I. Mazin, K. D. Belashchenko, V. P. Antropov, L. L. Boyer, Phys. Rev. Lett. **86** (2001) 4656
[435] K.-P. Bohnen, R. Heid, B. Renker, Phys. Rev. Lett. **86** (2001) 5771
[436] H. J. Choi, D. Roundy, H. Sun, M. L. Cohen, S. G. Louie, Nature **418** (2002) 758
[437] P. C. Canfield, G. Crabtree, Phys. Today **56** (March 2003) 34
[438] Y. Kong, O. V. Dolgov, O. Jepsen, O. K. Andersen, Phys. Rev. B **64** (2001) 020501
[439] G. Binnig, A. Baratoff, H. E. Hoenig, J. G. Bednorz, Phys. Rev. Lett. **45** (1980) 1352
[440] I. I. Mazin, O. K. Andersen, O. Jepsen, A. A. Golubov , O. V. Dolgov, J. Kortus, Phys. Rev. B **69** (2004) 056501
[441] H. J. Choi, D. Roundy, H. Sun, M. L. Cohen, S. G. Louie, Phys. Rev. B **69** (2004) 056502
[442] M. Naito, K. Ueda, Supercond. Sci. Technol. **17** (2004) R1
[443] W. Goldacker, this volume
[444] X. X. Xi, A. V. Pogrebnyakov, X. H. Zeng, J. M. Redwing, S. Y. Xu, Qi Li, Zi-Kui Liu, J. Lettieri, V. Vaithyanathan, D. G. Schlom, H. M. Christen, H. Y. Zhai, A. Goyal, Supercond. Sci. Technol. **17** (2004) S196

**ACKNOWLEDGEMENT**

Besides many others who helped us in the preparation of this overview we would like to thank Robert Eder, Christoph Meingast, Lothar Pintschovius, Christof Schneider and Stefan Schuppler for their critical reading of the manuscript and for many helpful comments.